\pdfoutput=1
\documentclass{aastex61}
\usepackage[T1]{fontenc}
\usepackage{ae,aecompl}
\usepackage{graphicx}
\usepackage[caption=false]{subfig}
\usepackage{epsf}
\usepackage{bm}
\usepackage{rotating}
\usepackage{color}
\usepackage{amsmath}

\usepackage{booktabs}
\usepackage{braket}
\usepackage{natbib}
\begin{document}
\title{The bispectrum of polarized galactic foregrounds}

\correspondingauthor{William R. Coulton}
\email{wcoulton@ast.cam.ac.uk}
\author{William R. Coulton}
\affiliation{Institute of Astronomy and Kavli Institute for Cosmology Cambridge, Madingley Road, Cambridge, CB3 0HA, UK}
\author{David N. Spergel}
\affiliation{Department of Astrophysical Sciences, Princeton University,  Peyton Hall, Princeton, NJ 08544, USA}
\affiliation{Center for Computational Astrophysics, Flatiron Institute,162 5th Avenue, 10010, New York, NY, USA}

\begin{abstract}
Understanding the properties of the galactic emission at millimetre wavelengths is important for studies of the cosmic microwave background (CMB). In this work we explore the bispectrum, the harmonic equivalent of the three point function, from galactic dust and synchrotron emission. We investigate these effects across a broad range of frequencies using the synchrotron dominated S-band Polarization All Sky Survey (SPASS) maps at 2.3 GHz, the \textit{Planck} satellite maps (30-857 GHz) and dust dominated Infrared Astronomical Satellite (IRAS) maps at 3 THz. We measure bispectra of total intensity fields, T, as well as the gradient, E, and curl modes, B, of the polarization field. We find that the synchrotron and galactic dust have strong temperature bispectra with significant contributions in the squeezed limit, which probes the correlations between two small scale modes with a large scale mode. Additionally, we find that the dust also has strong polarised bispectra that also peak in the squeezed configuration. We explore parity odd bispectra, such as BTT bispectra, and find strong parity odd bispectra for the galactic dust notably in BTT, BTE and BEE configurations. After masking bright sources, we find no evidence for polarised synchrotron bispectra and no evidence for cross bispectra between the dust and synchrotron emission. The strong foreground bispectra discussed here need to be carefully controlled to avoid biasing measurements of primordial non-Gaussianity. Finally we use these bispectra tools to test for residual foregrounds in the component separated \textit{Planck} maps and find no evidence of residual foregrounds. These tools will be useful for characterizing residual foregrounds in component separated maps, particularly for experiments with less frequency coverage than the \textit{Planck satellite}. 
 \end{abstract}
\section{Introduction}
One of the main goals of current and up-coming cosmic microwave background (CMB) surveys is to detect the imprint of primordial tensor modes. Primordial tensor modes are theorised to have been generated in the early universe during inflation \citep{Grishchuk1975,Starobinskij1979,Rubakov1982} and these tensor modes then leave an imprint on the CMB at the surface of last scattering \citep{Fabbri1983,Abbott1984}. Their contribution to the temperature anisotropies (T mode) and curl-free component of the polarised CMB (E mode) has found to be masked by the dominant scalar modes \citep{Spergel2007}. Instead cosmologist seek to measure the tensor modes by measuring the curl component of the polarised CMB (B mode) \citep{Seljak1997,Seljak1997b}. Scalar modes do not produce B mode polarisation and so the detection of the B modes from the surface of last scattering would be strong evidence for primordial tensor modes \citep{Polnarev1985,Zaldarriaga1997,Kamionkowski1997} .

Whilst the level of the primordial B mode signal is unknown, the bounds from current experiments mean that it is unlikely that primordial signals will be the dominant sky signal at any frequency \citep{BICEP22018}. Currently there are four known sources of signal which could mask the primordial signal. These foregrounds are: polarised dust emission, polarised synchrotron emission, gravitational lensing and patchy tau. First polarized dust emission arises from asymmetric dust grains that are aligned with the galactic magnetic field. The dust emission can be described by a modified black body
\begin{equation}
I_{\mathrm{dust}} \propto {\nu}^{\beta_{\mathrm{dust}}} B_\nu(T_{\mathrm{dust}}),
\end{equation}
where $B_\nu(T)$ is the \textit{Planck} function, $T_{\mathrm{dust}}$ is the dust temperature (and the \textit{Planck} best fit value is $T_{\mathrm{dust}} =19.6 K$), and $\beta_{\mathrm{dust}}$ is the dust spectral index, which is found to be $\beta_{\mathrm{dust}} =1.48\pm 0.01$ for temperature and $\beta_{\mathrm{dust}} =1.53 \pm 0.02$ for polarization \citep{Planck2018LIV}.  Dust emision is the dominant foreground at frequencies above $\sim 150$ GHz \citep{Draine2004,Draine2009}. Second, polarised synchrotron emission arises from electrons spiralling in the galactic magnetic field and is dominant at lower frequencies \citep{Kogut2007}. At first order the Galactic synchrotron radiation's spectral behaviour can be described by a power law
\begin{equation}
I_{\mathrm{sync}} \propto {\nu}^{\beta_{\mathrm{sync}}}
\end{equation}
where $\beta_{\mathrm{sync}}$ is the sky mean spectral index and the current best fit values are  $-3.09 \pm 0.05$ at 23 GHz \citep{Kogut2012} for temperature and $\beta_s=-3.13 \pm 0.13$ for polarisation \citep{Gold2011,Planck2018LIV}. However, there is strong evidence that the synchrotron spectral index, for both polarised and temperature emission, varies across the sky \citep{MivilleDeschenes2008,Fuskeland2014,Jew2017,Jew2019,Krachmalnicoff2018} and shows spectral steepening \citep{Kogut2012,Dickinson2018}. Third, gravitational lensing B modes are generated as light propagates from the surface of last scattering to the observer. The light is lensed by intervening matter and the lensing shears the E mode signal to B mode signal \citep{Seljak1996,Blanchard1987}. The final contribution, patchy tau, arises due to inhomegenous reionization, where spatial variation in the optical depth generates T, E and B mode anisotropies \citep{Weller1999,Hu2000}. However as this effect is small compared to current and upcoming B mode bounds \citep{Roy2018,Namikawa2018}, we ignore this effect in this work.
 
The primordial B mode signal distinguishes itself from these other sources as it has a blackbody spectrum with temperature $2.726 \pm0.010 $ K \citep{Mather1994} and is thought to have highly Gaussian fluctuations. Dust and synchrotron emission have distinctly different spectra and whilst gravitational lensing signals have an identical spectra, they are highly non-Gaussian. Many different approaches to foreground cleaning have been proposed, see \citet{Ichiki2014} for a review of some of these approaches; several of which have successfully been applied to \textit{Planck} data \citep{planck2014-a11,planck2013-p06}. Methods to remove lensing, known as delensing, are described in \citet{Seljak2004,Sehgal2017}. The properties of galactic foregrounds are less well constrained than the lensing foregrounds and are the focus of this work. 

To remove foregrounds with high precision the foregrounds need to be accurately characterised. Recent measurements have started to constrain the spatial and spectral properties of the dust and synchrotron. Planck measurements have characterised the distribution of polarisation fractions, their correlation with intensity measurements and studied how these properties related to the galactic magnetic field \citep{planck2014-XIX}. Further they found that the galactic dust E mode power has roughly a factor of two more power than the B mode and that the dust T-B power-spectrum is non-zero \citep{Planck2018LIV}. Work by \citet{planck2014-XIX,Rotti2016} have shown that the dust polarised emission varies significantly across the sky. Outside the galactic plane polarised synchrotron emission is mainly from large filaments with high polarisation fractions and these have been extensively studied \citep[e.g.][]{Kogut2007,planck2014-a31} including in the context of CMB B mode foregrounds \citep{Vidal2015}.  Building on the work of \citet{Kogut2007,Page2007}, \citet{Choi2015} explored the correlation between the synchrotron emission and the dust emission. Most recently \citet{Rana2018} explored the properties of the temperature bispectrum of synchrotron emission at 408 MHz. 

Complimenting these measurements there has been extensive theoretical work on the properties of the foregrounds. Due to shocks and anisotropic motions in the interstellar medium, it is expected that the dust properties will be highly non-Gaussian and anisotropic \citep[see e.g.][for a review]{Elmegreen2004}. Theoretical models \citep{Caldwell2017} and numerical magnetohydrodynamics \citep{Kritsuk2017,Vansyngel2017,Kandel2018}  have explored the physics behind the observed  ratio of dust EE to BB power. Work by \citet{Kamionkowski2014} suggested that the anisotropy of the galactic dust could be in the form of local hexadecapolar type deviations.  This was further explored in \citet{Philcox2018}, where this anisotropy was measured in simulations and the detectability of the local hexadecapole with future surveys was discussed.

Once foreground cleaned CMB maps are obtained tools are need to validate that any remaining signal is not residual foregrounds. The importance of validating cleaned maps was highlighted in recent work \citep[e.g.][]{Madhavacheril2018} that found the \textit{Planck} foreground cleaned temperature maps contain significant residual thermal Sunyaev Zel'dovich signal. Traditional approaches rely on cross correlations with foreground dominated maps to constrain residual foregrounds. Whilst cross correlations are a very useful tool, alone they may not be sufficient to validate a potential primordial signal, particularly if the power spectra have been used during the foreground removal process or if foregrounds de-correlate with frequency, though there is currently no evidence to show that they do \citep{Sheehy2018}.   \citet{Kamionkowski2014} and  \citet{Rotti2016} suggested using measurements of the anisotropy to characterise remaining foregrounds, as the primordial signal is expected to be isotropic but residual foregrounds are not. Recently \citet{vonHausegger2018} explored how skewness and kurtosis measurements can be used to constrain the residual foreground contamination. In \citet{planck2014-a18} various additional methods (skewness, Minkowski functionals and N-point functions) were used to search for deviations from statistical isotropy and homogeneity, which could be indicative of residual foregrounds.

Our work explores how the bispectrum can be used to characterise the foregrounds and test for residual signals, exploiting the fact that the primordial signal is expected to be highly Gaussian. The bispectrum is the harmonic equivalent of the three point function and vanishes for Gaussian signals. The bispectrum is the lowest order non-Gaussian statistic. We first seek to characterise the bispectrum of the galactic foregrounds using data from the \textit{Planck satellite}. Then we demonstrate how measurements of the bispectrum can be used to test for residual foregrounds in foreground-cleaned maps. This work builds on previous bispectrum measurements of galactic foregrounds \citep{Komatsu2002, Renzi2012}, which used bispectrum measurements of foregrounds to avoid biases in primordial non-Gaussianity estimators. Our examination of the bispectrum extends the work in \citet{planck2014-a18} by looking at the full bispectrum (as opposed to two subsets of the three point function), by including polarized maps and by correlating cleaned maps with foreground dominated maps. Understanding the bispectrum from galactic foregrounds, and especially any residual foregrounds in cleaned maps, is very important for constraints on primordial non-Gaussianity. As this work was being prepared for publication, \citet{Jung2018} published their work,  in which they examine the temperature bispectrum of galactic foregrounds and show that these foregrounds can significantly bias  primordial non-Gaussianity measurements. Our work is highly complementary to theirs, with this work including polarized bispectra and discussing their use in verifying cleaned maps.

We use a binned bispectrum approach \citep{Bucher2010,Bucher2016}. The binned bispectrum is blind approach that requires no theoretical model of the signal and can constrain non-smooth signals. The cost of this broad approach is that it can be less optimal constraints that more targeted approaches. The details of our implementation of the binned bispectrum are described in Section \ref{sec:binnedEstimators}. In Section \ref{sec:dataSets} we describe the data sets used in this work and briefly outline our analysis pipeline.  Our results are presented in Section \ref{sec:foregroundBispectra} and then are discussed, along with our conclusions, in Section \ref{sec:conclusions}. In Appendix \ref{app:EstError} we discuss the details of the estimator variance and elaborate on some of the analysis choices used in this work. 

\section{Bispectrum Estimator}\label{sec:binnedEstimators}
In this section we will briefly overview the binned bispectrum estimator for the parity even and odd cases. For both estimators we must decompose the maps into spherical harmonic components. The measurements of the stokes I component, $\Delta T(\mathbf{n})$, are decomposed using spin-zero harmonics as
\begin{equation}
\Delta T(\mathbf{n})=\sum\limits_{\ell,m} a_{\ell,m} \mathrm{Y}_{\ell,m}(\mathbf{n}) .
\end{equation}
When no mask is applied, the stokes Q and U polarisation components can be decomposed into spin two spherical harmonics
\begin{align}
(Q\pm iU) (\mathbf{n})= \sum\limits_{\ell,m} a_{\pm2,\ell,m} {}_{\pm2}Y_{\ell,m}(\mathbf{n}),
\end{align}
which can then be further decomposed into two scalar fields
\begin{align}
a_{E,\ell,m}=-(a_{2,\ell,m}+a_{-2,\ell,m})/2,\\
a_{B,\ell,m}=i(a_{2,\ell,m}-a_{-2,\ell,m})/2.
\end{align}
When a sky mask is applied the decomposition mixes E and B modes together. In this work we use pure E and B estimators, as described in \citet{Smith2007} and \citet{Grain2012}, to obtain maps which are free of this leakage. 

\subsection{Parity Even Estimator} \label{sec:parityEvenEst}
The bispectrum is given by the ensemble average of three spherical harmonic coefficients
\begin{align}
\langle a^{X}_{\ell_1,m_1}  a^{Y}_{\ell_2,m_2}  a^{Z}_{\ell_3,m_3} \rangle = B^{X,Y,Z}(\ell_1,m_1,\ell_2,m_2,\ell_3,m_3).
\end{align}
where $B^{X,Y,Z}(\ell_1,m_1,\ell_2,m_2,\ell_3,m_3) $ is the bispectrum between maps $X$,$Y$ and $Z$. Under the assumptions that this signal is statistically homogeneous and isotropic, the bispectrum can be decomposed as  \citep{Komatsu2001}
\begin{equation}
\langle a^{X}_{\ell_1,m_1}  a^{Y}_{\ell_2,m_2}  a^{Z}_{\ell_3,m_3} \rangle = \sqrt{\frac{(2\ell_1+1)(2\ell_2+1) (2\ell_3+1) }{4 \pi}}  \begin{pmatrix}
    \ell_1 & \ell_2 & \ell_3 \\
    m_1 & m_2 & m_3
  \end{pmatrix} \begin{pmatrix}
    \ell_1 & \ell_2 & \ell_3 \\
    0 & 0 & 0
  \end{pmatrix}
 b^{X,Y,Z}_{\ell_1,\ell_2,\ell_3},
\end{equation}
where the terms in brackets are Wigner 3j symbols and $ b_{\ell_1,\ell_2,\ell_3}$ is the reduced bispectrum. The Wigner 3j symbol is related to the Clebcsh-Gordan coefficients that describe the coupling of two angular momenta. The Wigner 3j symbol vanishes unless the triangle conditions are satisfied. They require that
\begin{align}
| \ell_a-\ell_b| \leq \ell_c \leq \ell_a + \ell_b .
\end{align}
This estimator is zero unless $\ell_1+\ell_2+\ell_3=\mathrm{even}$ and so is only sensitive to even parity configurations. For this reason we will refer to this estimator as the parity-even estimator. The foregrounds are neither isotropic nor homogeneous and by focusing on the reduced bispectrum we lose any anisotropic information (this is also an issue for most power spectra approaches). It is left to future work to explore the anisotropic contributions to the bispectra. 

We use a binned bispectrum method similar to that described in \citet{Bucher2016}. Using the fact that the Wigner 3j symbols can be evaluated using the Gaunt integral
\begin{align}
\mathcal{G}^{m_1,m_2,m_3}_{\ell_1,\ell_2,\ell_3}& = \int\mathrm{d}^2\mathbf{n} \mathrm{Y}_{\ell_1,m_1}(\mathbf{n}) \mathrm{Y}_{\ell_2,m_2}(\mathbf{n})  \mathrm{Y}_{\ell_3,m_3}(\mathbf{n})  \nonumber \\ &= \sqrt{\frac{(2\ell_1+1)(2\ell_2+1) (2\ell_3+1) }{4 \pi}}  \begin{pmatrix}
    \ell_1 & \ell_2 & \ell_3 \\
    m_1 & m_2 & m_3
  \end{pmatrix} \begin{pmatrix}
    \ell_1 & \ell_2 & \ell_3 \\
    0 & 0 & 0
  \end{pmatrix}.
\end{align}
we can estimate the reduced bispectrum as
\begin{align}\label{eq:reducedEstimator}
\hat{b}^{X,Y,Z}_{\ell_1,\ell_2,\ell_3}= \frac{1}{N_{\ell_1,\ell_2,\ell_3}}\sum\limits_{m} \int\mathrm{d}^2\mathbf{n} \mathrm{Y}_{\ell_1,m_1}(\mathbf{n}) \mathrm{Y}_{\ell_2,m_2}(\mathbf{n})  \mathrm{Y}_{\ell_3,m_3}(\mathbf{n}) a^{X}_{\ell_1,m_1}a^{Y}_{\ell_2,m_2}a^{Z}_{\ell_3,m_3},
\end{align}
where $N_{\ell_1,\ell_2,\ell_3}$ is a normalisation constant. The binned estimator is a simple modification of the above formula. The maps are filtered in harmonic space to contain only modes with $\ell$ within the bin. The filtered map with $\ell$ satisfying $\ell_i<\ell\leq \ell_{i+1}$ is denoted as $W^X_{i}(\mathbf{n})$ and is given explicitly by
\begin{equation}
W^X_{i}(\mathbf{n}) = \sum\limits_{\ell_i<\ell\leq \ell_{i+1}} \sum\limits_{m}{Y}_{\ell,m}(\mathbf{n})a^X_{\ell,m}.
\end{equation}
The binned bispectrum estimator is then given by
\begin{equation}\label{eq:binnedParityEven}
\hat{b}^{X,Y,Z}_{i,j,k}= \frac{1}{N'_{i,j,k}}  W^X_{i}(\mathbf{n}) W^Y_{j}(\mathbf{n}) W^Z_{k}(\mathbf{n}) .
\end{equation}
The estimator normalisation is given by 
\begin{align}
N'_{i,j,k}= \sum\limits_{\ell_i<\ell_1\leq \ell_{i+1}}  \sum\limits_{\ell_j<\ell_2\leq \ell_{j+1}}  \sum\limits_{\ell_k<\ell_3\leq \ell_{k+1}}\frac{(2\ell_1+1)(2\ell_2+1) (2\ell_3+1) }{4 \pi}  \begin{pmatrix}
    \ell_1 & \ell_2 & \ell_3 \\
    0 & 0 & 0
  \end{pmatrix} \begin{pmatrix}
    \ell_1 & \ell_2 & \ell_3 \\
    0 & 0 & 0
  \end{pmatrix}.
\end{align}
The Gaussian part of the estimator's variance is given by
{
\begin{align}\label{eq:parityEvenVar}
V^{X,Y,Z,X',Y',Z'}_{i,j,k} = \frac{1}{N'_{i,j,k} N'_{i,j,k}} \sum\limits_{\ell_i<\ell_1\leq \ell_{i+1}}  \sum\limits_{\ell_j<\ell_2\leq \ell_{j+1}}  \sum\limits_{\ell_k<\ell_3\leq \ell_{k+1}}& \frac{(2\ell_1+1)(2\ell_2+1) (2\ell_3+1) }{4 \pi}  \begin{pmatrix}
    \ell_1 & \ell_2 & \ell_3 \\
    0 & 0 & 0
  \end{pmatrix}^2 \nonumber \\ & \times C^{X,X'}_{\ell_1}C^{Y,Y'}_{\ell_2}C^{Z,Z'}_{\ell_3} g_{\ell_1,\ell_2,\ell_3},
\end{align}}
where $C^{X,Y}$ is the power spectrum between map $X$ and map $Y$ and $g_{\ell_1,\ell_2,\ell_3},$ is 6 if all its arguments are equal, 2 if any two of the arguments are equal and 1 if none of them are equal. When masks are applied to the data the variance of the estimator is altered by a factor of $f_{\rm{sky}}$.

\subsection{Parity Odd Estimator}
\citet{Planck2018LIV} recently measured a non-zero BT power spectrum for the dust. This is an odd-parity signal and motivated us to consider odd parity bispectra, those which have  $\ell_1+\ell_2+\ell_3=\mathrm{odd}$. Odd parity bispectra were originally investigated in the context of parity violating inflationary models and searches for primordial magnetic fields \citep{Shiraishi2012,Shiraishi2013}. If the foreground signals are isotropic, homogeneous and parity invariant signals, then we would expect to find only parity even T and E bispectra \citep{Shiraishi2011,Kamionkowski2011}. However, even in the absence of parity violating signals, bispectra involving odd numbers of B modes naturally have odd parity \citep{Meerburg2016}. 

The reduced bispectrum estimator described in Section \ref{sec:parityEvenEst} is only sensitive to bispectra with even parity, those that satisfy $\ell_1+\ell_2+\ell_3 =\mathrm{even}$. To measure parity odd bispectra we introduce a second estimator, hereafter called the parity-odd estimator, based on the work of  \citep{Shiraishi2014,Shiraishi2015}.  For notational convenience we define the following function

\begin{align}
h^{m_1,m_2,m_3}_{\ell_1,\ell_2,\ell_3}&=\sqrt{\frac{(2\ell_1+1)(2\ell_2+1) (2\ell_3+1) }{4 \pi}}  \begin{pmatrix}
    \ell_1 & \ell_2 & \ell_3 \\
    m_1 & m_2 & m_3
  \end{pmatrix} \begin{pmatrix}
    \ell_1 & \ell_2 & \ell_3 \\
    2 & -1 & -1
  \end{pmatrix} \left( 1 -(-1)^{\sum \ell_i}\right) \nonumber \\
  &=\int \mathrm{d}^2\mathbf{n} \left({}_{-2}Y_{\ell_1,m_1}(\mathbf{n}){}_{1}Y_{\ell_2,m_2}(\mathbf{n}){}_{1}Y_{\ell_3,m_3}(\mathbf{n})-{}_{2}Y_{\ell_1,m_1}(\mathbf{n}){}_{-1}Y_{\ell_2,m_2}(\mathbf{n}){}_{-1}Y_{\ell_3,m_3}(\mathbf{n})\right).
\end{align}
We define the odd parity reduced bispectrum as
\begin{align}
\langle a^X_{\ell_1,m_1} a^Y_{\ell_2,m_2} a^Z_{\ell_3,m_3}\rangle =\frac{1}{6}\left( h^{m_1,m_2,m_3}_{\ell_1,\ell_2,\ell_3}+ h^{m_3,m_1,m_2}_{\ell_3,\ell_1,\ell_2} + h^{m_2,m_3,m_1}_{\ell_2,\ell_3\ell_1}  \right) b^{odd,X,Y,Z}_{\ell_1,\ell_2,\ell_3}.
\end{align} 
We note that this definition is not unique \citep[as was discussed in][]{Shiraishi2014} and different definitions would result in scalings of the reduced bispectrum by factors of $\sim \sqrt\ell$. Now we can write down an estimator for the full odd parity bispectrum as
\begin{align}
\hat{b}^{odd,X,Y,Z}_{\ell_1,\ell_2,\ell_3} =\frac{1}{6 N^{o}_{\ell_1,\ell_2,\ell_3}}\left( h^{m_1,m_2,m_3}_{\ell_1,\ell_2,\ell_3}+ h^{m_3,m_1,m_2}_{\ell_3,\ell_1,\ell_2} + h^{m_2,m_3,m_1}_{\ell_2,\ell_3,\ell_1}  \right) a^X_{\ell_1,m_1}a^Y_{\ell_2,m_2}a^Z_{\ell_3,m_3}.
\end{align} 
We symmetrise with respect to $h^{m_1,m_2,m_3}_{\ell_1,\ell_2,\ell_3}$ so that our definition of $\hat{b}^{odd,X,X,X}_{\ell_1,\ell_2,\ell_3}$ is symmetric under the permutation of $\ell_1$, $\ell_2$ and $\ell_3$. Our odd parity estimator vanishes for bispectra that do not satisfy the triangle conditions or have $\ell_1+\ell_2+\ell_3 =\mathrm{even}$. In a similar manner to the parity even bispectrum we define a set of filtered maps. The filtered maps are obtained via a spin weighted spherical harmonic transform and are defined by
\begin{align}
{}_{s}W^X_i=\sum \limits_{\ell_i<\ell<\ell_{i+1}} \sum\limits_{m}{}_{s}Y_{\ell,m}(\mathbf{n})a^X_{\ell,m}.
\end{align}
Then the binned odd parity bispectrum estimator is given by
\begin{align}\label{eq:binnedOddParity}
\hat{b}^{odd,X,Y,Z}_{i,j,k} =\frac{1}{6 N^{o'}_{i,j,k}} \int\mathrm{d}^2\mathbf{n}\left({}_{-2}W^X_i {}_{1}W^Y_j {}_{1}W^Z_k-{}_{2}W^X_i {}_{-1}W^Y_j {}_{-1}W^Z_k+{}_{-2}W^X_k {}_{1}W^Y_i {}_{1}W^Z_j-{}_{2}W^X_k {}_{-1}W^Y_i {}_{-1}W^Z_j+ \right . \nonumber \\ \left . {}_{-2}W^X_j {}_{1}W^Y_k{}_{1}W^Z_i-{}_{2}W^X_j {}_{-1}W^Y_k {}_{-1}W^Z_i  \right).
\end{align}
In analogy to the parity even case, the normalization is 
\begin{align}
 N'^{o}_{i,j,k}=\frac{1}{6} \sum\limits_{\ell_i<\ell_1\leq \ell_{i+1}}  \sum\limits_{\ell_j<\ell_2\leq \ell_{j+1}}  \sum\limits_{\ell_k<\ell_3\leq \ell_{k+1}}\sum_{m_i}\left( h^{m_1,m_2,m_3}_{\ell_1,\ell_2,\ell_3}+ h^{m_3,m_1,m_2}_{\ell_3,\ell_1,\ell_2} + h^{m_2,m_3,m_1}_{\ell_2,\ell_3\ell_1}  \right)\left( h^{m_1,m_2,m_3}_{\ell_1,\ell_2,\ell_3}+ h^{m_3,m_1,m_2}_{\ell_3,\ell_1,\ell_2} + h^{m_2,m_3,m_1}_{\ell_2,\ell_3\ell_1}  \right) ,
\end{align}
and the variance is 
\begin{align}\label{eq:parityOddVar}
{V^o}^{X,Y,Z,X',Y',Z'}_{\ell_1,\ell_2,\ell_3'} = \frac{1}{ N'^{o}_{i,j,k} N'^{o}_{i,j,k}} \sum\limits_{\ell_i<\ell_1\leq \ell_{i+1}}  \sum\limits_{\ell_j<\ell_2\leq \ell_{j+1}}  \sum\limits_{\ell_k<\ell_3\leq \ell_{k+1}}\sum_{m_i} & \left( h^{m_1,m_2,m_3}_{\ell_1,\ell_2,\ell_3}+ h^{m_3,m_1,m_2}_{\ell_3,\ell_1,\ell_2} + h^{m_2,m_3,m_1}_{\ell_2,\ell_3\ell_1}  \right)^2 \nonumber \\ & \times C^{X,X'}_{\ell_1}C^{Y,Y'}_{\ell_2}C^{Z,Z'}_{\ell_3} g_{\ell_1,\ell_2,\ell_3}.
\end{align}

\subsection{Linear Term}\label{sec:linearTerm}
The variance of the estimators defined above are only correct in the case that the data is homogenous and isotropic. In the case of a real analysis this condition is broken by the detector noise, masking and the galactic foreground signals. To account for these effects the estimator must be altered to include a linear correction term.  As this term has been discussed thoroughly in the literature \citet{Creminelli2006,Yadav2008,Bucher2016} here we just summarise the results.
The parity even estimator given in Eq. \ref{eq:binnedParityEven} is altered to
\begin{align}
\hat{b}^{X,Y,Z}_{i,j,k}= \frac{1}{N'_{i,j,k}}\int\mathrm{d}^2\mathbf{n}  W^X_{i}(\mathbf{n}) W^Y_{j}(\mathbf{n}) W^Z_{k}(\mathbf{n})-W^X_{i}(\mathbf{n}) \langle {W^{(G)}}^Y_{j}(\mathbf{n}) {W^{(G)}}^Z_{k}(\mathbf{n}) \rangle \nonumber \\ -W^Y_{j}(\mathbf{n}) \langle {W^{(G)}}^X_{i}(\mathbf{n}) {W^{(G)}}^Z_{k}(\mathbf{n}) \rangle -W^Z_{k}(\mathbf{n}) \langle {W^{(G)}}^X_{i}(\mathbf{n}) {W^{(G)}}^Y_{j}(\mathbf{n}) \rangle,
\end{align}
and the parity odd estimator given in Eq. \ref{eq:binnedOddParity} is altered in an identical manner.

Computation of the linear term requires a large number of simulations. In this work we explore many different configurations of maps and this would require a very large number of simulations. However, the cuts we use, which are described in Section \ref{sec:pipeline}, mean that the linear term from the noise and mask cuts has only a small effect on our estimator, decreasing the variance by $\sim 10 \%$.  This is because the linear term from the mask an anisotropic noise most significantly affect squeezed configurations \citep{Creminelli2006,Bucher2016} and these configurations are masked with our low $\ell$ cut. For this reason we do not calculate the linear term for the results discussed below. There are two an important caveats here. Firstly the foregrounds are anisotropic and their potential impact on the linear term is discussed in Section \ref{sec:signalInhom}. Secondly, we tested the importance of the linear term using \textit{Planck} single frequency sky simulations \citep{planck2014-a14} and using component separate simulations. The component separation method used was the Spectral Matching Independent Component Analysis SMICA method. The SMICA technique is a component separation method that is capable of minimising the variance of a linear combination of frequency channels in the spatial, harmonic, or hybrid domains. In extends beyond a standard ILC by fitting a model for the frequency map weights (as opposed to using the observed covariance). However in our analysis we also use other component separated maps. The noise in these maps is likely very different from the single frequency and SMICA maps and without noise simulations from the component separation pipelines we are unable to accurately asses the importance of this noise. With that said, all the maps used here passed our null test and we are satisfied that the linear term for our configurations can be ignored even for these caveated cases, though in future work will further investigate this.

\section{Data Sets and Pipeline}\label{sec:dataSets}
\subsection{Data sets}
For this work we used data from the Planck 2015 data release \citep{planck2014-a12,planck2014-a11}. For our fiducial studies of the foregrounds we use the Commander synchrotron map and the 353 GHz map as a tracer of the dust. For a detailed description of the Commander component separation method we refer the reader to \cite{eriksen2006,eriksen2008} and here we very briefly overview the method. Commander is a Bayesian, parametric map-based method which separates a set of different frequency maps into their components.  The map components are modelled by an amplitude for each pixel and a spectral parameterisation. The amplitudes and spectral indexes in each pixel are then obtained by an MCMC Gibbs sampling algorithm. The Commander method uses temperature data from the \textit{Planck} Satellite, WMAP and a reprocessed version of the 408 MHz radio continuum all-sky map HASLAM map and only \textit{Planck} data for the polarization maps \citep{haslam1982,bennett2012,Remazeilles2015}.  The effective beam beam full width half-maximum (FWHM )of the 353 GHz maps is 4.86 arcmin.

We also investigate the frequency dependence of the dust by investigating correlations between the \textit{Planck} 353 GHz and the \textit{Planck} Collaboration's reprocessing of the IRAS and IRIS maps \citep{planck2013-p06b}, hereafter we refer to this as the IRIS map. This temperature map is obtain from combining the processing of the Infrared Astronomical Satellite (IRAS) and COBE/DIRBE from \citet{Schlegel1998} with a reprocessing of this data-set by \citet{Miville2005}. This combined map provides the best measure of the galactic dust at $100 \mu$m by utilizing more accurate large scale features from the \citet{Schlegel1998} with the high resolution small scale modes from \citet{Miville2005}. The effective beam FWHM of the IRIS map is 4.3 arcmin.

For our synchrotron analysis we used the data from the S-band Polarization All Sky Survey (S-PASS) \citep{Carretti2019}. The SPASS observed the southern sky at Dec$ <-1^\circ$ at 2.3 GHz in temperature and polarization. This survey has low noise (with a signal-to-noise ratio larger than 3 on around 94$\%$ of the Q and U pixels) and high resolution, with a 8.9 arcmin  FWHM, and so provides a matching data-set to the dust study.  After masking the galaxy we can only probe a relatively small fraction of the sky and so we complement this analysis with the higher noise and lower resolution Commander maximum likelihood synchrotron maps. The Commander temperature map has a reference frequency of 408 MHz and is smoothed such that the effective beam (FWHM) is 60 arcmin, and the Commander synchrotron polarization maps have reference frequencies of 30GHz and are smoothed to have an effective FWHM of 40 arcmin.

In Section \ref{sec:cleanedBispectra}, where we investigate cleaned CMB maps, we used maps produced by \textit{Planck}'s four different cleaning methods: Commander, SMICA,  Spectral Estimation Via Expectation Minimization (SEVEM) and Needlet Internal Linear Combination (NILC). The Needlet method works in the needlet domain to reconstruct the CMB anisotropies, whilst minimising the variance, and the SEVEM method is a template fitting method that removes foreground templates to construct a CMB map, with the templates constructed from combinations of the individual frequency channels. These four methods approach the task of foreground cleaning from four different directions that range from data orientated, minimal assumption methods, such as NILC and SEVEM, to parametric methods, such as Commander. For the purpose of this work it is sufficient to note that the different assumptions, approaches and spaces (i.e. wavelet, pixel and harmonic space) in which these methods work mean that these methods are a good representation of the different cleaning methods available in the literature. For more detailed information on the foreground separation methods we refer the reader to \citet{planck2014-a12} for a general overview, \citet{delabrouille2009} for NILC, \citet{fernandez2012} for SEVEM and \citet{cardoso2008}  for SMICA.

\subsection{Pipeline}\label{sec:pipeline}
Before estimating the bispectrum from the maps we preform several pre-processing steps. Firstly we apply a galactic sky mask. Our fiducial mask is the $60\%$ sky mask provided by the \textit{Planck} experiment but in Section \ref{sec:foregroundBispectra} we investigate the effect of the sky cut by using the $20\%$, $40\%$ and $60\%$ \textit{Planck} masks \citep{planck2014-a01}.  The mask is apodized with a $3$ degree Gaussian to reduce the effect of mode coupling. We apply the mask as without it our results would be entirely dominated by the bright signal from the galactic plane and because we are interested in characterizing the foregrounds in sky regions of interest to cosmology. Next we apply the \textit{Planck} point source mask and a second galactic mask that masks the brightest $2\%$ of the sky within in the 60\% sky mask. We discuss the motivation behind this mask in Section \ref{sec:signalInhom}. We then iteratively infill these masks to apodize them. \citet{Gruetjen2015}  showed that infilling point sources allowed nearly optimal (i.e. minimum variance) efficient bispectrum estimation, see \citet{Bucher2016} for a more detailed discussion of the effect of infilling on the binned bispectrum. We then use the libsharp \citep{libsharp} and HEALPIX \citep{gorski2005} packages combined with the pure E/B decomposition method from \citet{Grain2012}  to obtain the spherical harmonic coefficients. 
Finally as the maps released as part of the \textit{Planck} 2015 data release have had the modes with $\ell<30$ in polarization masked due to residual systematics \citep{planck2014-a08}, we filter our maps for both polarization and temperature to remove all modes with $\ell<40$. 

The error bars on our measurements are computed using Gaussian simulations that have processed with the same pipeline, including same masks and cuts. These error-bars show good agreement with error-bars computed using Eqs. \ref{eq:parityEvenVar} and \ref{eq:parityOddVar}. We validated this procedure with two tests; firstly we applied our method to a set of end-to-end noise simulations from the \textit{Planck} satellite, secondly we examined bispectra constructed from the difference of the half ring maps; these maps should have the same variance properties as the data without any bispectrum signal.

\subsection{Signal Inhomogeneity and non-Gaussianity} \label{sec:signalInhom}
We have made three suboptimal analysis choices: our low $\ell$ cut in temperature, the second dust mask described in Section \ref{sec:pipeline} and choosing to use the 353 GHz maps rather than the Commander dust map, which should be a cleaner map of the dust. These choices were necessary as without them we found we were unable to accurately model the estimator's variance. The dust is highly anisotropic and highly non-Gaussian and we found without these cuts the estimator variance was larger than the Gaussian prediction. There are two possible causes, firstly as the dust is anisotropic it could contribute to the linear term, see Section \ref{sec:linearTerm}. Currently we are not able to calculate the dust's contributions to the linear term as this requires simulating the anisotropic distribution of the dust. The second possibility is that, as the dust is highly non-Gaussian, there could be significant contributions to the variance from non-Gaussian contributions to the six point function, which we are currently unable to model.  Accurately calculating the linear term and non-Gaussian variance is the subject of on-going work and is necessary to obtain optimal results. The 2\% dust mask was chosen as it was found that with this mask, and the low $\ell$ cut, we can model the variance of our estimator, using the empirically measured power spectra, to within $\sim10\%$. We examine this in more detail, and present a potential alternative to these cuts, in Appendix \ref{app:EstError}. 
Similarly we found that the high level of non-Gaussianity in the synchrotron temperature map results in a significantly larger estimator variance than expected. We found this was not easily suppressed via masking or $\ell$ cuts and so we opt to only use configurations involving one temperature leg (with the exception of the TTT bispectrum) as these configurations can be well modelled.

\section{Foreground Bispectra}\label{sec:foregroundBispectra}

\begin{figure}
\begin{minipage}{.47\textwidth}
  \centering
    \includegraphics[width=.95\textwidth]{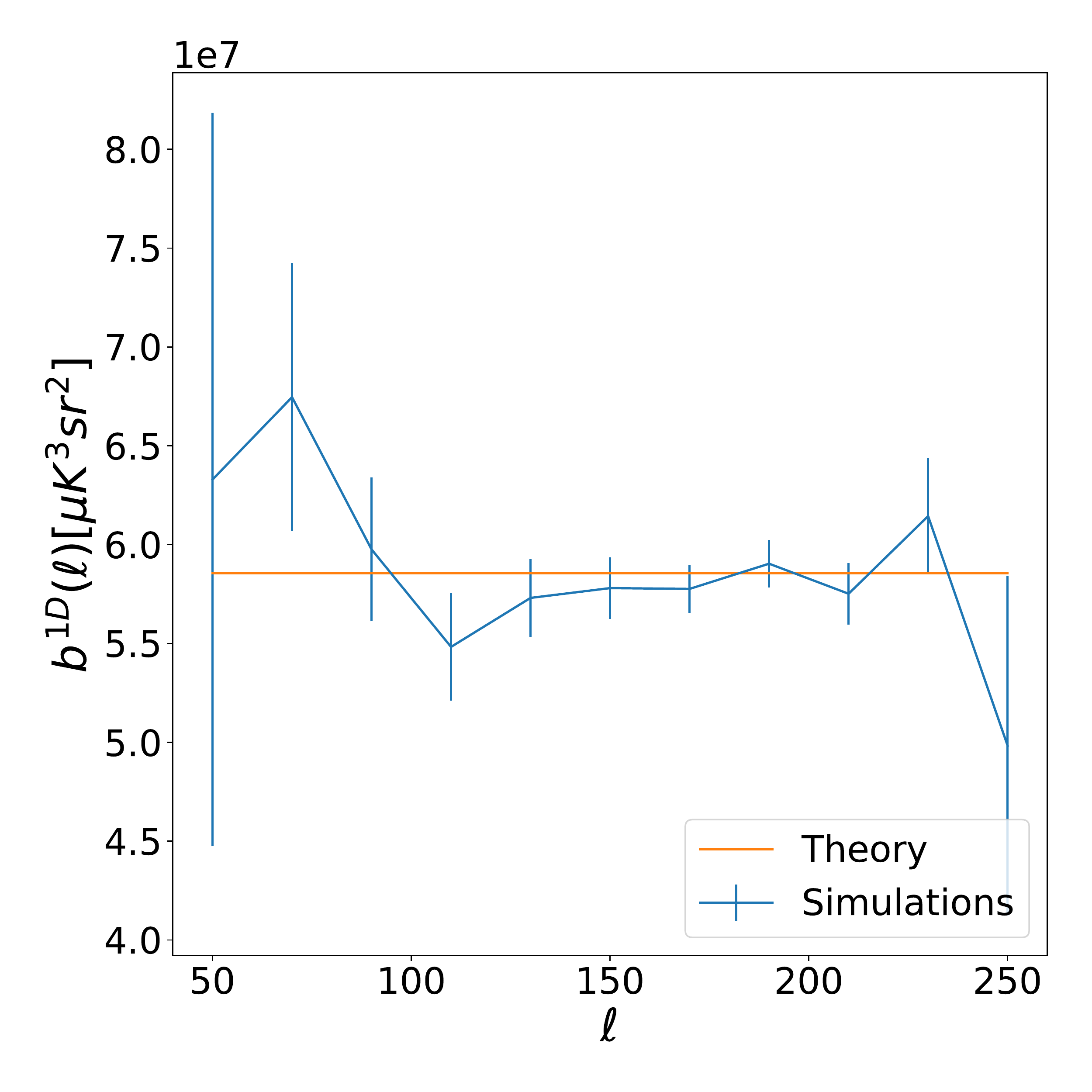}
    \caption{The average 1-D bispectrum of 100 realizations of a Poisson random field compared with the theoretical value.}
    \label{fig:1DPois}
\end{minipage}%
    \qquad
\begin{minipage}{.47\textwidth}
    \centering
    \includegraphics[width=.95\textwidth]{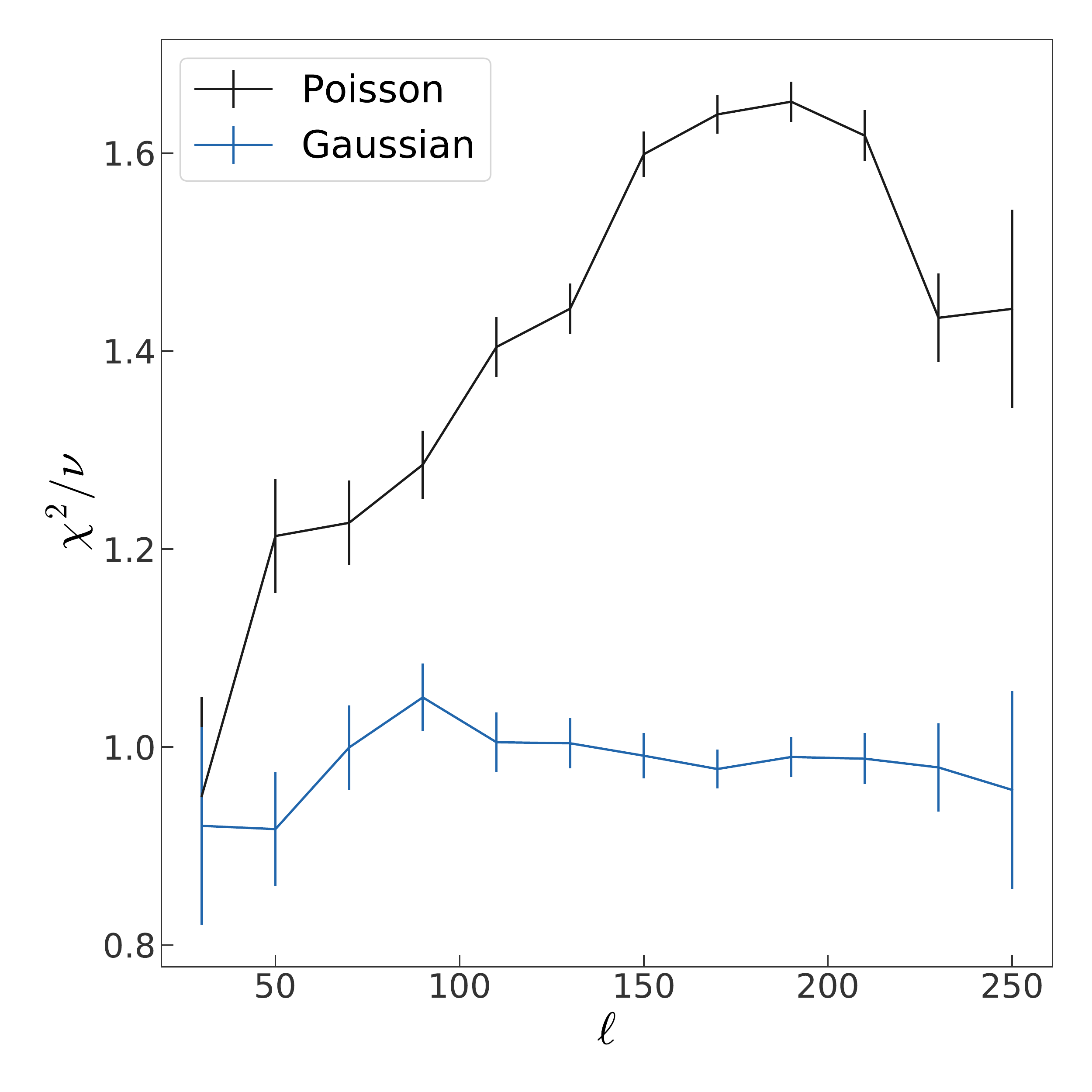}
\caption{The average 1-D chi-squared of 100 realizations of a Poisson noise and 100 realizations of a Gaussian CMB.}
    \label{fig:1DPoisGaus_chisquared}
\end{minipage}%
\end{figure}

\begin{figure}
  \centering
    \includegraphics[width=.45\textwidth]{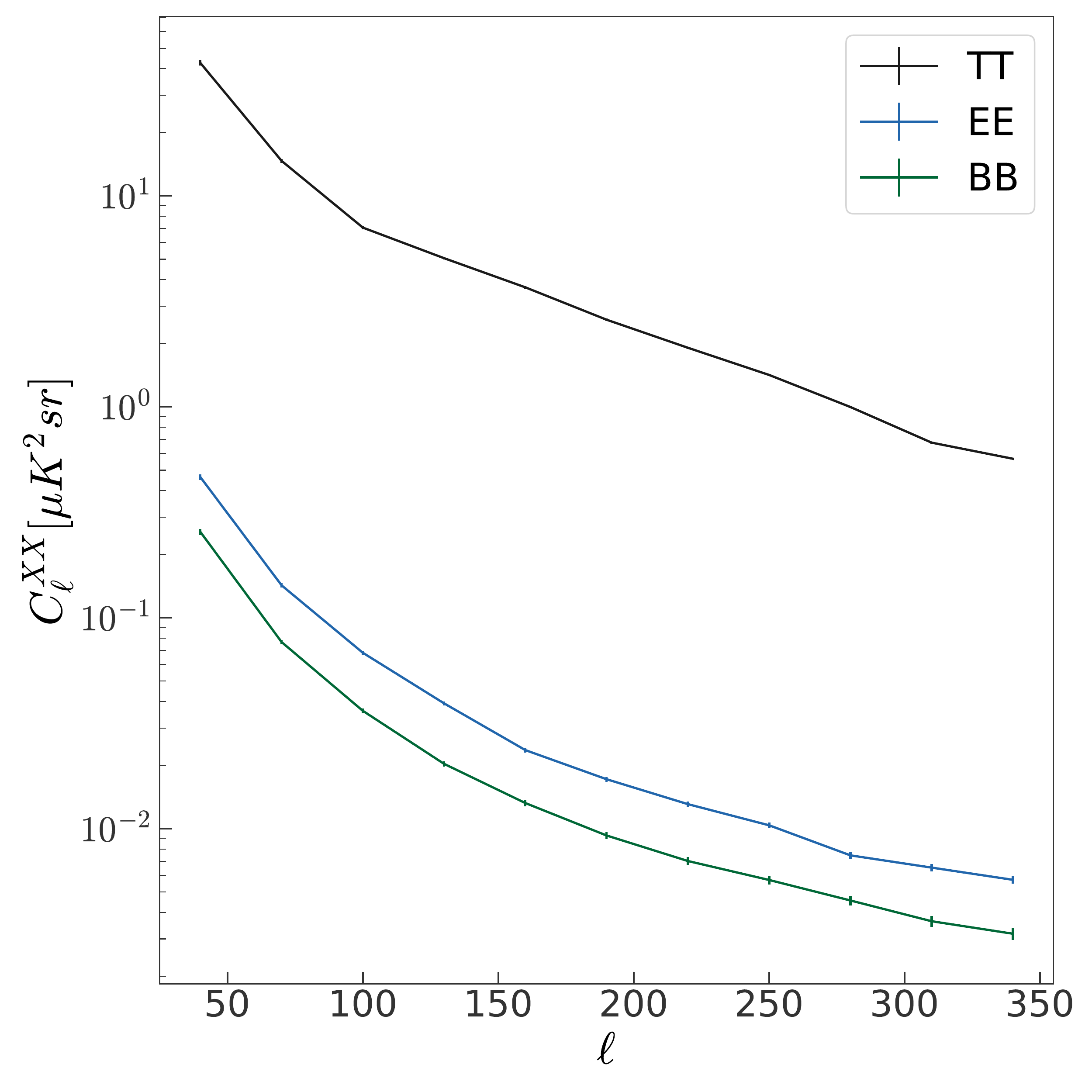}
\caption{The 353 GHz TT, EE and BB pseudo-Cl power-spectrum obtained from $60\%$ of the sky.. }
    \label{fig:353_powspec_gm4}
\end{figure}

The bispectrum is a three dimensional function and for visualisation we use two projections of the data for ease of visualisation.  The first is the two dimensional bispectrum defined as
\begin{align}\label{eq:2dBispectrum}
\hat{b}^{2D:\, X,(Y,Z)}_{i,I}=\frac{1}{N}\sum\limits_{2\ell_{I}^2<\ell_j^2+\ell_k^2 \leq2\ell^2_{I+1}} \hat{b}^{X,Y,Z}_{i,j,k}
\end{align}
and its associated two dimensional chi-squared
\begin{align}
{\chi^2}^{2D:\,X,(Y,Z)}_{i,I}=\frac{1}{N}\sum\limits_{2\ell_{I}^2<\ell_j^2+\ell_k^2 \leq2\ell^2_{I+1}}\hat{b}^{X,Y,Z}_{i,j,k}(V^{X,Y,Z,X,Y,Z})^{-1}_{i,j,k}\hat{b}^{X,Y,Z}_{i,j,k},
\end{align}
where $\ell_i$ is the $\ell$ of the centre of the i$^{th}$ bin and N is the number of modes in each bin. In our notation the capital subscript indices indicate the scale of the averaged leg and the superscript indices in brackets $(Y,Z)$ indicate which fields have been average over. The two dimensional bispectrum enables us to explore some of the shape dependence of the bispectra. In this work we will consider many different bispectra and to compress the data further we use a second projection, to a one dimensional bispectrum, to allow an overview of the many bispectra.
The one dimensional bispectrum defined as
\begin{align}
\hat{b}^{1D:\, X,Y,Z}_I=\frac{1}{N} \sum\limits_{3\ell_{I}^2<\ell_i^2+\ell_j^2+\ell_k^2\leq3\ell^2_{I+1}} \hat{b}^{X,Y,Z}_{i,j,k},
\end{align}
and the one dimensional chi-squared
\begin{align}
{\chi^2}^{1D:\,X,Y,Z}_I=\frac{1}{N}\sum\limits_{3\ell_{I}^2<\ell_i^2+\ell_j^2+\ell_k^2\leq3\ell_{I+1} }\hat{b}^{X,Y,Z}_{i,j,k}(V^{X,Y,Z,X,Y,Z})^{-1}_{i,j,k}\hat{b}^{X,Y,Z}_{i,j,k}.
\end{align}
While the 1-D bispectrum is useful for visualisation purposes, it can obscure the signal as the variance for each bin is set by the noisiest bin included in the sum. The one dimensional  $\chi^2$ is based on the Gaussian covariance and assumes that the bispectra bins are independent. The application of a mask introduces mode coupling between the bispectra bins violating the independence assumptions, however we choose the width of our bins such that the bin to bin coupling constrained to be $<5\%$ such that this assumption is approximately true. For the largest mask, $f_{\mathrm{sky}}=60\%$, we use bin widths starting from $\Delta \ell=35$ and for the smallest mask, $f_{\mathrm{sky}}=20\%$, we use bin widths starting  from $\Delta \ell=45$. Note that we deconvolve the beams for all the bispectra presented in this work.

Before presenting our results, we first present a simple application of the 1-D statistics presented above in order to aid the interpretation of our results and as a simple verification of our pipieline. We apply these 1-D statistics to a set of 100 Gaussian realizations of the CMB and 100 maps with Gaussian CMB and with Poisson point sources. In Figure \ref{fig:1DPois} we plot the average 1D bispectrum from the 100 Poisson maps and  we plot the theoretical expectation (which is just a constant). In Figure \ref{fig:1DPoisGaus_chisquared} we plot the average 1D chi-squared for the 100 Poisson maps and for the 100 Gaussian maps. As expected the Gaussian maps $\chi^2$ is consistent with 1 across all the bins, which says there is no evidence of non-Gaussianity. This is not the case for the Poisson maps which deviate significantly from 1.

\begin{figure}
\subfloat[353 GHz 1-D parity even bispectrum]{
  \centering
    \includegraphics[width=.47\textwidth]{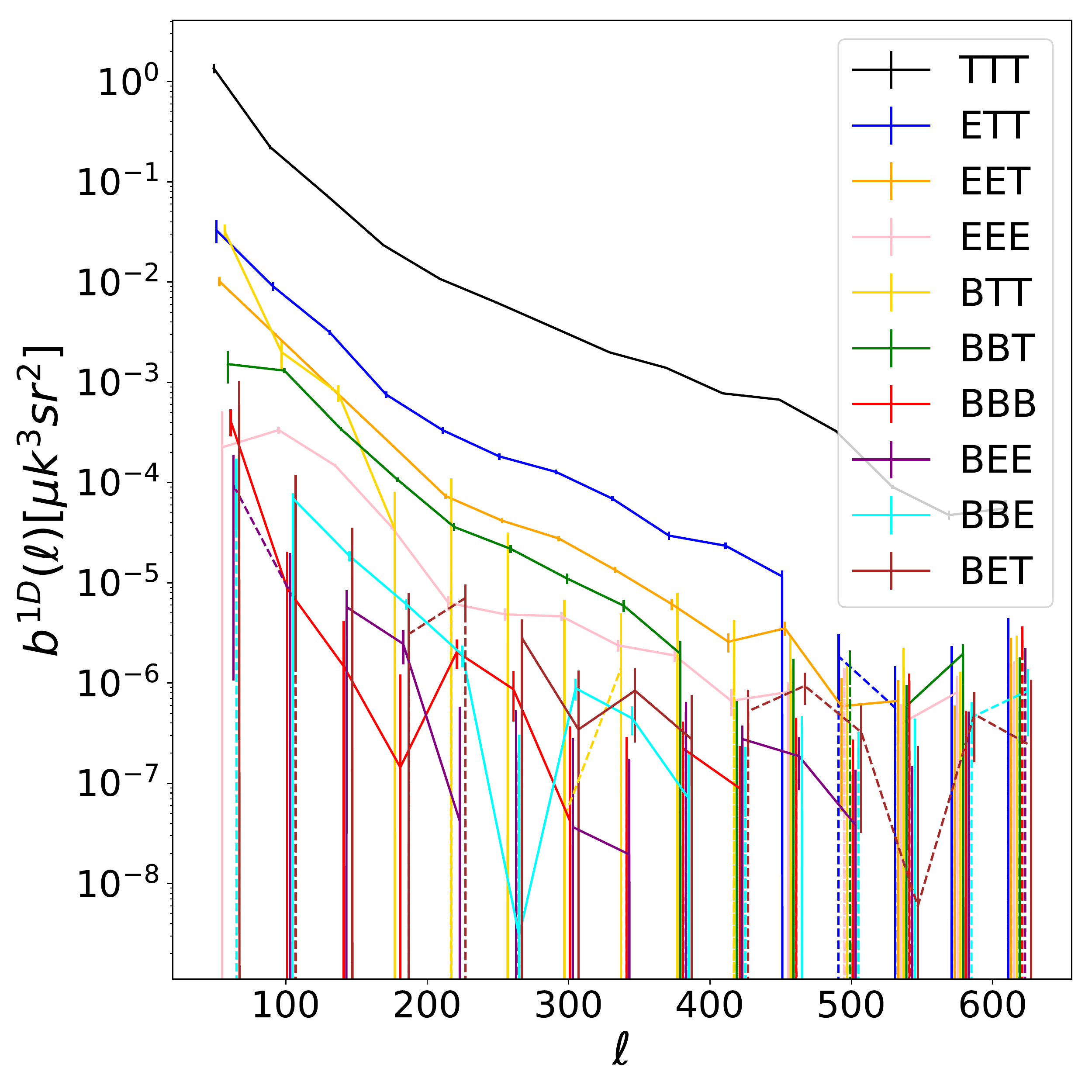}
    \label{fig:dust1Dbispec}}
    \qquad
\subfloat[353 GHz 1-D parity even Chi-squared]{
    \centering
    \includegraphics[width=.47\textwidth]{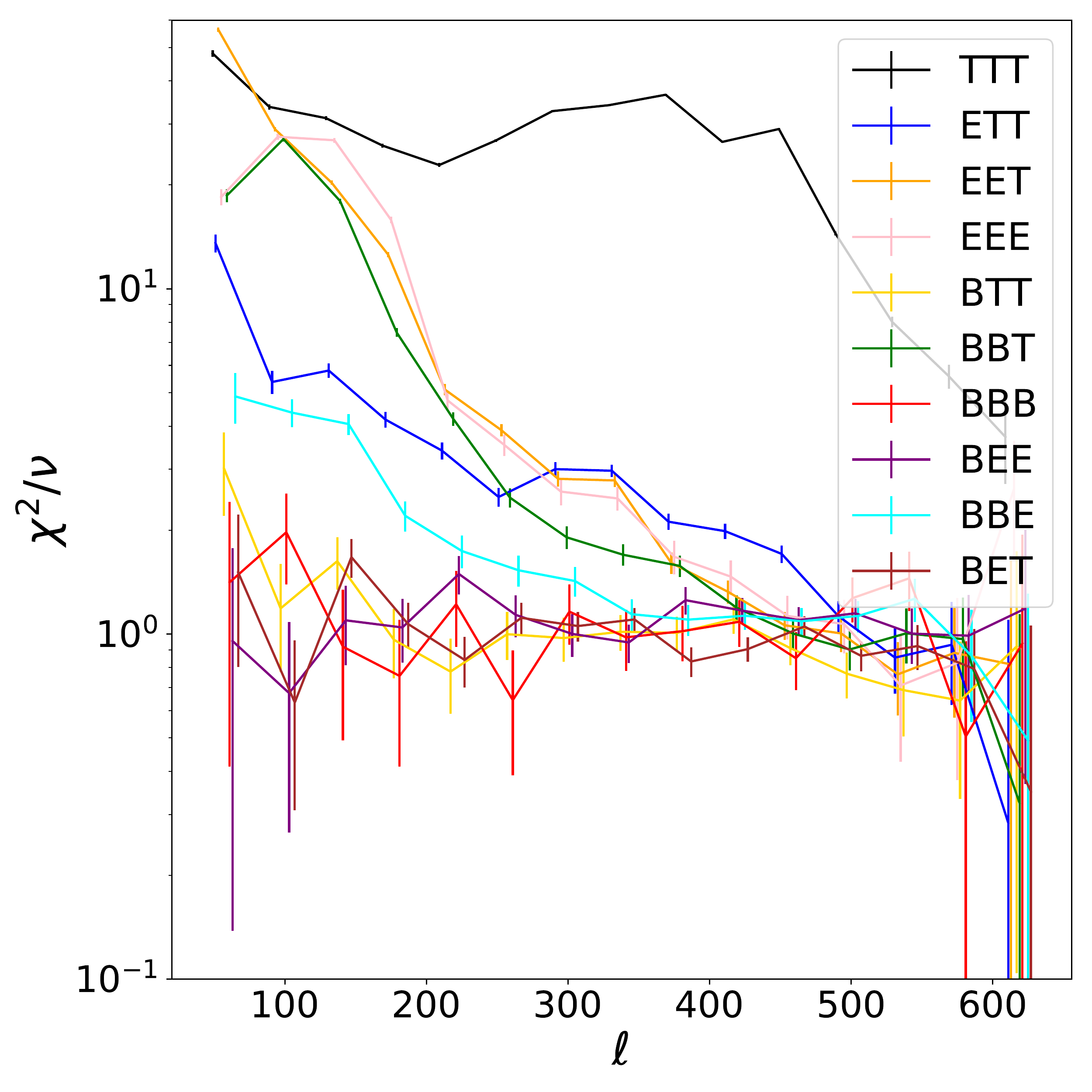}
    \label{fig:dust1Dchisquared}}
\caption{The parity even 1-D bispectrum and 1-D chi-squared for the 353 GHz map. We used $60\%$ of the sky for these measurements. The dotted lines represent negative values of the bispectrum.}
\end{figure}

\begin{figure}
\begin{minipage}{.47\textwidth}
  \centering
    \includegraphics[width=.95\textwidth]{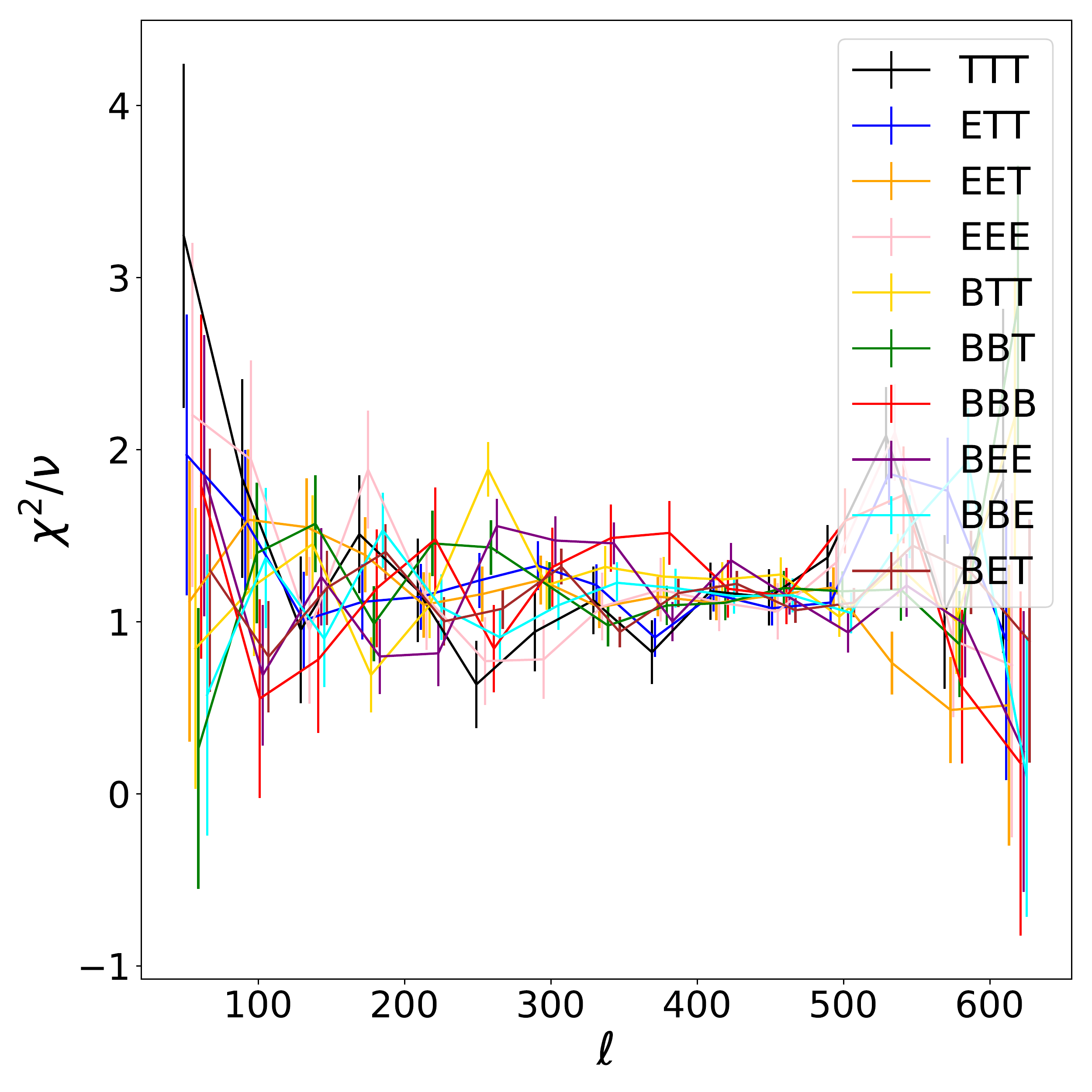}
    \caption{1-D chi-squared from half-ring difference maps  for $60\%$ of the sky.}
    \label{fig:DustNoise1Dchisquared}
\end{minipage}%
    \qquad
\begin{minipage}{.47\textwidth}
    \centering
    \includegraphics[width=.95\textwidth]{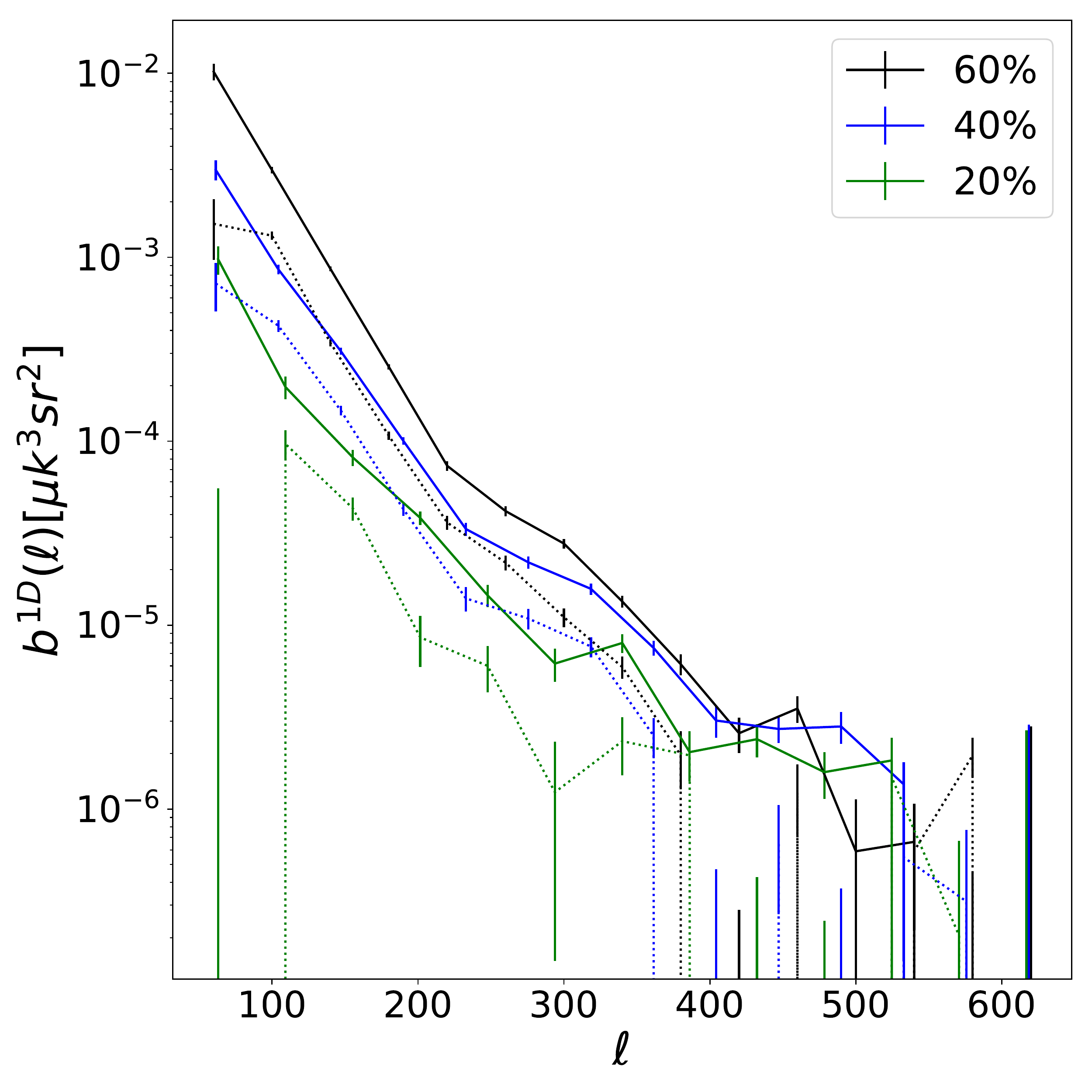}
\caption{EET (solid lines) and BBT (dotted lines) 1-D 353 GHz bispectra for different fractions of the sky.}
    \label{fig:fSky_BBT_EET}
\end{minipage}%
\end{figure}

\begin{figure}
\centering
\subfloat[2D Bispectrum: $\ell L^2 b^{2D:\, T,(T,T)}_{\ell,L}$]{
  \centering
    \includegraphics[width=.45\textwidth]{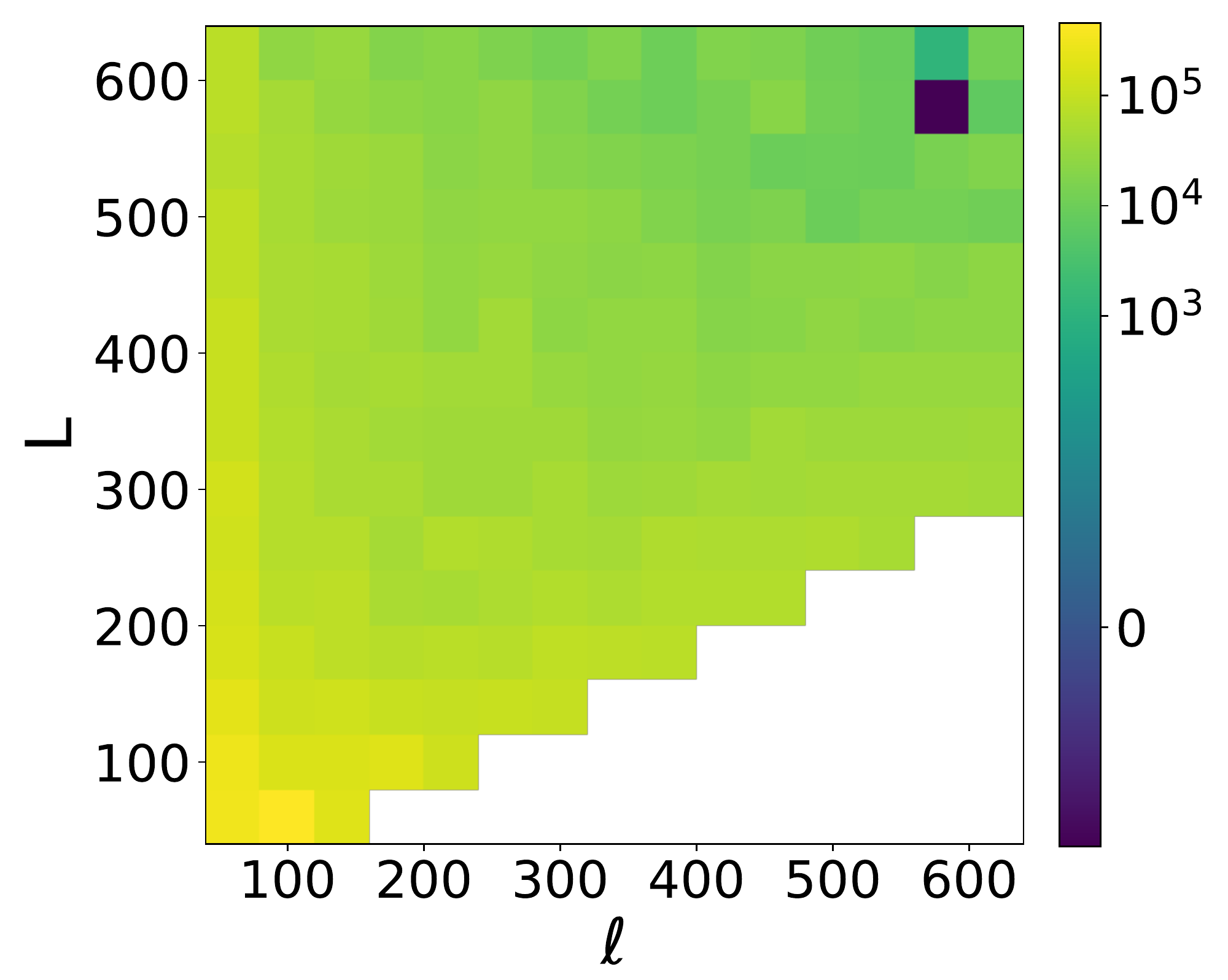}
    \label{fig:353_2d_TTT}}
    \qquad
\subfloat[2D  chi-squared:$ {\chi^2}^{2D:\, T,(T,T)}_{\ell,L}$]{
  \centering
    \includegraphics[width=.45\textwidth]{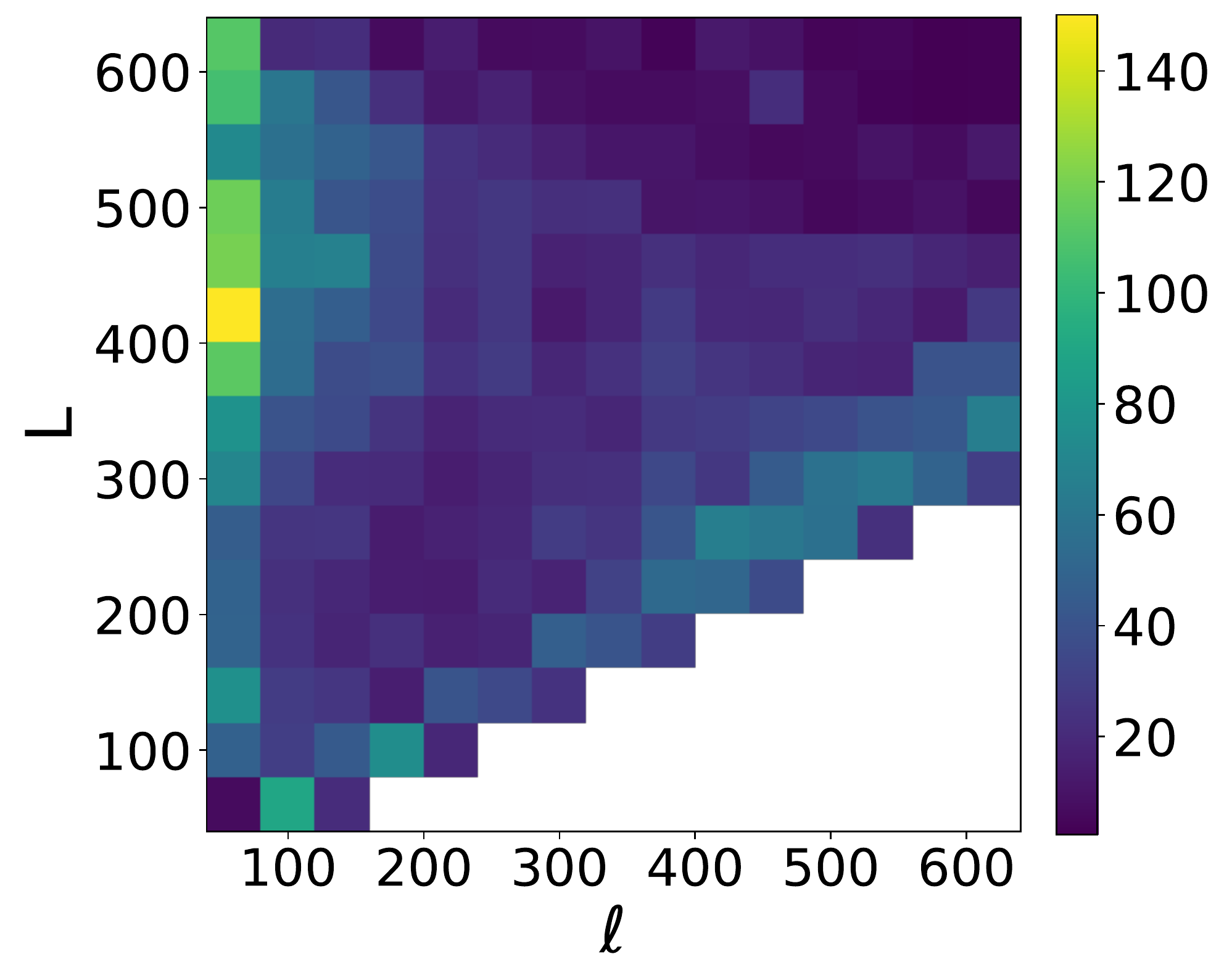}
    \label{fig:353_2d_TTT_chi}}
    
\subfloat[2D Bispectrum: $\ell L^2 b^{2D:\, T,(B,B)}_{\ell,L}$]{
    \centering
    \includegraphics[width=.45\textwidth]{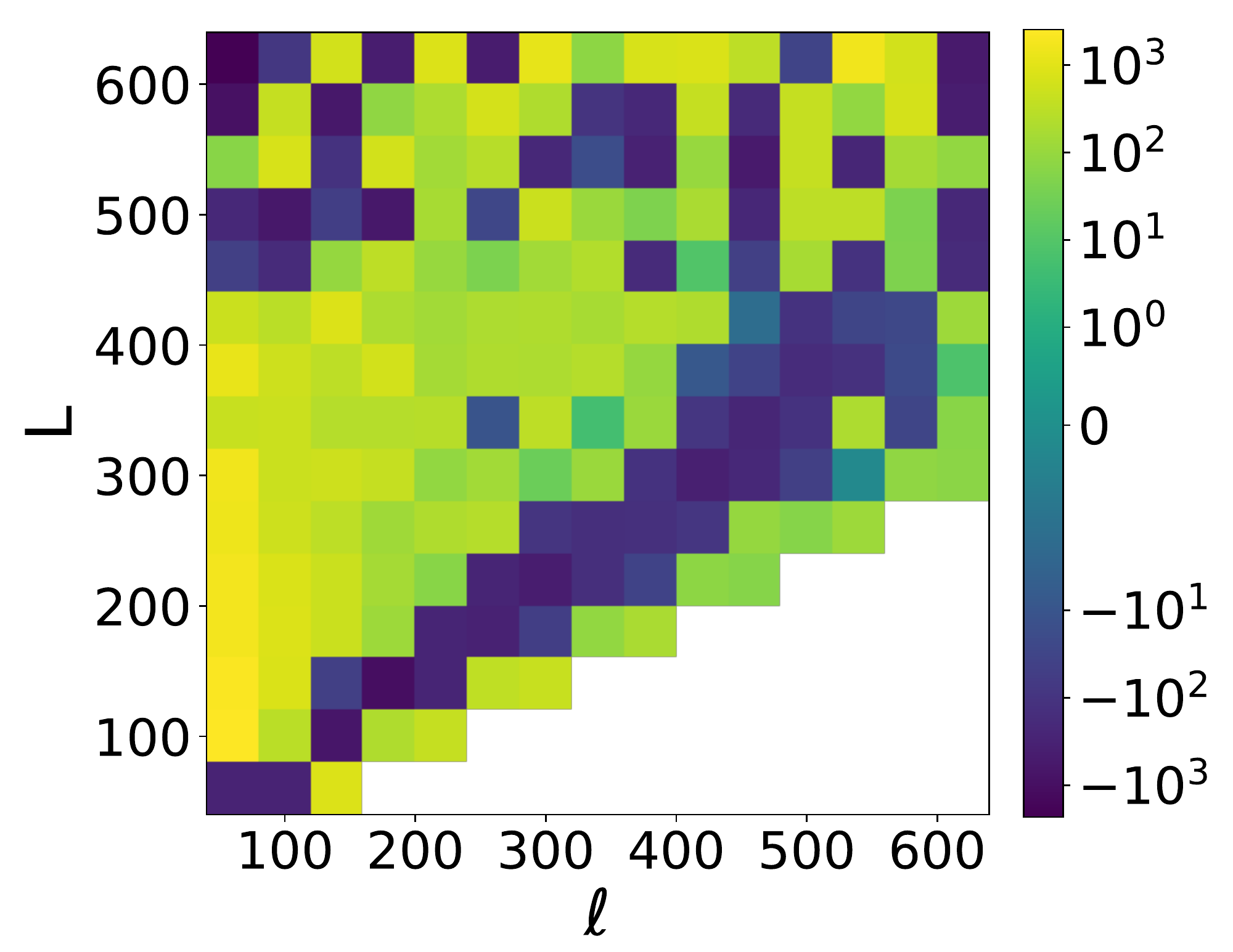}
    \label{fig:353_2d_TBB}}
    \qquad
\subfloat[2D  chi-squared:$ {\chi^2}^{2D:\, T,(B,B)}_{\ell,L}$]{
    \centering
    \includegraphics[width=.45\textwidth]{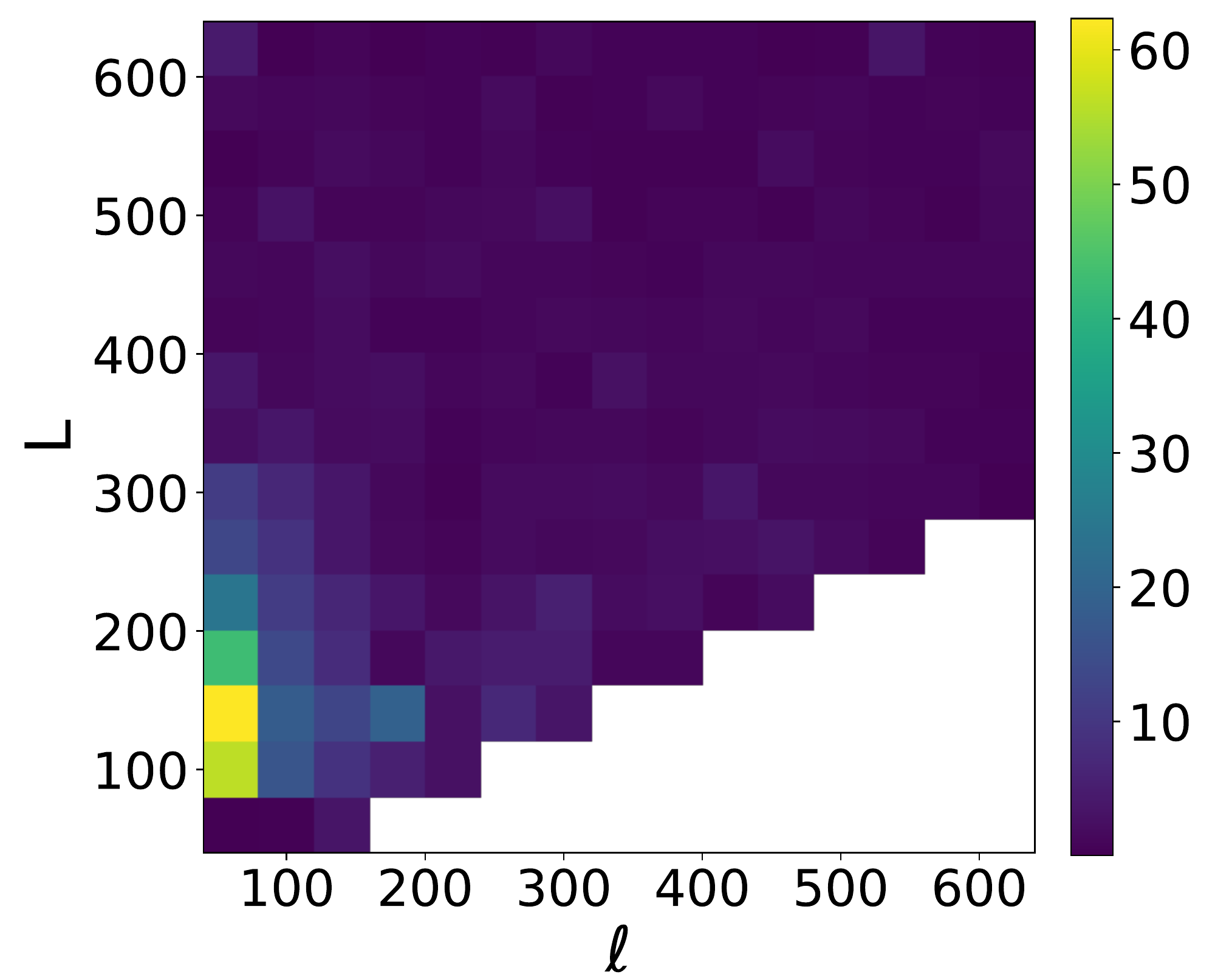}
    \label{fig:353_2d_TBB_chi}}
    
\subfloat[2D Bispectrum: $\ell L^2 b^{2D:\, E,(E,E)}_{\ell,L}$]{
    \centering
    \includegraphics[width=.45\textwidth]{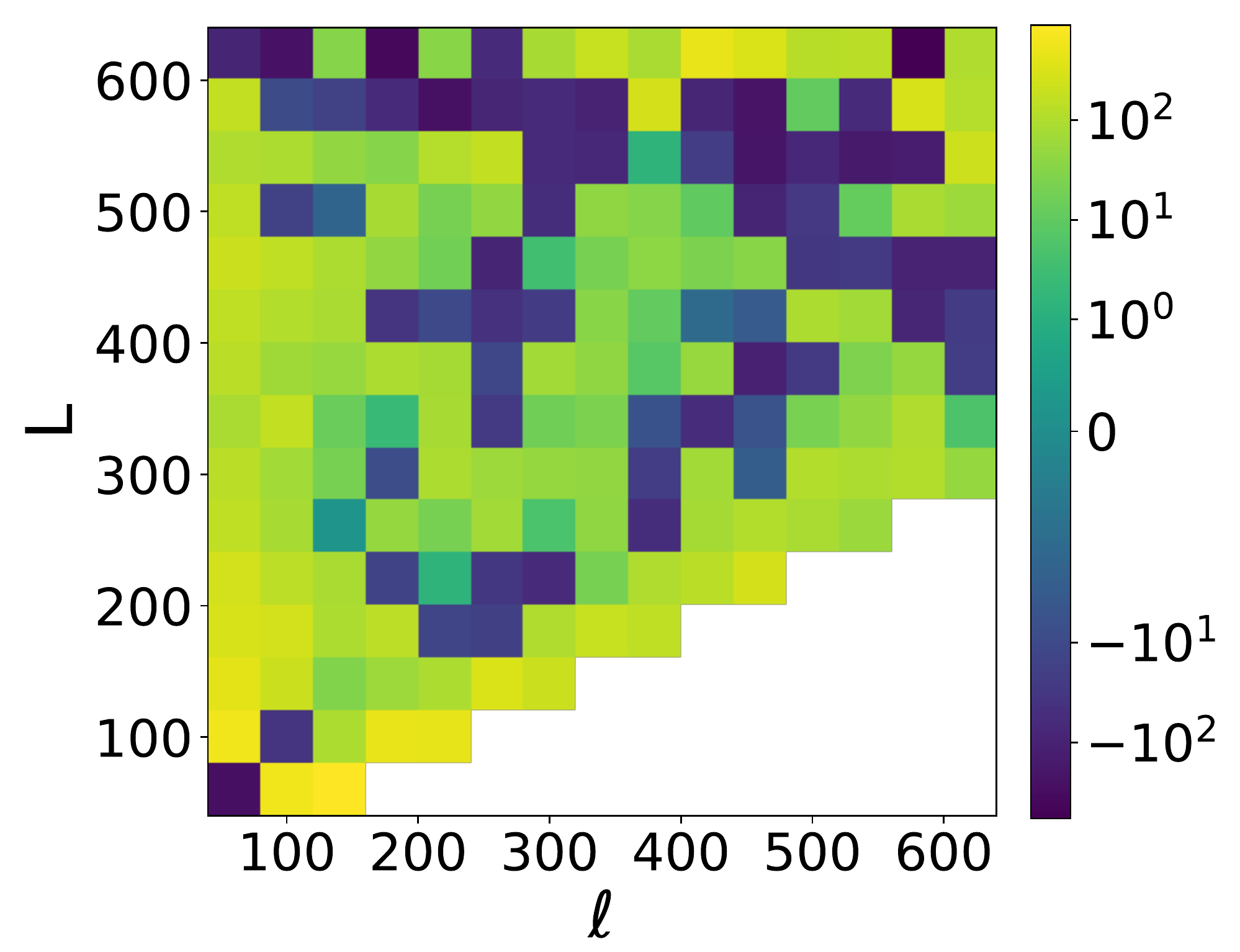}
    \label{fig:353_2d_EEE}}
    \qquad
\subfloat[2D  chi-squared:$ {\chi^2}^{2D:\, E,(E,E)}_{\ell,L}$]{
    \centering
    \includegraphics[width=.45\textwidth]{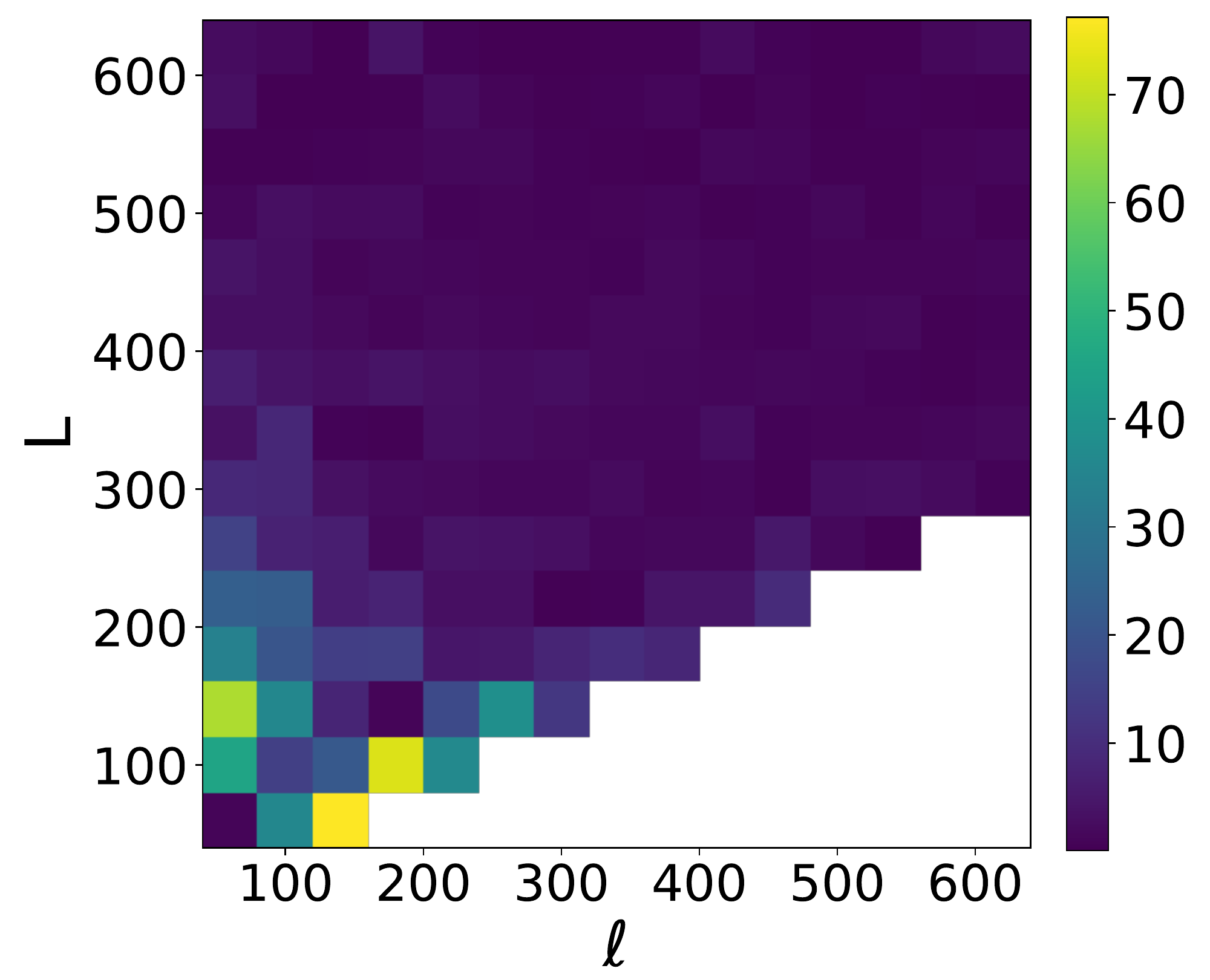}
    \label{fig:353_2d_EEE_chi}}   

\caption{The parity even 2D bispectrum and 2D chi-squared for the 353 GHz dust map. We used $60\%$ of the sky for these measurements. The 2D bispectrum is defined in Eq. \ref{eq:2dBispectrum} and involves averaging configurations over two legs whilst holding one leg fixed. We use the following notation $b^{2D:\,X,(Y,Z)}_{\ell,L} $where the field $X$ is held fixed at scale $\ell$ and we average over fields $Y,Z$ with average scales $L$. In these plots the y axis is the scale of the averaged leg and the x axis is the fixed leg. Configurations close to the y axis probe squeezed configurations, configurations in the lower right probe flattened configurations and configurations in-between probe many configurations with the dominant contribution from equilateral-like configurations.} 
\label{fig:2D_353plots}
\end{figure}

\begin{figure}

\begin{minipage}{.47\textwidth}
  \centering
    \includegraphics[width=.95\textwidth]{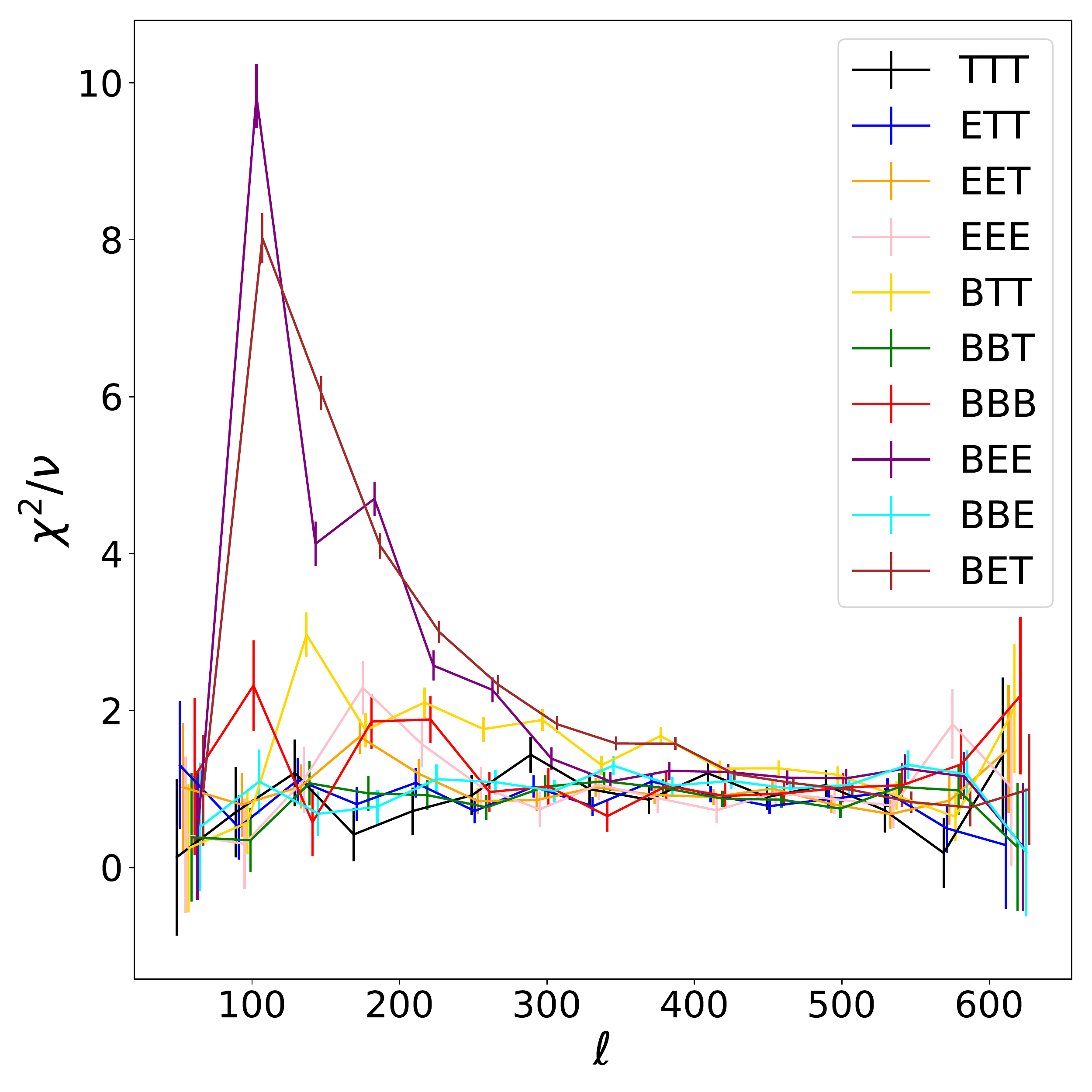}
\caption{The parity odd  1-D chi-squared for the 353 GHz maps. $60 \%$ of the sky was used for these measurements. }
    \label{fig:dust1Dchisquared_odd}
\end{minipage}%
    \qquad
\begin{minipage}{.47\textwidth}
  \centering
  \includegraphics[width=.95\textwidth]{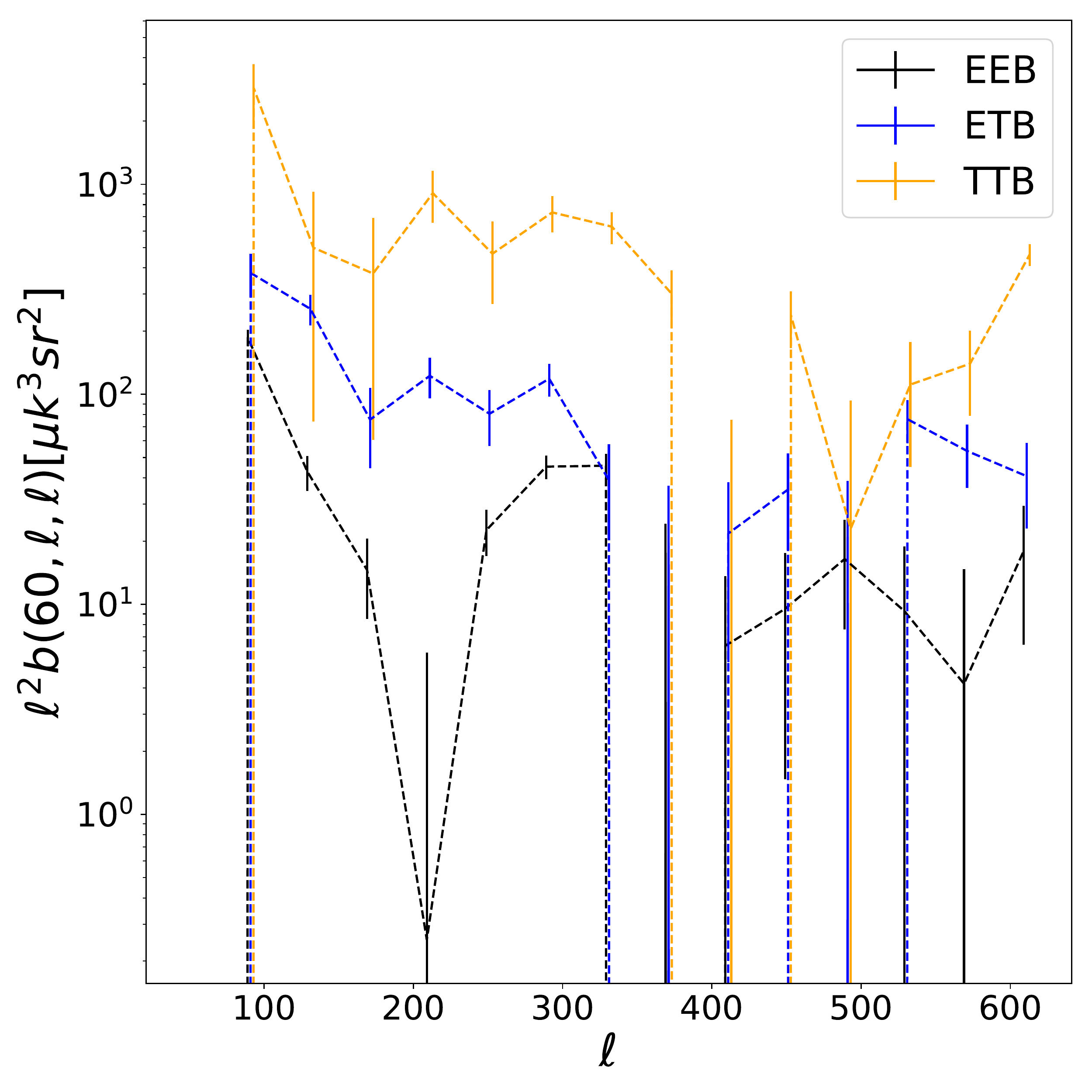}
    \caption{Squeezed slice of 353 GHz parity odd bispectrum from $60\%$ of the sky. For these configurations we hold the B leg of the bispectrum fixed on the lowest $\ell$ bin and vary the other two legs.}
    \label{fig:353_squeezed}
\end{minipage}
\end{figure}

\begin{figure}
\centering
\subfloat[2D Bispectrum: $\ell L^2 b^{2D:\, B,(T,T)}_{\ell,L}$]{
  \centering
    \includegraphics[width=.45\textwidth]{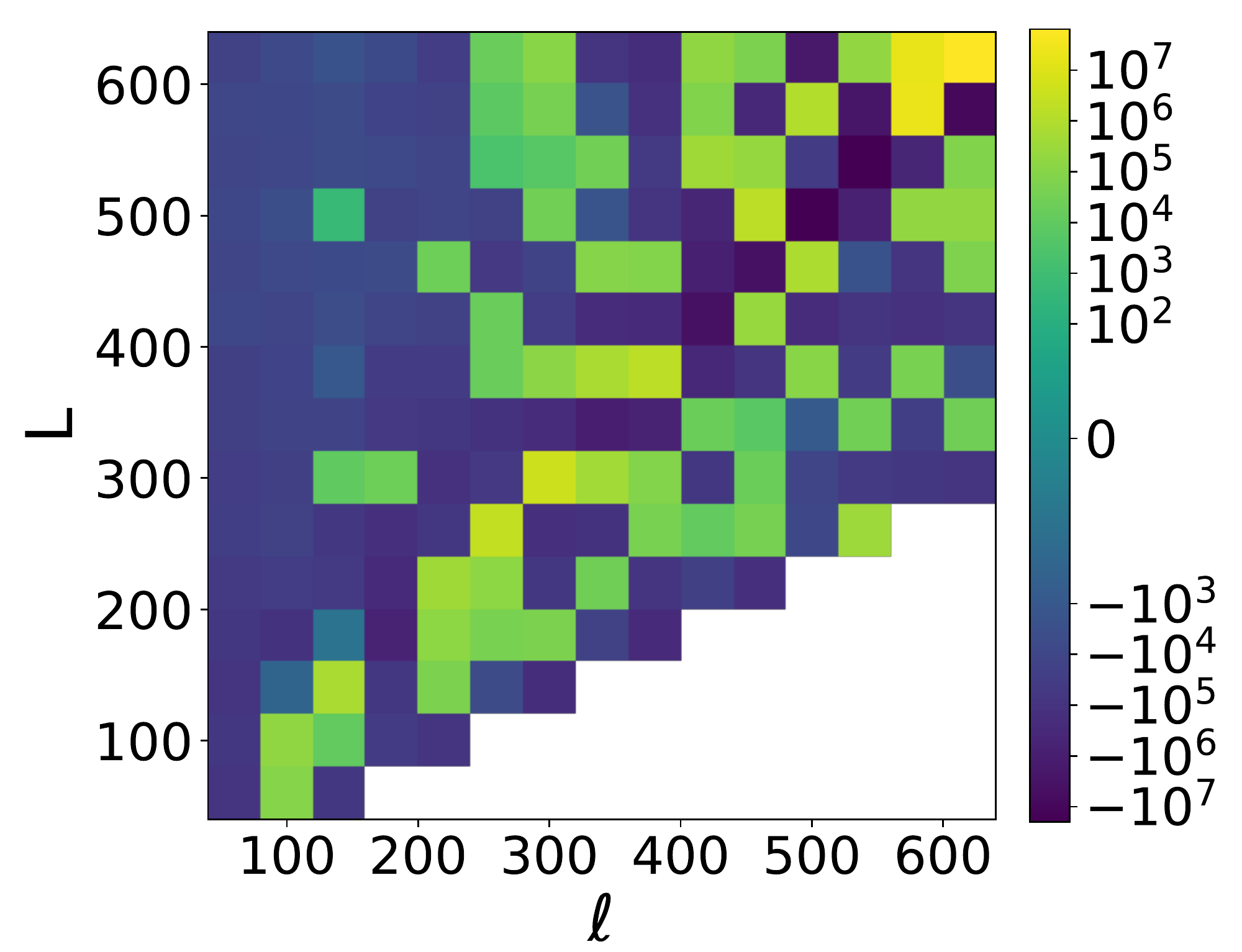}
    \label{fig:353_2d_BTT}}
    \qquad
\subfloat[2D  chi-squared:$ {\chi^2}^{2D:\, B,(T,T)}_{\ell,L}$]{
  \centering
    \includegraphics[width=.45\textwidth]{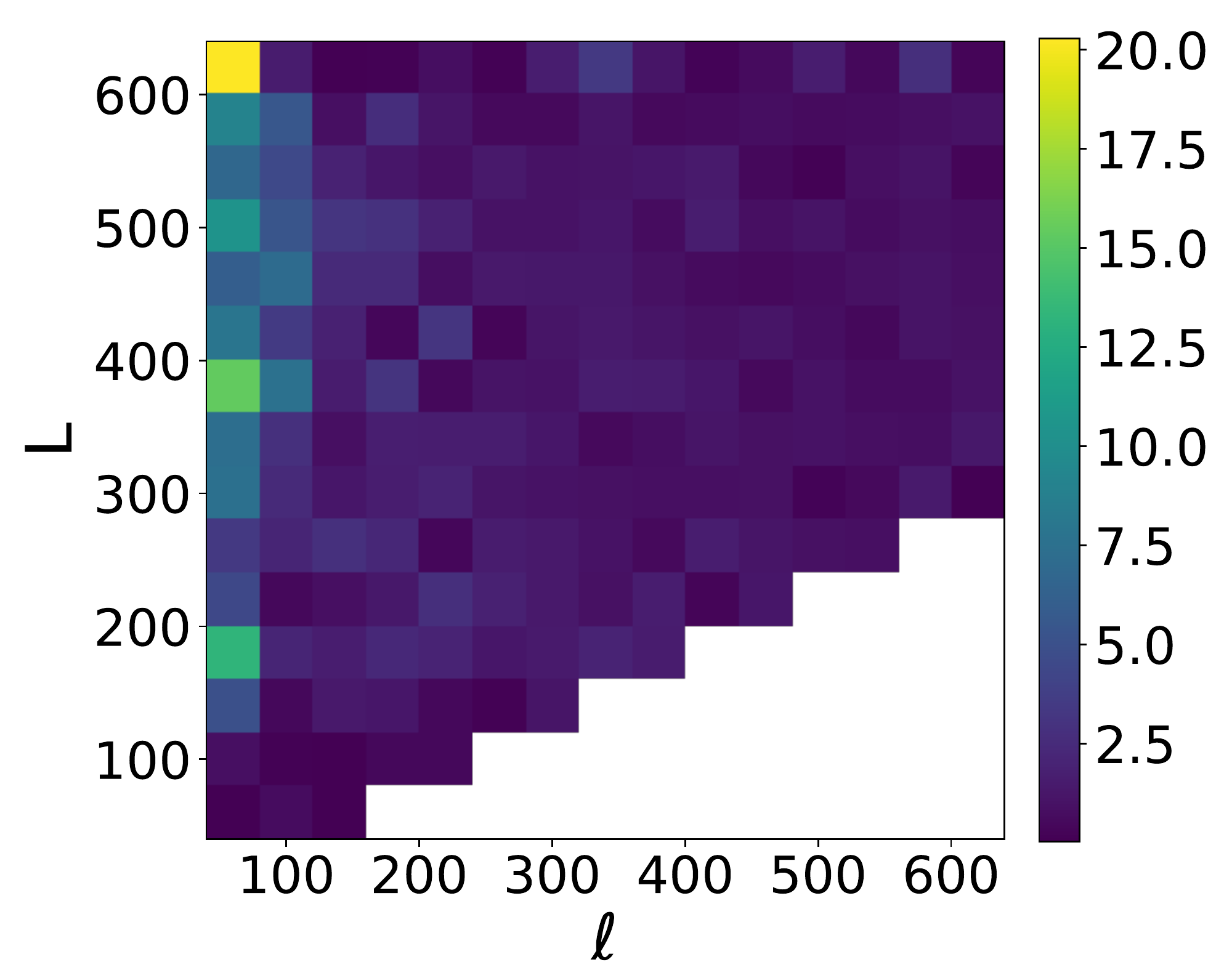}
    \label{fig:353_2d_BTT_chi}}
    
\subfloat[2D Bispectrum: $\ell L^2 b^{2D:\, B,(T,E)}_{\ell,L}$]{
    \centering
    \includegraphics[width=.45\textwidth]{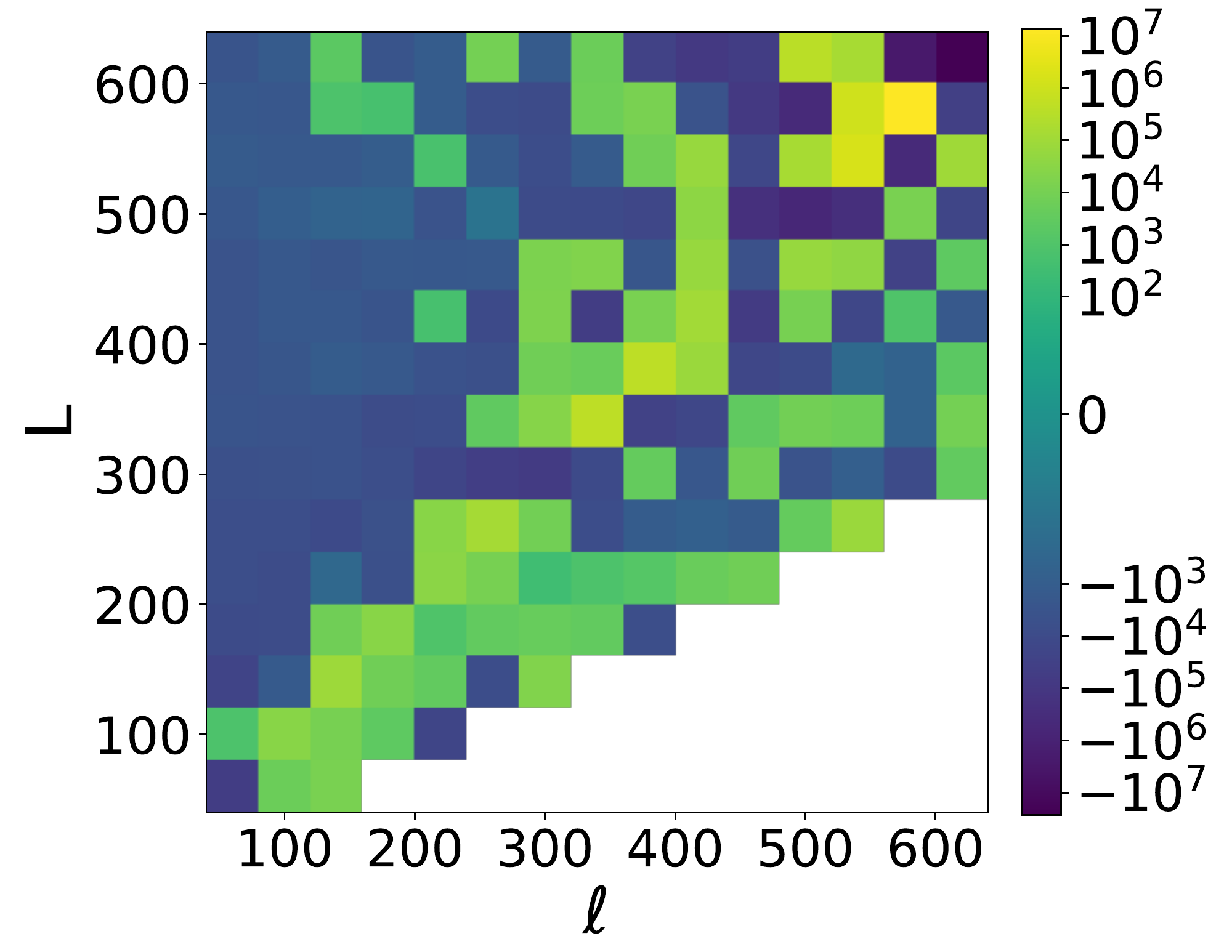}
    \label{fig:353_2d_BTE}}
    \qquad
\subfloat[2D  chi-squared:$ {\chi^2}^{2D:\, B,(T,E)}_{\ell,L}$]{
    \centering
    \includegraphics[width=.45\textwidth]{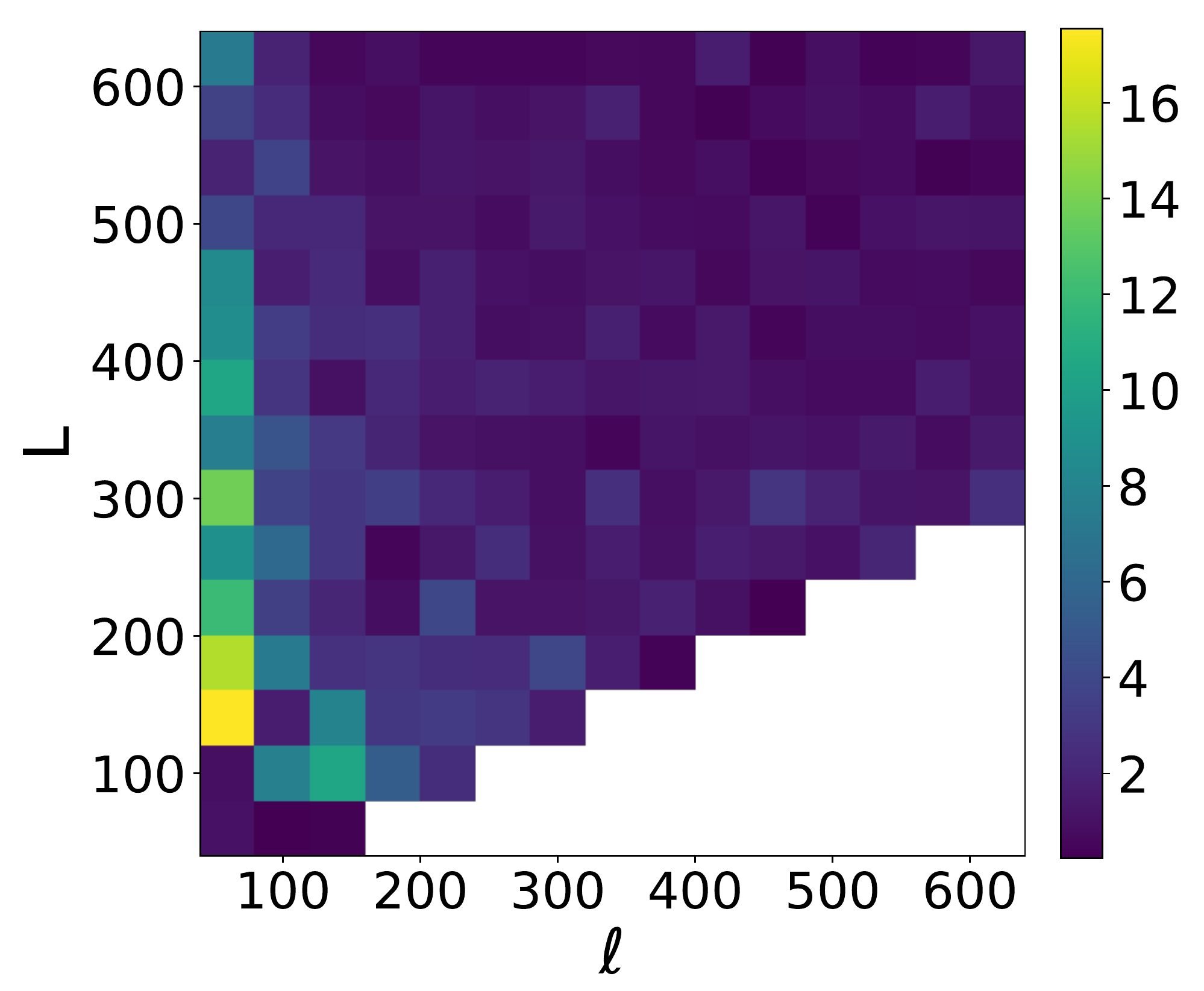}
    \label{fig:353_2d_BTE_odd}}

\caption{The parity odd 2D bispectrum and 2D chi-squared for the 353 GHz dust map. We used $60\%$ of the sky for these measurements. The 2D bispectrum is defined in Eq. \ref{eq:2dBispectrum} and involves averaging configurations over two legs whilst holding one leg fixed. We use the following notation $b^{2D:\,X,(Y,Z)}_{\ell,L} $where the field $X$  is held fixed at scale $\ell$ and we average over fields $Y,Z$ with average scales $L$. In these plots the y axis is the scale of the averaged leg and the x axis is the fixed leg.} 
\label{fig:2D_353plots_odd}
\end{figure}
\subsection{Dust Bispectra}\label{sec:dustBispectrum}
\subsubsection{353 GHz  Bispectra}\label{sec:dus353tBispectrum}
It is useful to first review the 353 GHz power spectrum. In Figure \ref{fig:353_powspec_gm4} we plot TT, EE and BB pseudo-Cl power-spectra for the 353 GHz maps. We see that the BB mode power is a factor of $\sim 2$ less than the EE power and than the power spectra are well described by power laws $C_\ell \propto \ell^\alpha$ with $\alpha \sim -2.5$ for all three spectra. Now we move to the bispectrum results. In Figure \ref{fig:dust1Dbispec} we present the 1-D parity even binned bispectrum of the \textit{Planck }353 GHz maps. In this projection, we see a strong signal for several bispectra combinations, notably TTT, ETT, BBT, BBE, EET and EEE. The remaining combinations seem to show no clear signal. By examining the 1-D parity even chi-squared, shown in Figure \ref{fig:dust1Dchisquared}, we see that the other configurations are consistent with zero non-Gaussianity. 

Having observed non-zero bispectrum it is necessary to verify that is a sky signal. To do this we use the difference of the two-half ring data splits \citet{planck2014-a09}. This combination should have no signal and only noise. In Figure \ref{fig:DustNoise1Dchisquared}  we plot the 1-D chi-squared for the noise splits. We see no strong evidence of non-Gaussianity in any of the splits. We also have analysed the Planck 353 GHz noise simulations and found that the levels of non-Gaussianity in these maps are consistent with zero.  Secondly we wish to verify this is galactic dust and not another source of non-Gaussianity, such as extra-galactic point sources \citep{Argueso2003,Lacasa2014,Crawford2014,coulton2017}. To do this we plot the signal as a function of sky fraction. The masks are constructed from the 353 GHz maps such that the cleanest parts of the sky are selected, thus as we reduce the used fraction of sky we should see a reduction in the size of the bispectrum signal. In Figure \ref{fig:fSky_BBT_EET} we plot the BBT and EET bispectrum signals as a function of the sky cut and find that as the fraction of the sky is reduced the signals are reduced in amplitude. 

Having verified that this the signal from galactic dust we now explore the shape dependence for TTT, TBB and EEE. In Figure \ref{fig:2D_353plots} we present the 2D bispectra for these configurations as well as the corresponding 2D chi-squared. Generally, we find that the dust bispectrum is strongest in the squeezed and folded configurations. It should be noted that the noise in the polarized maps limits our ability to probe these bispectra to significantly higher $\ell$ as well as reduces the configurations that we can probe (equilateral configurations at similar scales to squeezed configurations will be significantly noisier).

In \citet{planck2014-XXX} it was found that the dust EE power was twice that of the BB power. In Figure \ref{fig:fSky_BBT_EET} we find that the amplitude of the EET bispectrum is greater than the BBT bispectrum for all sky cuts by a factor of $\sim 2$. \textit{Planck} found evidence for non zero BT power-spectrum signal, a parity invariance violating signal, thus it is of interest to apply our parity even bispectra estimator to the configurations BTT, BTE, BEE and BBB, which should be zero in the case of no parity violation. We find no evidence for non-zero BTT, BTE,BEE or BBB bispectra with our parity even estimator, thus, for these sky cuts, there is no evidence of parity violation at the bispectrum level. We then consider the parity odd 353 GHz bispectrum and the resulting 1-D chi-squared are shown in Figure \ref{fig:dust1Dchisquared_odd}. We find measurable levels of parity odd non-Gaussianity for many of the bispectrum configurations, most significantly the BET, BEE and BTT.  We find no evidence for parity violating non-Gaussianity in our parity-odd estimator measurements, which would appear as non-zero TTT, TTE etc bispectra. In Figure \ref{fig:2D_353plots_odd} we explore the shape dependence of these parity odd bispectra. These bispectra are also strongly peaked in the squeezed limit and show a strongly negative signal.

The dust power spectra are well characterised by a power laws \citep[$C_\ell \propto \ell^\alpha$ with $\alpha \sim -2.5$ ][]{Planck2018LIV}  and we explore the scale dependence of the bispectrum in Figure \ref{fig:353_squeezed} where we plot the squeezed slice of three parity odd dust bispectra. The 1-D bispectrum plotted elsewhere merges many different configurations and can hide the scale dependence of the bispectrum, and thus for a cleaner probe of the scale dependence we plot the squeezed slice. We see that the BTT, BET and BEE bispectra are roughly described by $b(\ell,\ell,\ell) \propto \ell^{\alpha}$ where $\alpha \sim -2$, higher signal to noise measurements are necessary to obtain a more precise relation.

Beyond characterizing these bispectra, we would like to understand the features that cause them. Fully understanding their origin is a complex task that should be explored with simulations and so here we only briefly explore some of the possible implications of these observed bispectra. The filamentary structure seen in the temperature and polarization maps likely leads to the folded and equilateral bispectrum shapes seen in TTT and EEE as these bispectrum configurations corresponds to elongated over-dense regions. As was discussed in \citet{planck2014-XXX}, the amplitude of the dust polarisation power spectrum at $\ell=80$ is correlated with the dust intensity. This should result in squeezed TEE and TBB bispectra and we observe a strong squeezed signal for these configurations. Physically this arises as lines of sight with high levels of dust intensity also have high levels of polarization. We observe a strong positive squeezed ETT bispectrum, which implies that the small scale temperature power is correlated with the large scale E mode power. Finally we also observe a strong negative parity odd BTT (and a weaker but still negative BTE) bispectrum. To understand these parity odd bispectra we consider these bispectra in the flat-sky limit, where building an intuition for these bispectra is simpler.  In the flat-sky limit, assuming isotropy, homogeneity and that the underlying physics is parity conserving, bispectra between temperature and E mode fields can be written as
\begin{align}\label{eq:flatSky_TTT}
\langle a^{X}(\mathbf{\ell_1})a^{Y}(\mathbf{\ell_2})a^{Z}(\mathbf{\ell_3})= \delta^{(2)}(\mathbf{\ell_1}+\mathbf{\ell_2}+\mathbf{\ell_3}) b(\ell_1,\ell_2,\ell_3),
\end{align}
where $X,Y,Z \in \{T,E\}$ and the reduced bispectrum $ b(\ell_1,\ell_2,\ell_3)$ which contains the physical information, depends only on the magnitude of the Fourier modes. The parity odd bispectrum between one B mode and two T/E fields has the following form \cite{Meerburg2016}
\begin{align}
\langle a^{X}(\mathbf{\ell_1})a^{Y}(\mathbf{\ell_2}))a^{B}(\mathbf{\ell_3})= \delta^{(2)}(\mathbf{\ell_1}+\mathbf{\ell_2}+\mathbf{\ell_3}) \left(\hat{\mathbf{\ell}}_1 \times  \hat{\mathbf{\ell}}_3   b(\ell_1,\ell_2,\ell_3)+ \hat{\mathbf{\ell}}_2 \times  \hat{\mathbf{\ell}}_3   b(\ell_2,\ell_1,\ell_3) \right).
\end{align}
Unlike the T and E mode only bispectra, bispectra involving B fields depend on the cross product between the B field and the T / E mode field. This means these bispectra explicitly probe correlations between B modes and perpendicular T/E modes. For the BTT (as well as BTE and BEE bispectra), this implies that there is a significant correlation between the variation of the small scale T/E power with the gradient of the perpendicular B modes.

\begin{figure}
\centering
\subfloat[2D Bispectrum: $\ell L^2 b^{2D:\, T,(T,T)}_{\ell,L}$]{
  \centering
    \includegraphics[width=.45\textwidth]{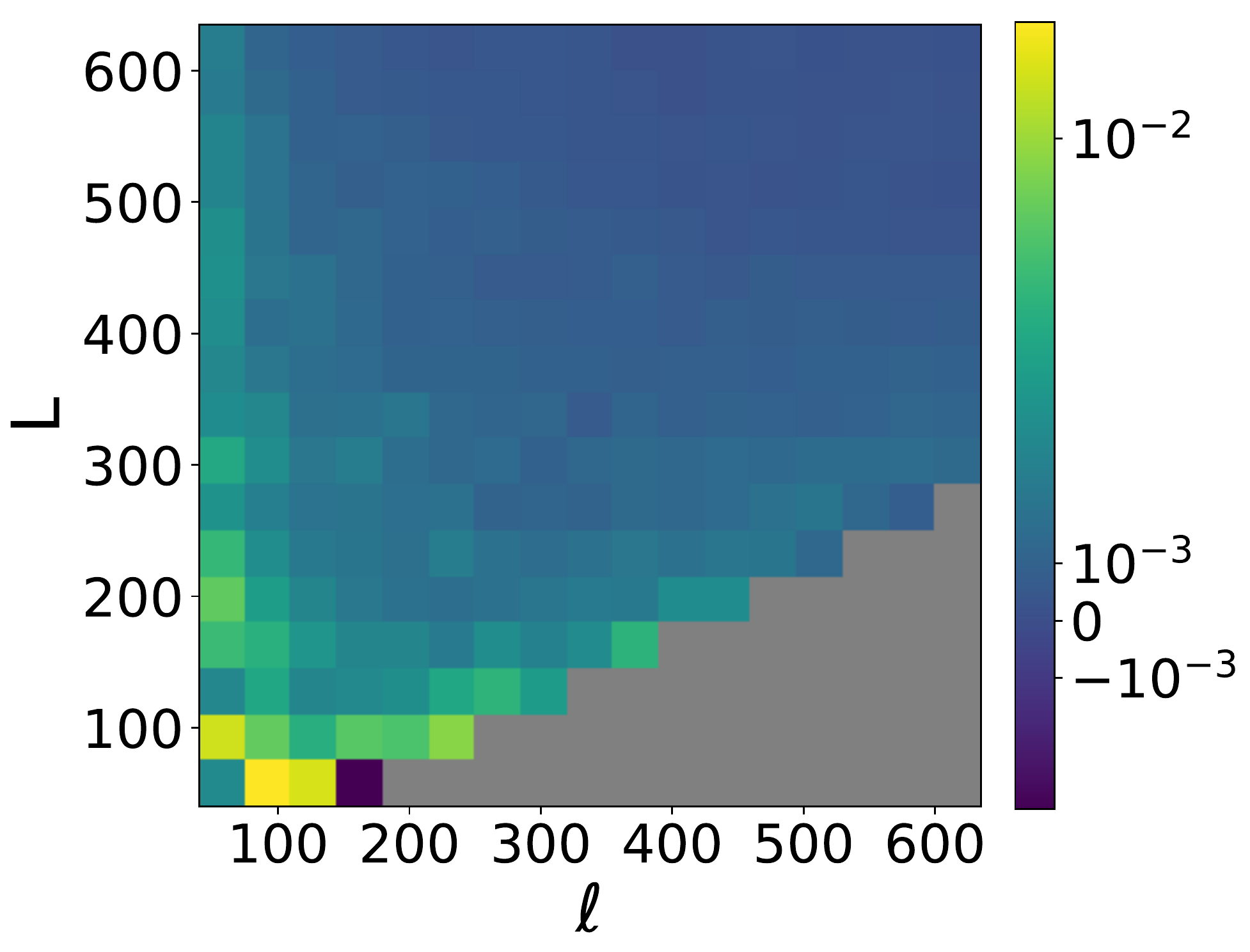}
    \label{fig:IRIS_2d_TTT}}
    \qquad
\subfloat[2D  chi-squared:$ {\chi^2}^{2D:\, T,(T,T)}_{\ell,L}$]{
  \centering
    \includegraphics[width=.45\textwidth]{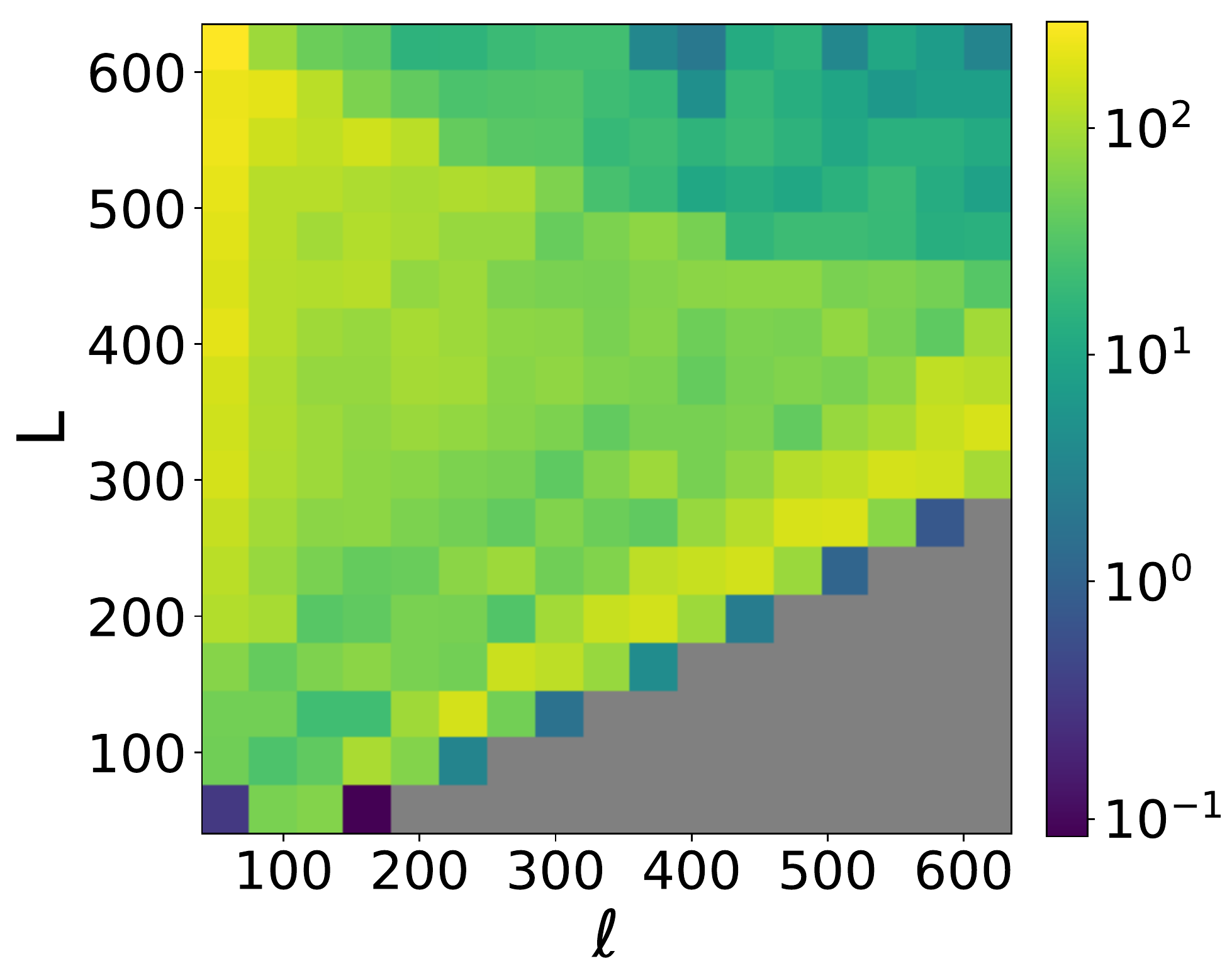}
    \label{fig:IRIS_2d_TTT_chi}}
    
\caption{ The parity even 2D bispectrum and reduced chi-squared of the IRIS map from 60$\%$ of the sky.}

\end{figure}

\begin{figure}
  \centering
    \includegraphics[width=.60\textwidth]{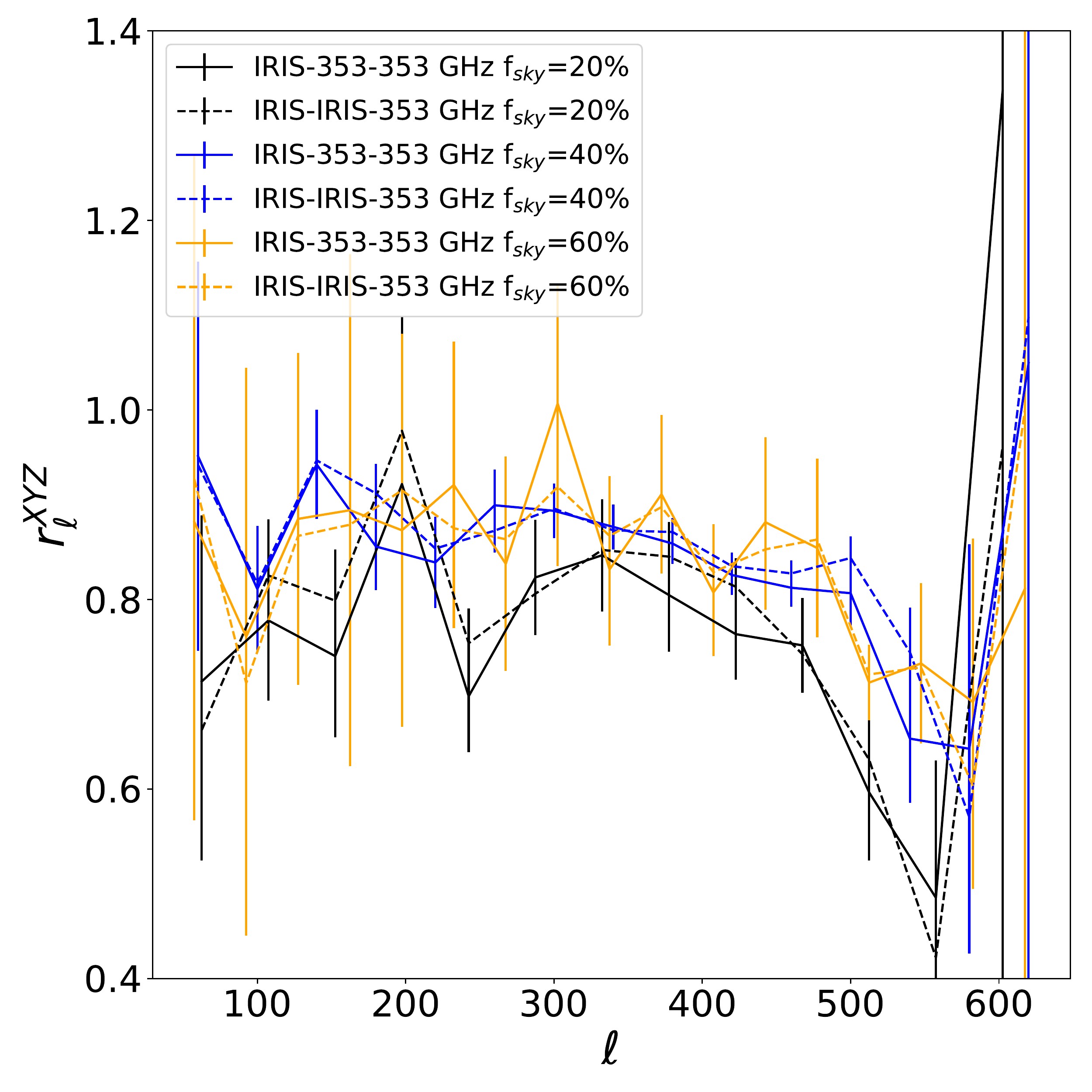}
   \caption{The parity even bispectrum correlation coefficient between the IRIS dust map and the 353 GHz map for three sky masks.}
    \label{fig:IRIS_353_correlation}
\end{figure}

\subsubsection{Frequency Evolution}
We then examine the frequency evolution of the dust bispectrum by investigating the IRIS map, described in Section \ref{sec:dataSets}, along with correlations with the 353 GHz map. We restrict this analysis to temperature as the IRIS map is a temperature only map. We find that the IRIS map has a strong bispectrum signal and in Figures \ref{fig:IRIS_2d_TTT} and \ref{fig:IRIS_2d_TTT_chi} we examine the bispectrum. We find a qualitatively similar features in the IRIS bispectrum and 353 GHz bispectrum; a signal with a strong squeezed limit and similar scale dependence in the equilateral configuration.

To quantitatively examine the degree of correlation we introduce a bispectrum cross-correlation coefficient as
\begin{equation} \label{eq:bispectrumCorCoef}
r^{XYZ} = \frac{b^{XYZ}}{(b^{XXX}b^{YYY}b^{ZZZ})^{\frac{1}{3}}}.
\end{equation}
This can be used on the full bispectrum or on the 1D or 2D quantities. In order to evaluate the errors on this quantity we propagate the errors using the method described in Appendix B of \citet{Villaescusa2018}. In Figure \ref{fig:IRIS_353_correlation} we plot the bispectrum correlation coefficient for three different sky masks. We see a generally strong correlation between the two maps. This correlation is not perfect as there is spatial variation in the dust temperature spectral index \citep[see e.g. ][]{planck2013-p06b} This results in a de-correlation of the bispectra, as the relative brightness of regions changes as a function of frequency. In Figure \ref{fig:IRIS_353_correlation} we see there is a weak evidence for variation in the correlation coefficient as a function of sky, with the smallest sky region seeming to show more de-correlation than the largest regions. This increased de-correlation is likely due to the increasing importance of anisotropies in the cosmic infrared background (CIB) in low dust regions. The CIB de-correlates with frequency as this is composed of the emission from galaxies with similar rest frame spectra, but located at different redshifts and thus having different observed frequency dependencies. \citet{planck2013-XVII} showed that the CIB was only $29\%$ correlated between 353 GHz and 100 $\mu$m. The impact of the CIB becomes more important at smaller scales and perhaps explains the slight decrease in the correlation coefficient at smaller scales.

\begin{figure}
\subfloat[SPASS 1D bispectrum: $ b^{1D:\, T,T,T}_{\ell,L}$]{
  \centering
    \includegraphics[width=.45\textwidth]{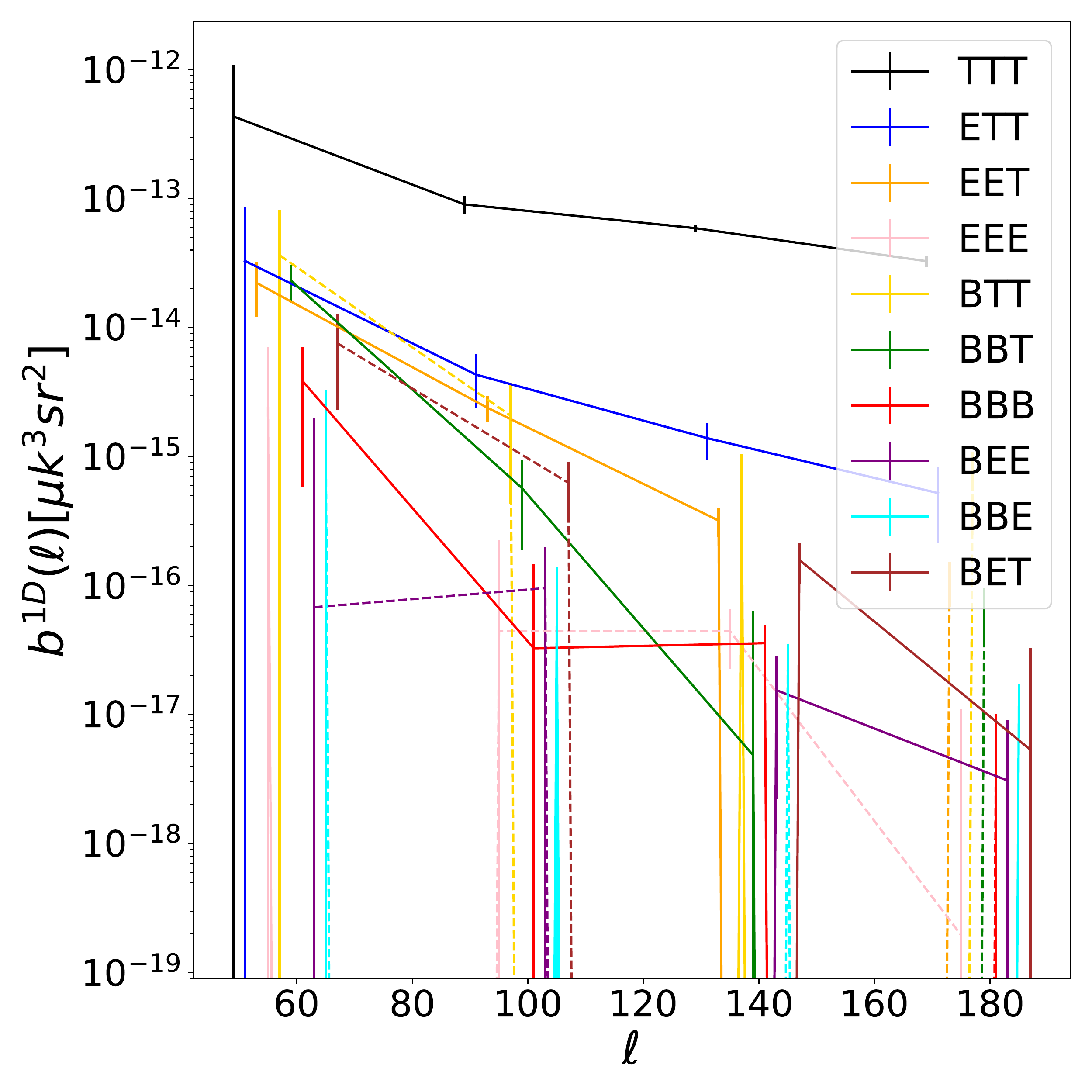}
    \label{fig:spass1Dbispec}}
    \qquad
\subfloat[SPASS 1D Chi-squared: $ {\chi^2}^{1D:\, T,T,T}_{\ell,L}$ ]{
    \centering
    \includegraphics[width=.45\textwidth]{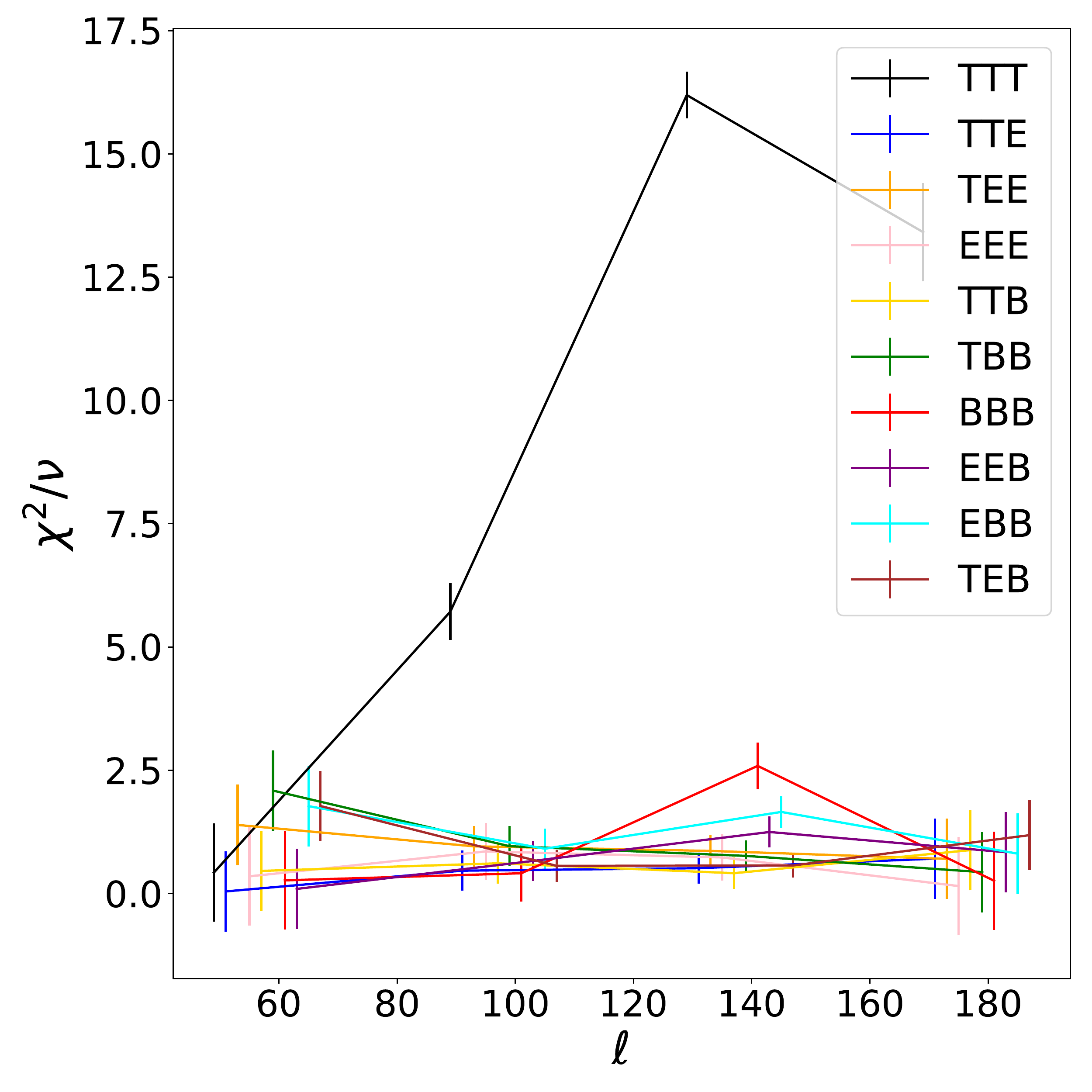}
    \label{fig:spass1Dchisq}}

\caption{The parity even 1D bispectrum and 1D chi-squared for the SPASS maps.}
\end{figure}

\begin{figure}
\subfloat[Synchrotron 2D bispectrum: $ b^{2D:\, T,(T,T)}_{\ell,L}$]{
  \centering
    \includegraphics[width=.45\textwidth]{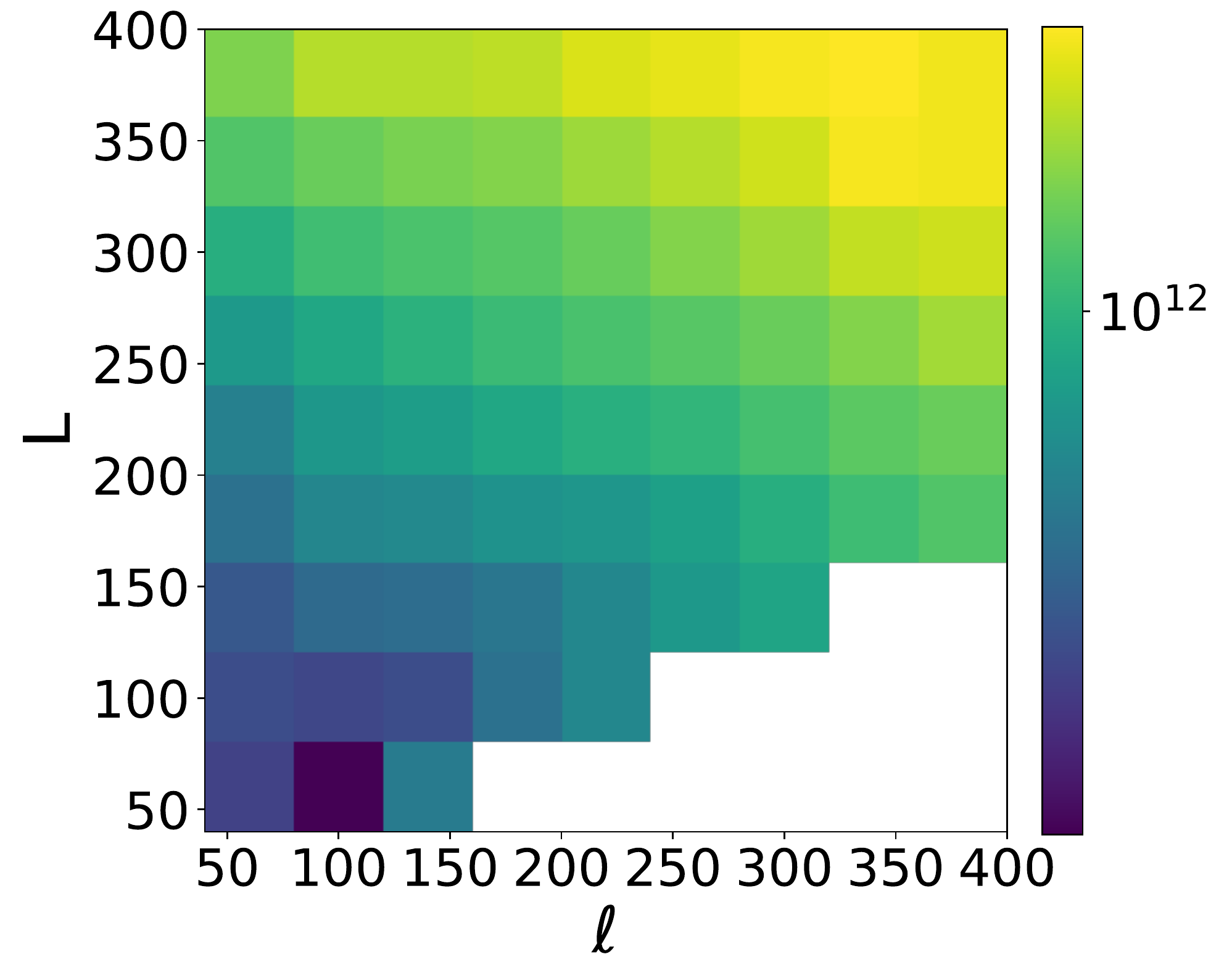}
    \label{fig:sync2Dbispec_TTT}}
    \qquad
\subfloat[Synchrotron 2D Chi-squared: $ {\chi^2}^{2D:\, T,(T,T)}_{\ell,L}$ ]{
    \centering
    \includegraphics[width=.45\textwidth]{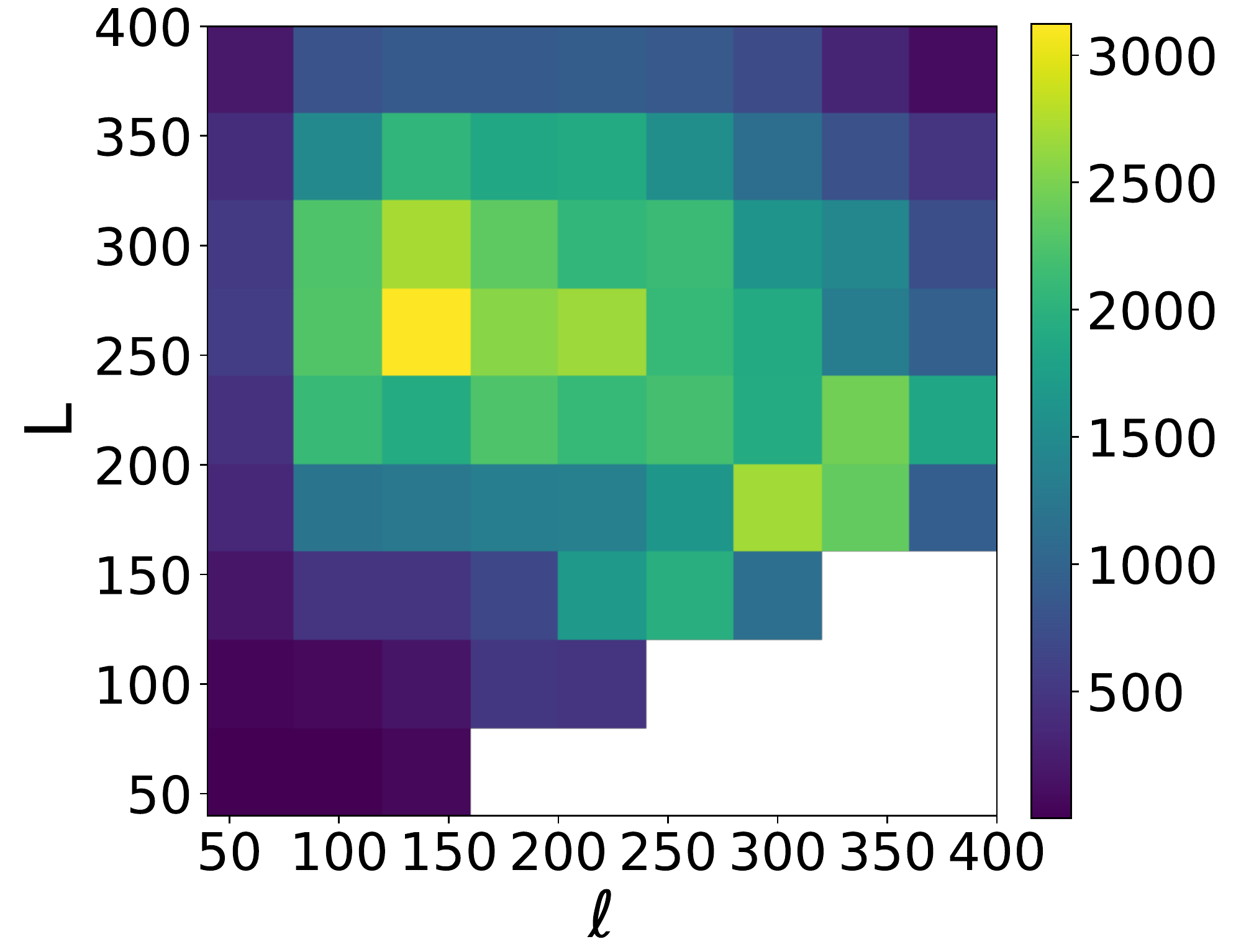}
    \label{fig:sync2Dchisquared_TTT}}

\caption{The parity even 2D bispectrum and 2D chi-squared for the Commander synchrotron temperature map. We used $60\%$ of the sky for these measurements.}
\end{figure}

\begin{figure}

\begin{minipage}{.47\textwidth}
  \centering
    \includegraphics[width=.95\textwidth]{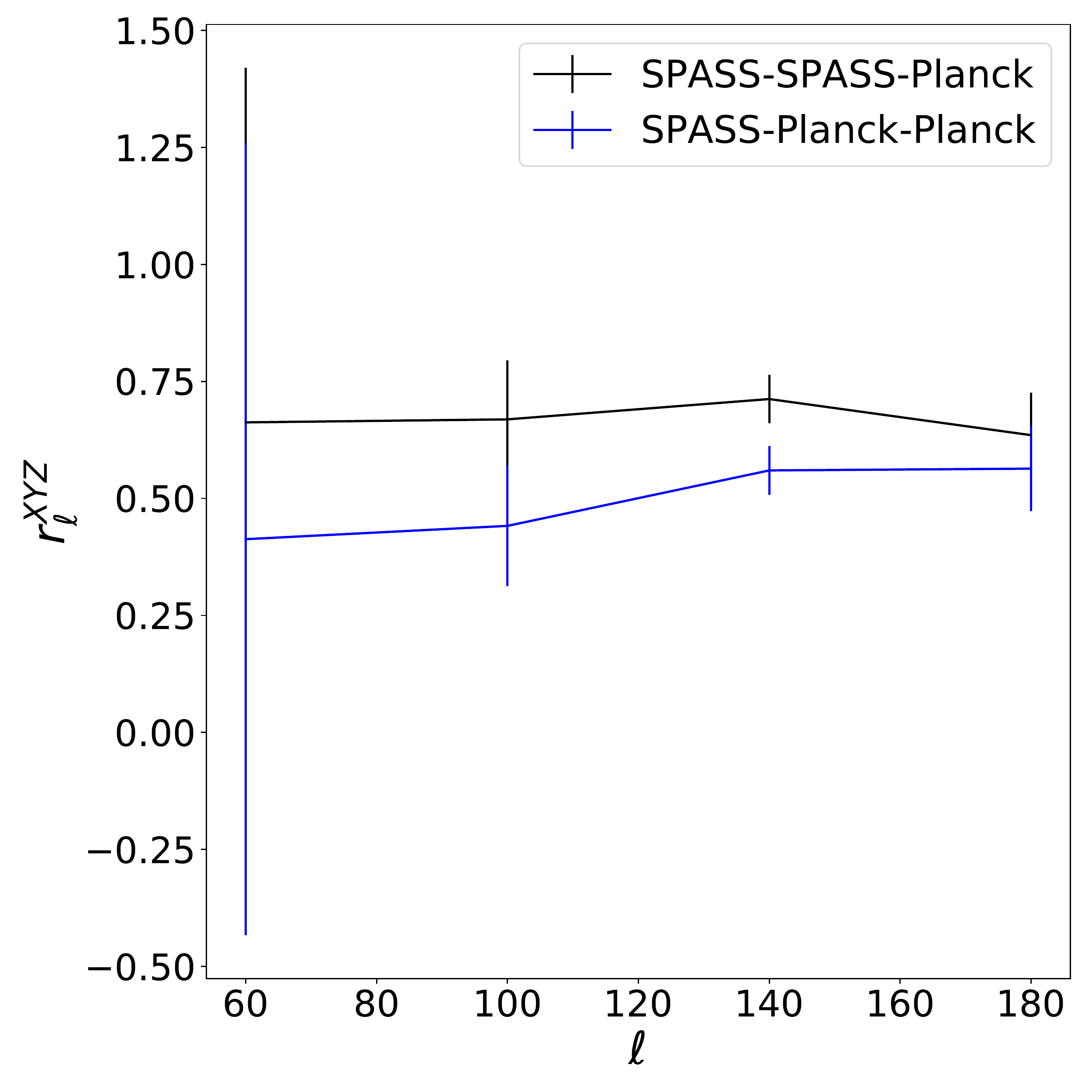}
   \caption{Correlation coefficient between the SPASS and Commander synchrotron temperature parity even bispectra.}
    \label{fig:spass_sync_cross}
\end{minipage}%
    \qquad
\begin{minipage}{.47\textwidth}
  \centering
    \includegraphics[width=.95\textwidth]{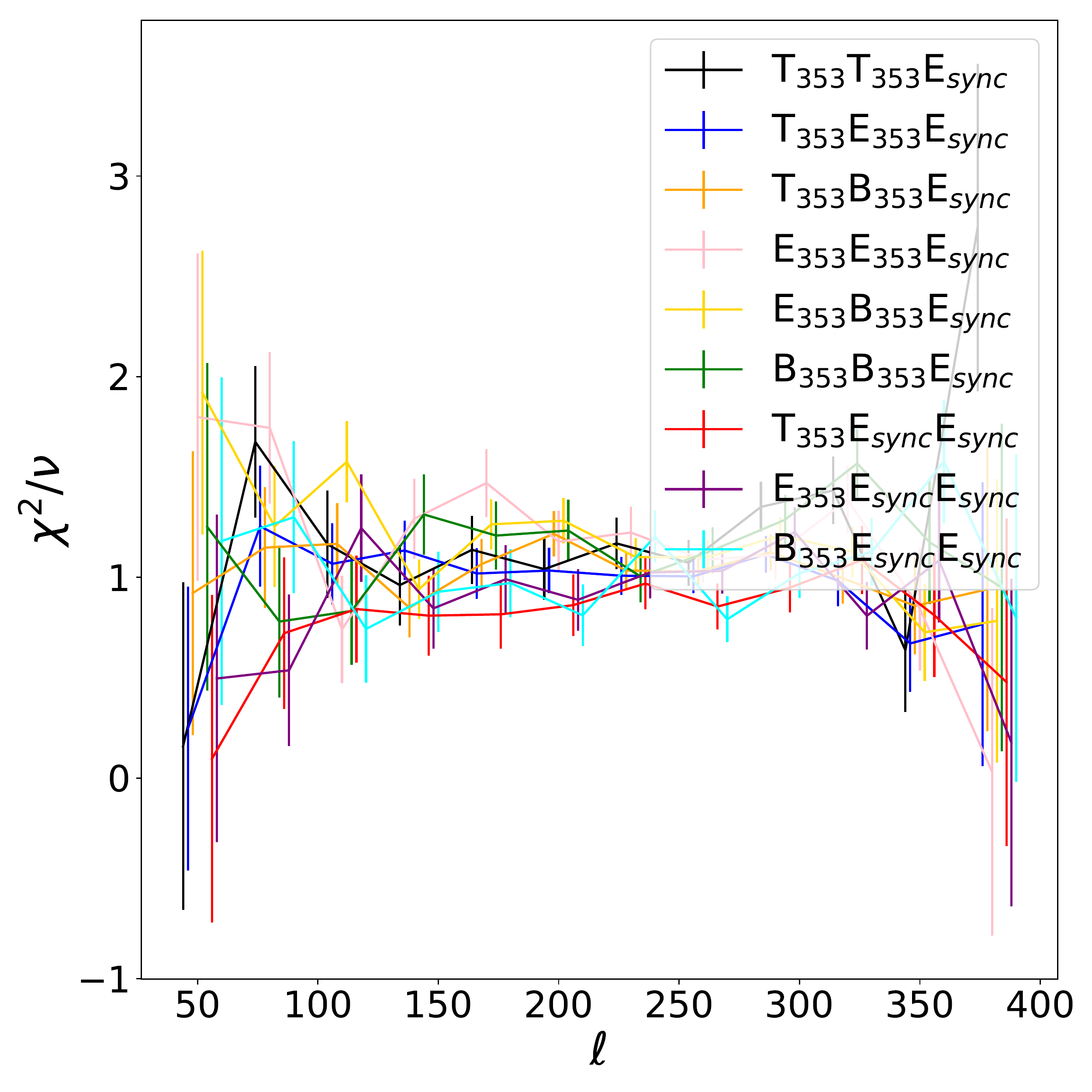}
   \caption{The 1-D bispectrum chi-squared for 353 GHz T, E and B maps cross the Commander synchrotron E mode for $60\%$ of the sky.}
    \label{fig:Dust_sync_1Dchisquared}
\end{minipage}
\end{figure}

\subsection{Synchrotron Bispectra}
\subsubsection{SPASS Synchrotron Bispectra}
To analyse the SPASS data we make four modifications to the pipeline described in Section \ref{sec:pipeline}. Firstly we alter the \textit{Planck} sky mask to be restricted to the region observed by SPASS. Secondly we introduce a mask to mask \textit{Centaurus A}, the \textit{ Large Magellanic Cloud} and \textit{Fornax A } as described in \citet{Krachmalnicoff2018}. Next we restrict our multipole range to $40 \le \ell \le 200$. This is motivated by the results from \citet{Krachmalnicoff2018} that the maps are dominated by point sources above $\ell>200$; this is also seen in our analyses. The high significance of the measurements of these point sources means they dominate the bispectrum and have significant higher order correlations, that contribute to the bispectrum errors via the connected four and six point functions.

The final change is to mask the point sources detected in the SPASS polarization maps. Without masking we see strong evidence of all the bispectra configurations, including parity violating bispectra such as BBB. These bispectra are all dominated by point sources. Uncorrelated polarised point sources should have no bispectra in configurations with an odd number of polarization fields, such as TTE or EEE, as the point sources should cancel each other. However, that is only true for averaged quantities, any single realisation can have a non-zero bispectrum, particularly when the source number counts has an extended high flux tail. For example, this arises if only a small number of bright polarized sources are dominating the bispectrum. In order to study the bispectrum of the diffuse galactic synchrotron emission we need to mask the point sources.  For the intensity map this is done using an iterative in-painting as discussed in Section \ref{sec:pipeline}. For polarization this in-painting procedure leads to mixing of E and B power and has the potential to mix E and B bispectra. To avoid this we just mask these point sources and then apodize them to prevent leakage of power. This apodization is performed with 3 degree Gaussian (as is used for the Galactic mask) and if many point sources were masked this would result a significant fraction of the map being masked. Thus we can only mask the brightest sources found in the \citet{Lamee2016}  polarization catalog and map. Even so with the apodized masking of the three extended sources and the bright sources we have a sizeable mask and potentially suppress non-Gaussianity. In future work we will explore more efficient methods to mask polarized point sources without apodization. In temperature we mask all sources with flux greater than 200mJy and which are detected in the maps or in the \citet{Meyers2017} catalog at more than $4\sigma$.

In Figures \ref{fig:spass1Dbispec} and \ref{fig:spass1Dchisq} we plot the parity even 1D bispectrum and chi-sqared for the SPASS maps. We find strong evidence for a temperature bispecturm which exhibits a weak scale dependence. This likely arises from a large scale bispectrum from the diffuse emission combined with emission from point sources on smaller scales, as is seen in the power-spectrum  \citep{Krachmalnicoff2018}. We see no evidence, after masking the brightest sources, for polarized bispectra or cross correlations.

\subsubsection{Planck synchrotron maps}
The \textit{Planck} synchrotron maps are provided at lower resolution than the dust maps and so we consider a reduced $\ell$ range ($\ell \leq 400$) for this analysis. In Figure \ref{fig:sync2Dbispec_TTT} and \ref{fig:sync2Dchisquared_TTT} we present the 2-D parity even binned bispectrum and 2-D chi-squared of the \textit{Planck} Commander temperature synchrotron map. We find that there is strong evidence for the TTT bispectrum. As was discussed in Section \ref{sec:signalInhom}, we find that the estimator variance for configurations involving multiple synchrotron temperature maps is larger than expected. This is attributed to the large non-Gaussianity and inhomogeneity of the synchrotron emission and the origins of the extra variance are discuss in Appendix \ref{app:EstError}. In light of this, the chi-squared shown in Figure \ref{fig:sync2Dchisquared_TTT} can be taken as measure of how non-Gaussian the synchrotron is, but not a measure of the significance of the detection of the bispectrum. To avoid this increased variance we only considered polarization configurations involving one temperature leg and find no evidence for bispectrum involving the synchrotron E and B maps. We also search for, and find no evidence for any odd-parity non-Gaussianity.  Examining the shape of the synchrotron temperature bispectrum we see that it has most significance in the equilateral limit. As was noted in \citet{Jung2018} and \citet{planck2014-a12}, this is caused by the contribution of unresolved point sources and is evidenced by our examination of the higher resolution SPASS maps. 

\subsubsection{SPASS - Commander cross correlations}

Given the two measures of the synchrotron emission we can investigate the evolution of the bispectrum as a function of frequency. Using the bispectrum correlation coefficient defined in Eq. \ref{eq:bispectrumCorCoef} we investigate the degree of correlation between the SPASS and Commander synchrotron maps. To do this we apply the SPASS mask to the \textit{Planck} synchrotron map and use the same point source masking level of 1Jy - dictated by the \textit{Planck} resolution. The resulting correlation coefficient is plotted in Figure \ref{fig:spass_sync_cross}. We see that these two maps are relatively strongly correlated at the bispectrum level. The synchrotron bispectrum has contributions from the diffuse emission, which has a variable spectral index \citep[see e.g.][]{Guzman2011}, and from radio point sources, which can be broadly divided into two populations -flat spectrum and steep spectrum radio sources \citep[see e.g.][]{Massardi2010}. The different spectral indices of these components, combined with the spatial variations, explains the imperfect correlation between these bispectra and is a possible explanation for the difference seen in the two bispectra configurations.

\subsection{Dust - Synchrotron Cross Bispectra}

In \citet{Choi2015} they studied the correlation between the dust and synchrotron polarisation signals. Physically this correlation arises as both effects are influenced by the galactic magnetic field. Analogously to that we investigated the dust-synchrotron bispectrum using the 353 GHz map and either the Commander synchrotron map or the SPASS maps. We find that the dust-synchrotron bispectrum is consistent with zero for all of the masks. The 1-D chi-squared for the parity even E mode bispectra are shown in Figure \ref{fig:Dust_sync_1Dchisquared}.   We find no evidence for a 353-synchrotron bispectrum.  We note that revisiting this correlation with full-sky, SPASS quality-maps at frequencies used in CMB analyses would be interesting. Though, as was described above, a new method for treating polarized point sources will be required.

\begin{figure}
\subfloat[353 GHz T and B map cross lower frequency B map parity-even bispectrum ]{
  \centering
    \includegraphics[width=.47\textwidth]{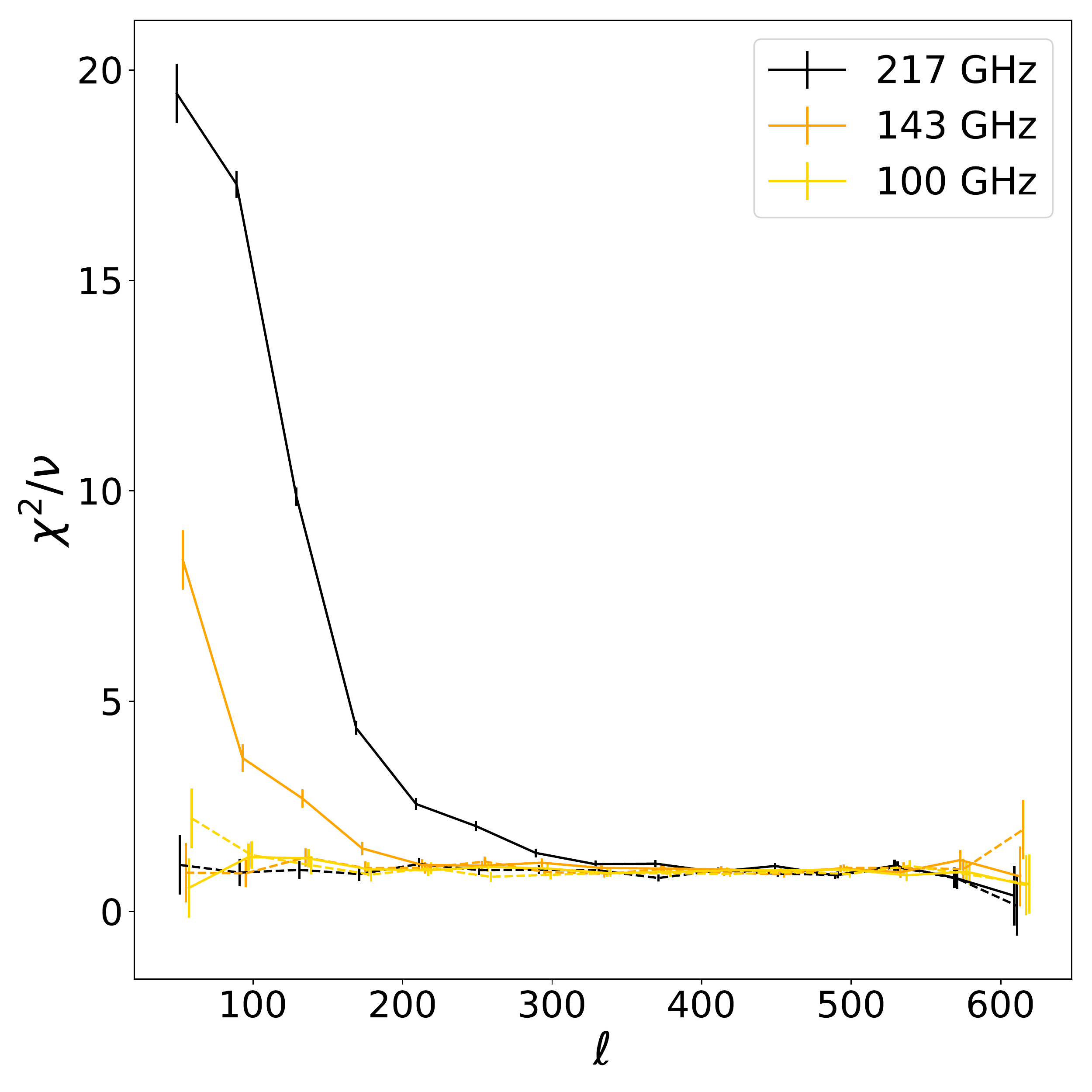}
    \label{fig:SingleFreqTest_gm4_even}}
    \qquad
\subfloat[Two 353 GHz E maps cross lower frequency B map parity-odd bispectrum ]{
    \centering
    \includegraphics[width=.47\textwidth]{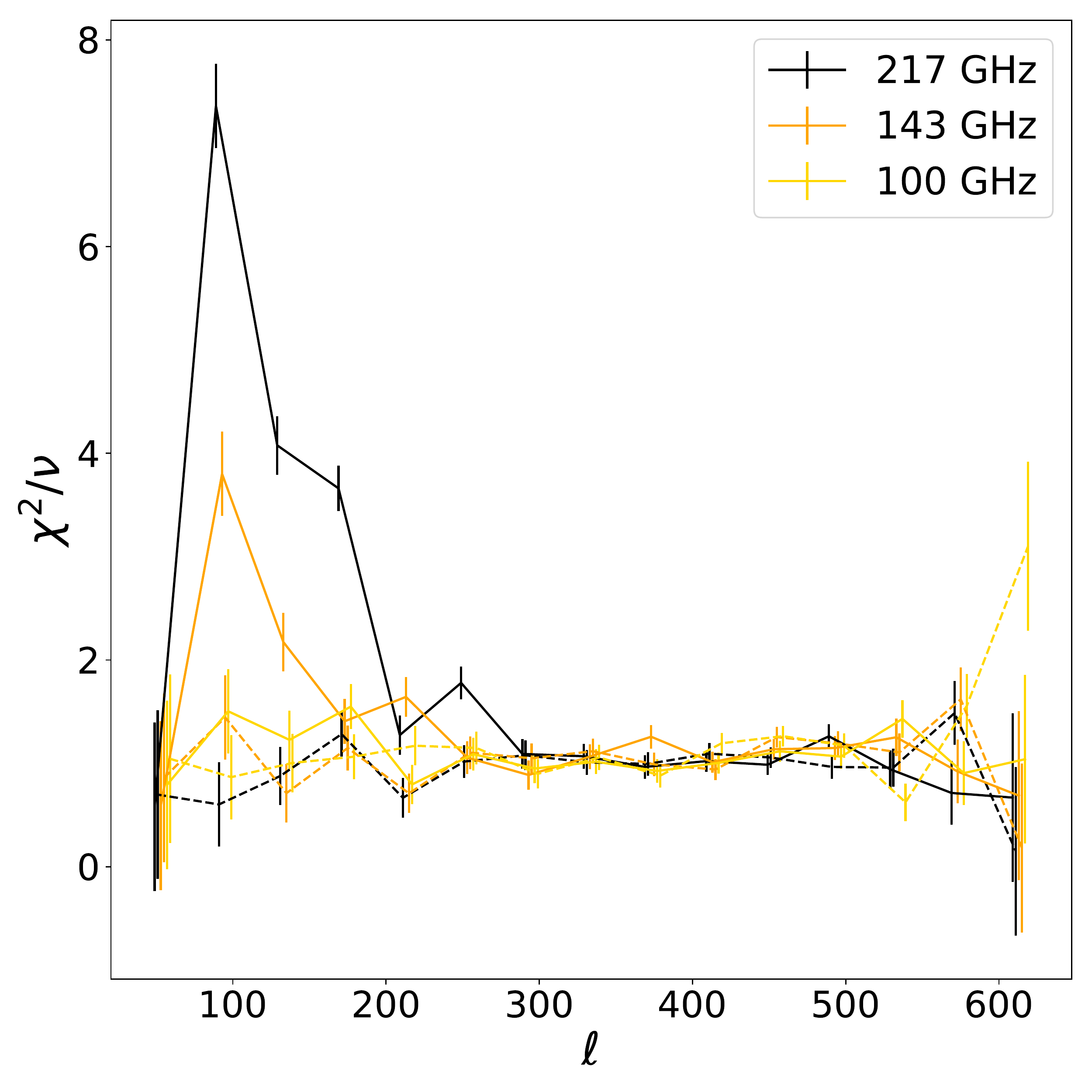}
    \label{fig:SingleFreqTest_gm4_odd}}
    \caption{The 1-D chi-squared for the cross bispectra between the 353 GHz maps and the Planck 100,143 and 217 GHz B mode map for  $60\%$ of the sky. The solid lines are the signal and the dashed lines are noise maps constructed from half-ring maps.}
\end{figure}

\begin{figure}
\subfloat[Parity Even]{
  \centering
    \includegraphics[width=.47\textwidth]{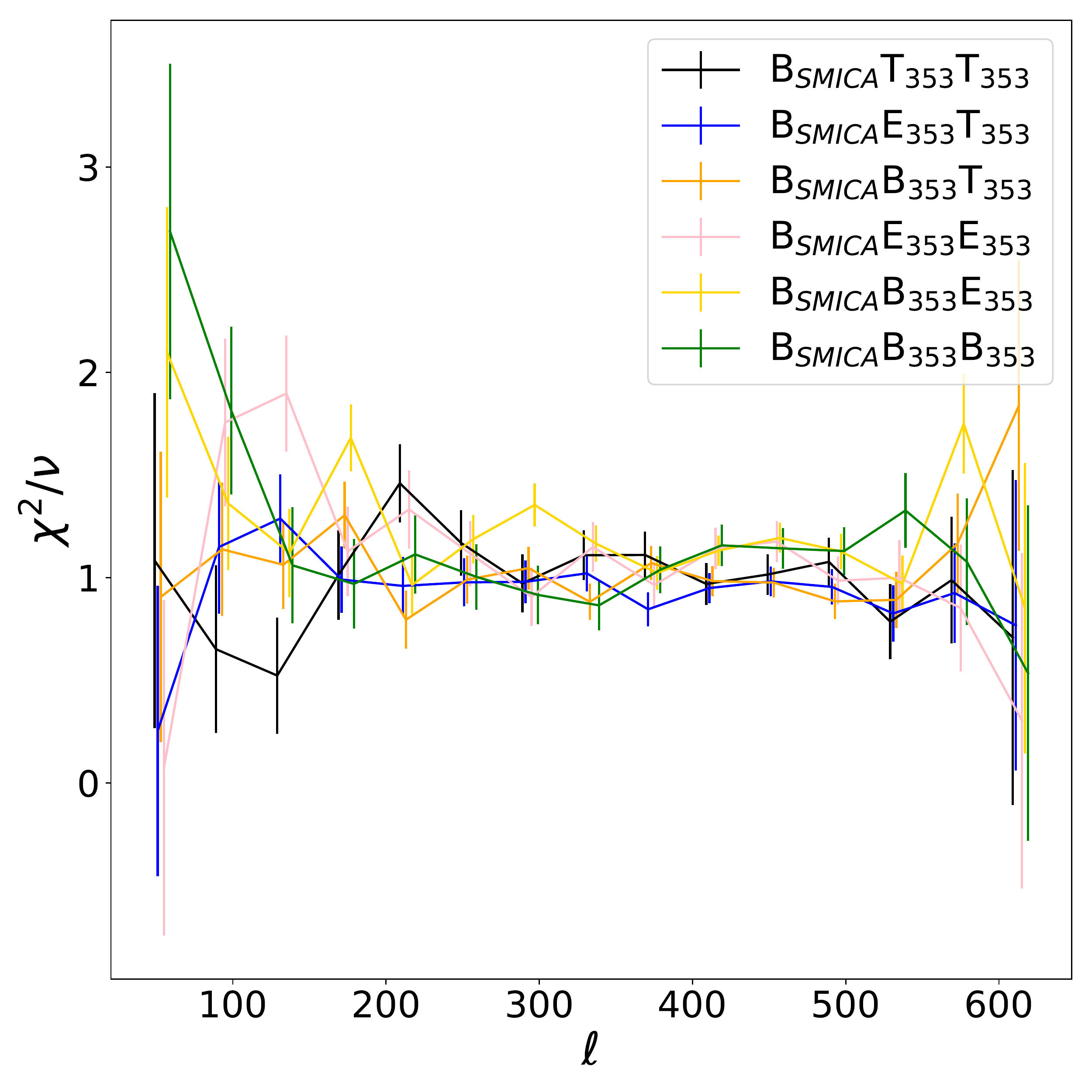}
    \label{fig:dust_smica_1Dchisquared}}
    \qquad
\subfloat[Parity Odd]{
    \centering
    \includegraphics[width=.47\textwidth]{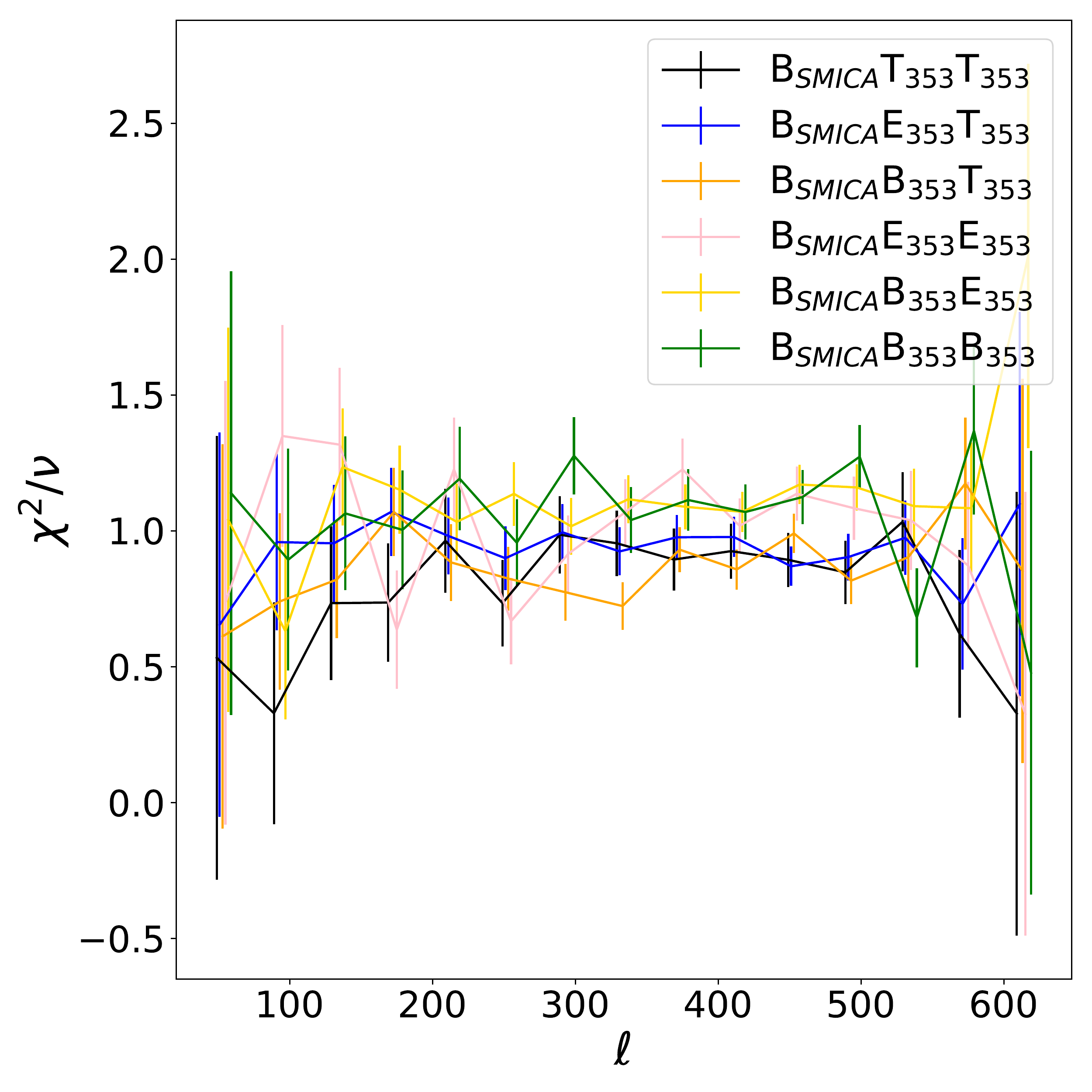}
    \label{fig:dust_smica_1Dchisquared_odd}}
    \caption{The 1-D chi-squared for the cross bispectra between the 353 GHz T, E and B map and the SMICA B mode map for $60\%$ of the sky.} 
\end{figure}

\begin{figure}
\subfloat[Parity even cross-bispectrum between the 353 GHz B and T maps and the component separated B map ]{
  \centering
    \includegraphics[width=.47\textwidth]{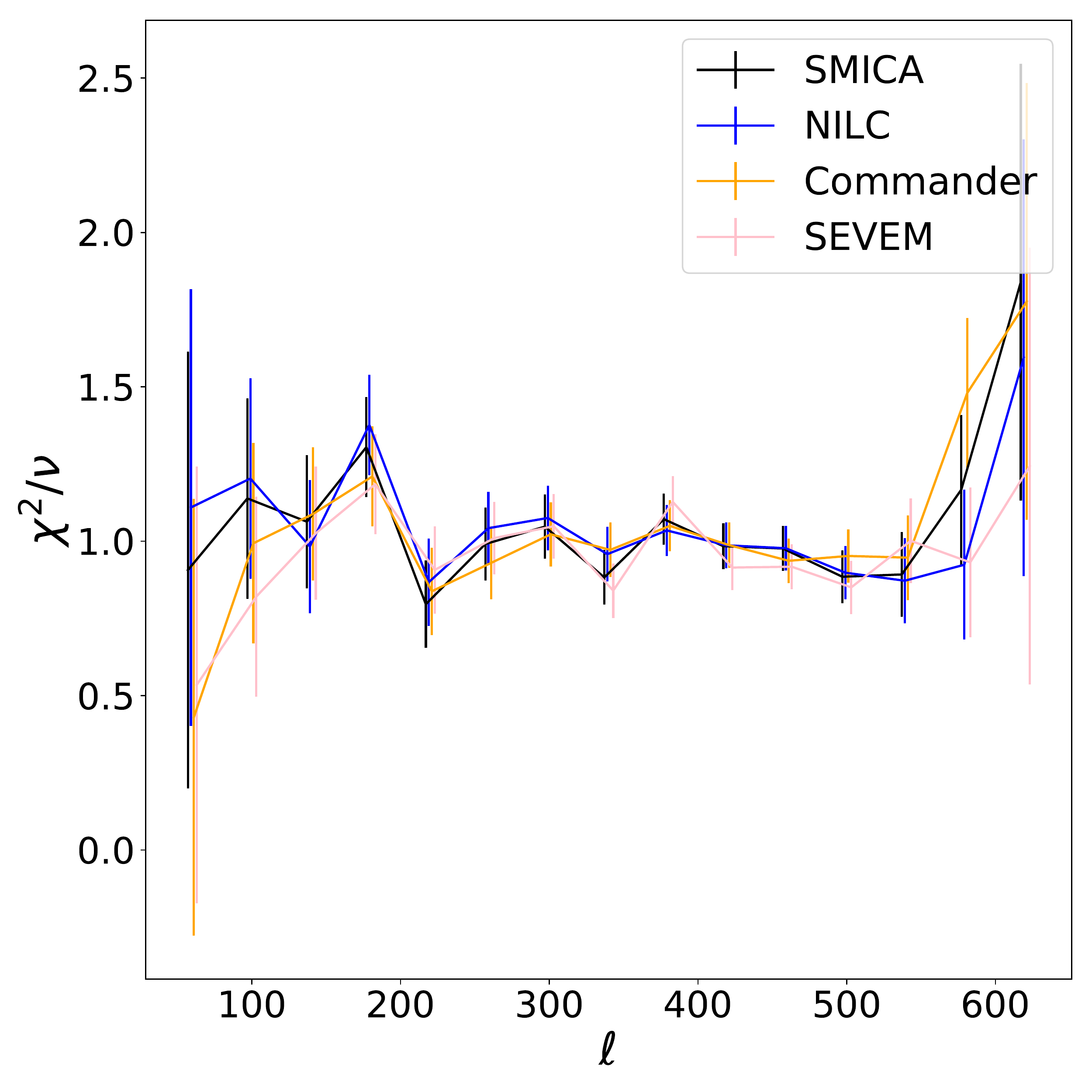}
    \label{fig:multiMethod}}
    \qquad
\subfloat[Parity odd cross-bispectrum between two 353 GHz E maps and the component separated B map ]{
    \centering
    \includegraphics[width=.47\textwidth]{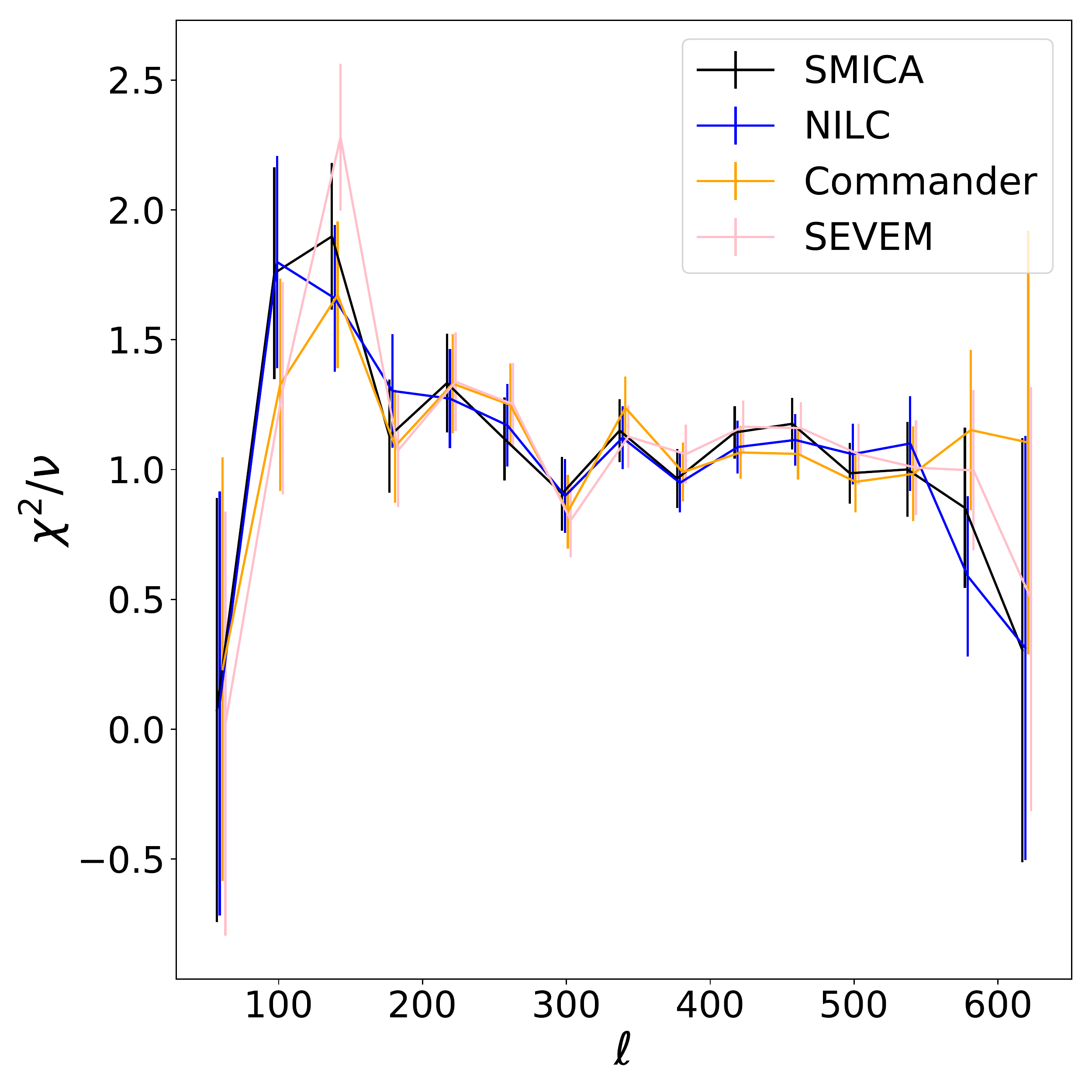}
    \label{fig:multiMethod_odd}}
    \caption{The 1-D chi-squared for the cross bispectra between 353 GHz B and E maps and the B mode map from the different cleaning maps for $60\%$ of the sky} 
\end{figure}

\begin{figure}
\subfloat[SMICA T map]{
  \centering
    \includegraphics[width=.47\textwidth]{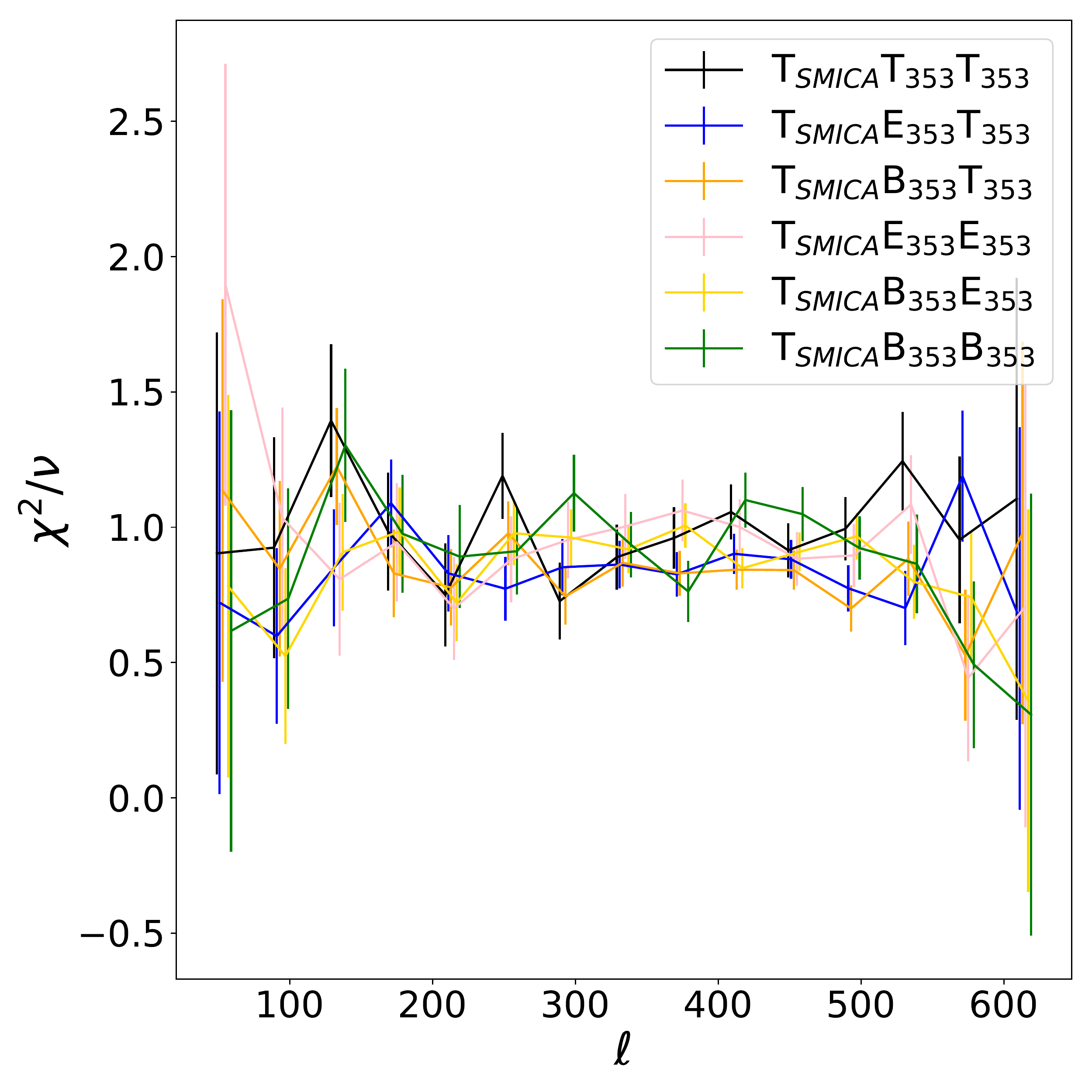}
    \label{fig:dust_smica_TEBT_comp}}
    \qquad
\subfloat[SMICA E map ]{
    \centering
    \includegraphics[width=.47\textwidth]{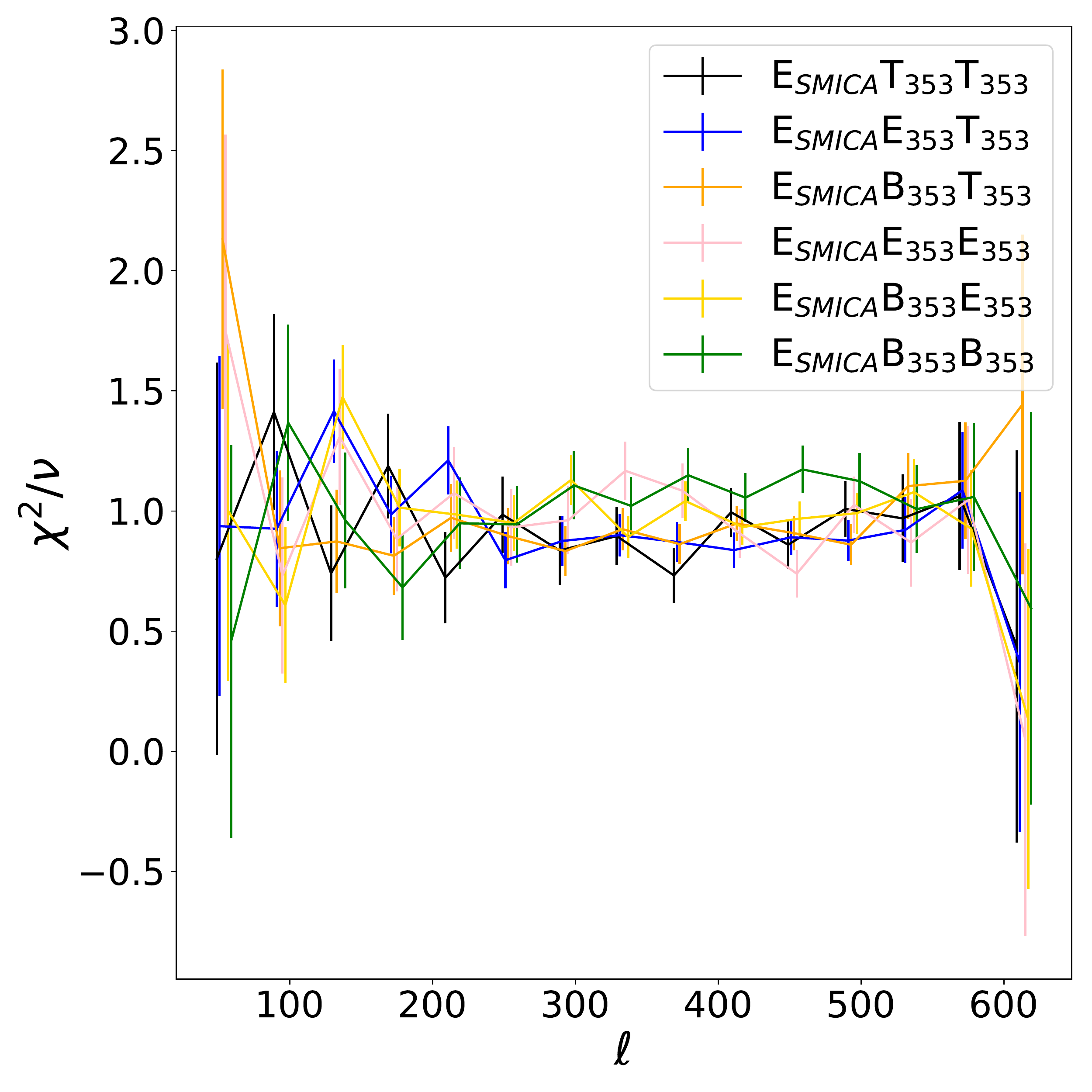}
    \label{fig:dust_smica_TEBE_comp}}
    \caption{The 1-D chi-squared for parity even cross bispectra between 353 GHz T, E and B maps and SMICA  T and E maps on $60\%$ of the sky.} 
\end{figure}

\subsection{Cleaned Map Bispectra}\label{sec:cleanedBispectra}

Motivated by the signals seen above, we investigated whether bispectra can be used to search for residual foregrounds. Before considering foreground cleaned maps it is necessary to see if we see a bispectrum in the \textit{Planck} single frequency maps. The bispectrum in these maps will be significantly reduced when compared to the results in Section \ref{sec:dustBispectrum}. In Figures \ref{fig:SingleFreqTest_gm4_even} and \ref{fig:SingleFreqTest_gm4_odd} we examined the 1-D chi-squared for cross bispectra  between the 353 GHz T, E and B and the B mode maps from the 100 GHz, 143 GHz and 217 GHz channels. Using the strong foregrounds in the 353 GHz maps we look for correlations in the B map. We find strong evidence for the parity even BTB bispectrum and odd parity EEB bispectrum at 143 and 217 GHz, with hints of a signal in the 100 GHz map.  

The signal seen in Figures \ref{fig:SingleFreqTest_gm4_even} and \ref{fig:SingleFreqTest_gm4_odd}  means that the bispectrum could be used to test for residual foregrounds. In Figures \ref{fig:dust_smica_1Dchisquared} and \ref{fig:dust_smica_1Dchisquared_odd} we plot the cross bispectra between the 353 GHz T,E and B maps with the SMICA B mode map. As can be seen there is no strong evidence of a residual bispectrum. As discussed in Appendix \ref{app:EstError} we have neglected the linear term in this analysis and this results in an underestimation of the variance of $~10\%$ and so our error bars are slightly underestimated. In Figures \ref{fig:multiMethod} and  \ref{fig:multiMethod_odd} we explore the residual correlations for the four different component separation methods. We find high levels of consistency between the foreground cleaning methods, with none of the methods showing evidence of residual foregrounds. Similarly we can test for residuals foregrounds in the temperature and E mode polarisation maps and the results are shown in Figures \ref{fig:dust_smica_TEBT_comp} and \ref{fig:dust_smica_TEBE_comp}. We see no evidence for residual foregrounds.

\section{Discussion and Conclusions}\label{sec:conclusions}
The results presented in Section \ref{sec:foregroundBispectra} show that there are large levels of non-Gaussianity in the galactic foregrounds, particularly the galactic dust. It also suggests that these bispectra could be useful for cross checking residual foregrounds in cleaned temperature maps. Whilst we found no strong evidence for residual foreground non-Gaussianity in the component separated maps, the work presented here is the first step. To further investigate the consistency with the Gaussian expectation we need to more accurately model the signal variance. Ideally we would also like to push to larger sky areas and to lower $\ell$, however further work is required to understand the excess variance described in detail in Appendix \ref{app:EstError}.

The strong dust and synchrotron bispectra observed here need to be accounted for when searching for primordial non-Gaussianity. This is particularly important as these bispectra peak in the squeezed configurations and so could bias measurements of local type non-Gaussianity. The large polarization signals means that, unlike extragalactic terms discuss in \citet{Hill2018}, these biases cannot be avoided by solely using polarization data. The lack of evidence for bispectra in the cleaned maps means that the \textit{Planck} non-Gaussianity results \citep{planck2014-a19} should be unaffected by residual Galactic foregrounds (which, for the temperature foregrounds, has been thoroughly examined in \citet{Jung2018}). In future searches for scalar-tensor non-Gaussianity \citep{Meerburg2016} these contaminants will need to be tightly controlled and the tools discussed here will be very useful for validating those results.

In this paper we have focused on characterising the bispectra signals and discussing the utility of the bispectrum to constrain residual foregrounds, however there is far more information available. \citet{planck2014-XIX} examined the physical origins of the polarized dust signal. Bispectrum measurements can also be used to study the physical properties of the dust and complement the results of  \citet{planck2014-XIX}. In particular,  in \citet{Burkhart2009} they demonstrate first using simulations, and then in \citet{Burkhart2010} using measurements of the Small Magellanic Cloud, that column density bispectra can be used to constrain Alfv\'en and sonic Mach number.  Thus in the future bispectrum measurements could potentially be used to constrain the magnetic and kinetic properties of the interstellar medium.

\section{Acknowledgements}
The authors would like to thank Anthony Challinor, Adri Duivenvoorden, Joanna Dunkley, Daan Meerburg and Blake Sherwin for useful discussions. This work has made use of S-band Polarisation All Sky Survey (S-PASS) data. The authors would also like to thank the referee for their useful comments and great suggestions.

\appendix
\section{An examination of the Bispectrum Estimator Variance}\label{app:EstError}

In this Appendix we explore more the comments made in Section \ref{sec:signalInhom}. To begin with we review the contributions to the estimator variances before examining the larger than naively expected observed variance. We then discuss our approach to mitigate this problem, before finally presenting an alternative method to alleviate the problem.
\subsection{Review of Estimator Variance}

In this section we limit the discussion to the parity even estimator, for conciseness, however this discussion generalises trivially to the parity odd estimator. The formulae for the estimator variance given in Section \ref{sec:binnedEstimators} are valid either for a full sky measurement of a homogenous field or for a measurement of an inhomogeneous field with the linear used in the estimator \citep{Babich2005,Komatsu2002}. In the case when the linear term is neglected, or cannot be calculated, there are the additional terms, which have the schematic form
\begin{align}\label{eq:inhomContr}
V^{\mathrm{additional}}_{i,j,k}\propto & \prod\limits_{1\leq a \leq 6} \,\sum\limits_{m_a}\, \sum\limits_{\ell_{i_a}<\ell_{a}\leq\ell_{i_a+1}}\mathcal{G}^{m_1,m_2,m_3}_{\ell_1,\ell_2,\ell_3} \mathcal{G}^{m_4,m_5,m_6}_{\ell_4,\ell_5,\ell_6} \left[ \langle a_{\ell_1,m_1}a_{\ell_2,m_2}\rangle \langle a_{\ell_3,m_3}a_{\ell_4,m_4}\rangle \langle a_{\ell_5,m_5}a_{\ell_6,m_6}\rangle \right. \nonumber \\ & \left. + \text{ cyclic permutations}\right].
 \end{align}
 For the homogeneous and isotropic case  $\langle a_{\ell_1,m_1}a_{\ell_2,-m_2}\rangle =  (-1)^{m_2}  C_{\ell_1} \delta_{\ell_1,\ell_2} \delta_{m_1,m_2}$, the following Wigner 3j property
 \begin{align}
 \sum\limits_m (-1)^{\ell-m}  \begin{pmatrix}
    \ell & \ell & \ell' \\
    m & -m & m'
  \end{pmatrix} = \sqrt{2 \ell+1} \delta_{\ell',0}\delta_{m',0} \, ,
\end{align} 
means that these terms are proportional to the monopole, which is set to zero. For anisotropic fields it is no longer the case that $ \langle a_{\ell_1,m_1}a_{\ell_2,-m_2}\rangle = (-1)^{m_2} C_{\ell_1} \delta_{\ell_1,\ell_2} \delta_{m_1,m_2}$, there will be off diagonal terms. These new contributions to the variance can be  significant \citep{Bucher2016}.
 
Secondly, when the non-Gaussianty is strong there can be non-Gaussian contributions to the covariance matrix. So far the contributions we have considered are the Gaussian (or disconnected) components of the variance, which can be calculated via Wick's Theorem. However for non-Gaussian fields there can also be other components (the connected or non-Gaussian contributions), see for example \citet{kayo2013} for an overview of the non-Gaussian contributions to the bispectrum covariance matrix. These have the schematic form of
 \begin{align}\label{eq:nonGausContr}
 V^{\mathrm{non-Gaussian}}_{i,j,k}\propto & \prod\limits_{1\leq a \leq 6} \, \sum\limits_{m_a}\, \sum\limits_{\ell_{i_a}<\ell_{a}\leq\ell_{i_a+1}}\mathcal{G}^{m_1,m_2,m_3}_{\ell_1,\ell_2,\ell_3} \mathcal{G}^{m_4,m_5,m_6}_{\ell_4,\ell_5,\ell_6} \left[ C_{\ell_1,\ell_4} T_{\ell_2,\ell_3,\ell_5,\ell_6} + B_{\ell_1,\ell_2,\ell_4} B_{\ell_3,\ell_5,\ell_6} + S_{\ell_1,\ell_2,\ell_3,\ell_4,\ell_5,\ell_6}  \right. \nonumber \\ & \left. + \text{ cyclic permutations}\right].
 \end{align}
 where $T_{\ell_a,\ell_b,\ell_c,\ell_d} $ is the connected trispectrum, $B_{\ell_a,\ell_b,\ell_c}$  is the bispectrum and $S_{\ell_a,\ell_b,\ell_c,\ell_d,\ell_e,\ell_f}$ is the connected six point function. For a weakly non-Gaussian field these will be subdominant to the connected terms, but that is not necessarily the case for a strongly non-Gaussian field.
 
 \subsection{Application to dust bispectra}
  \begin{figure}
\begin{minipage}{.47\textwidth}
  \centering
    \includegraphics[width=.95\textwidth]{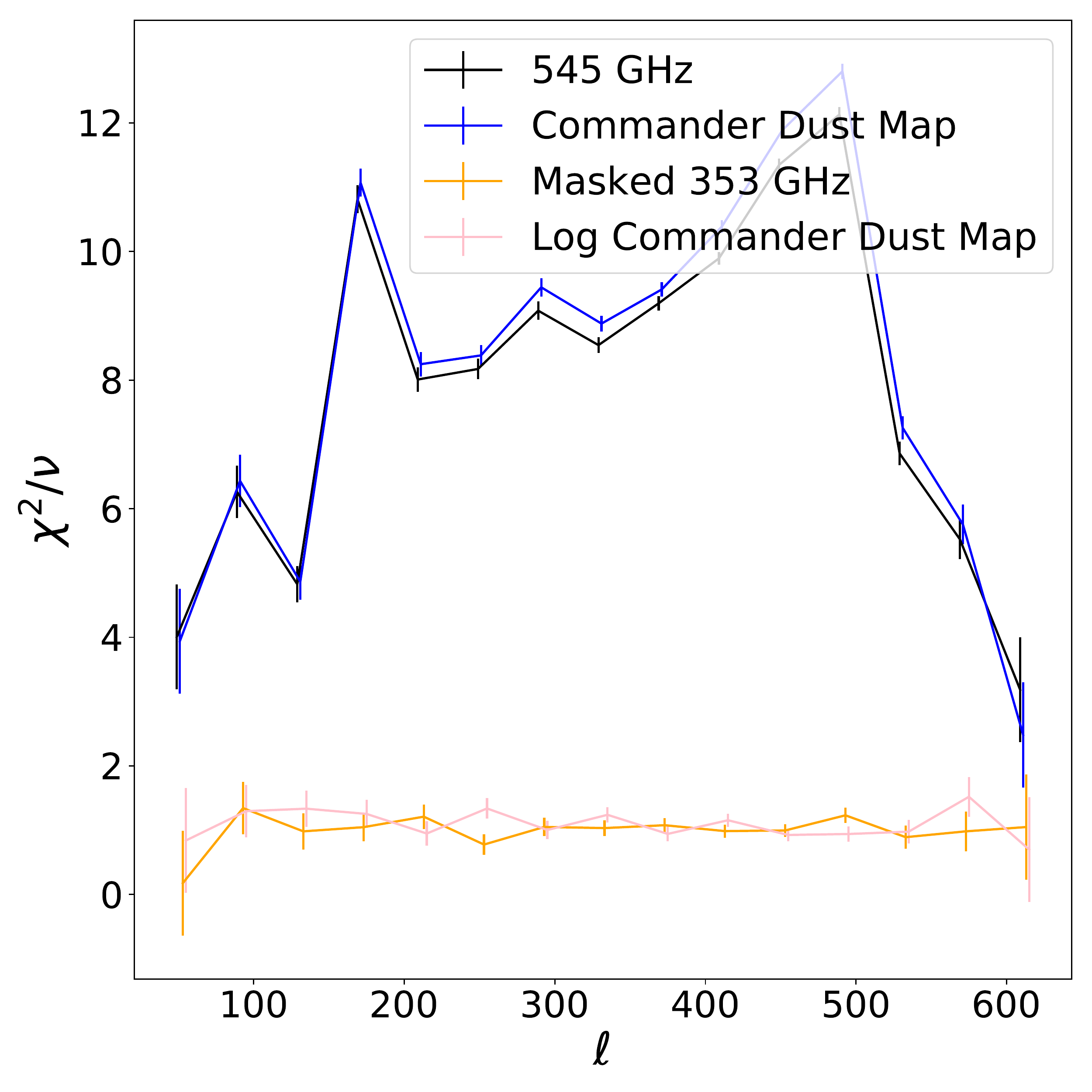}
\caption{The 1D chi-squared for parity even measurements between a simulated Planck 353 GHz E mode noise map and two temperature maps. We compare the results from four different versions of the temperature map: the Planck 545 GHz map, the Commander dust map, the masked 353 GHz map used in this analysis and the logarithm of the Commander temperature map.}
    \label{fig:NoiseNullTests}

\end{minipage}%
    \qquad
\begin{minipage}{.47\textwidth}
    \centering
    \includegraphics[width=.95\textwidth]{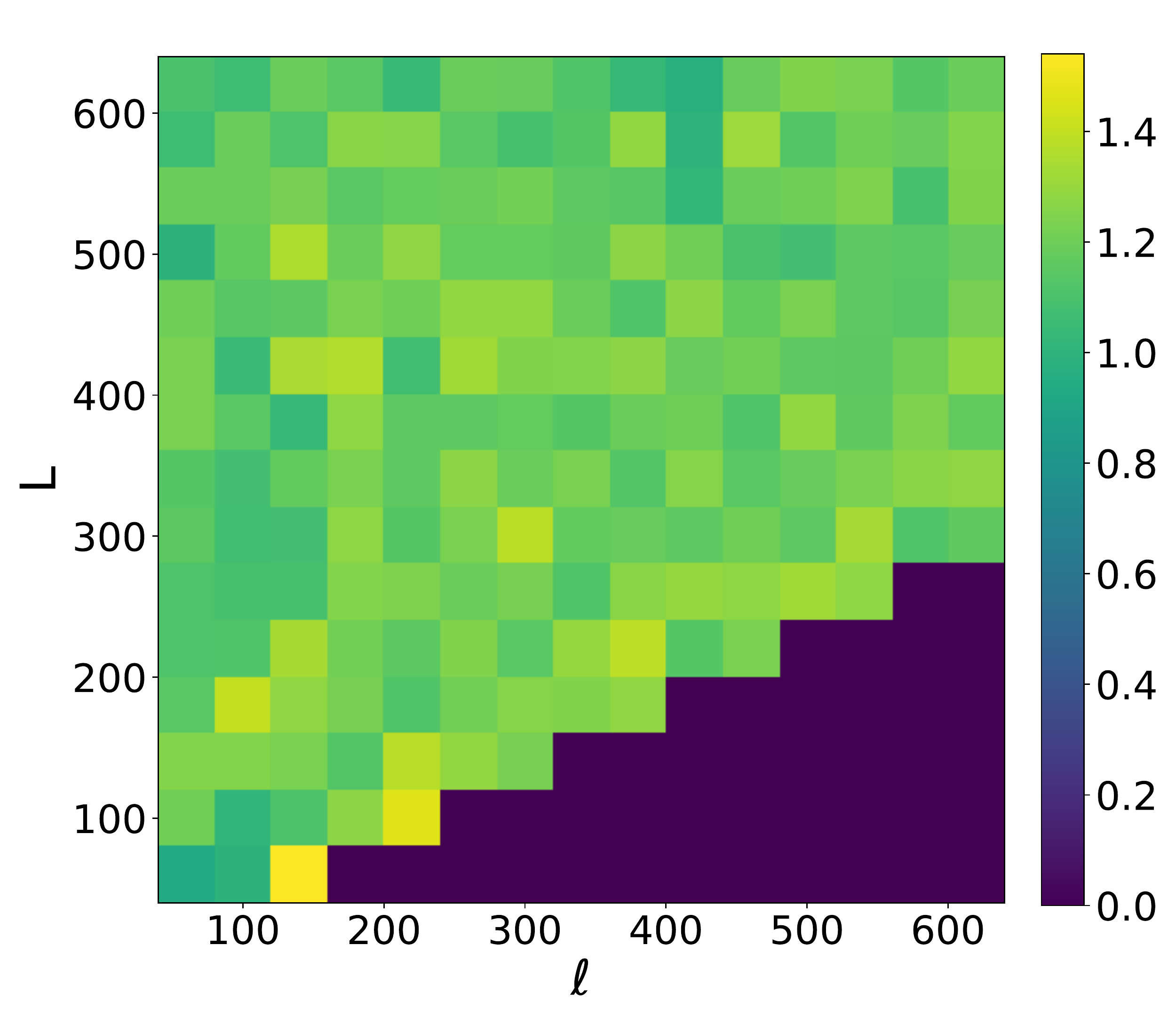}
\caption{The average BBT 2D chi-squared,$ {\chi^2}^{2D:\, B,(T,B)}_{\ell,L}$, of 30 Planck SMICA noise simulations using $60\%$ of the sky.}
    \label{fig:chisquaredAvSMICA}
\end{minipage}%
\end{figure}
 
\begin{figure}
     \centering
    \includegraphics[width=.50\textwidth]{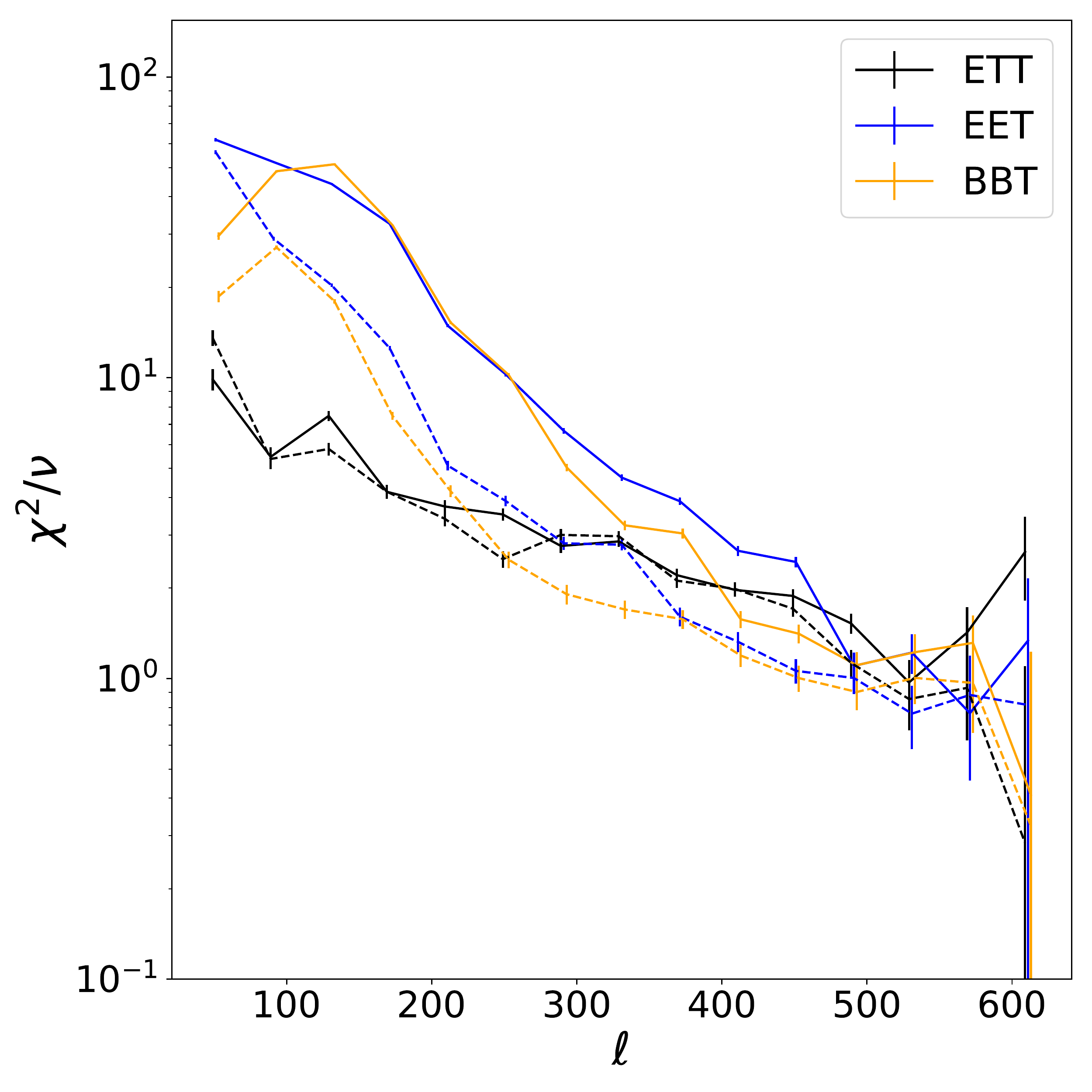}
\caption{The 1D chi-squared for the Commander dust logT, E and B maps, where we use the log of the temperature map. The dashed lines are the results using the 353 GHz T,E and B maps with the mask described in Section \ref{sec:pipeline}.}
    \label{fig:chisquared_logT_vs_353}
\end{figure}
 
 In Figure \ref{fig:NoiseNullTests} we plot the results from a simple null test. We compute the bispectrum, and the 1D chi-squared, between two 545 GHz temperature data maps and a simulated Planck 353 GHz noise map. As there is no correlation between the simulated map and the data we expect a vanishing bispectrum and a 1D chi-squared consistent with 1. For the 545 GHz map we do not see this, instead we see a non zero signal. The covariance used to calculate the 1D chi-squared was obtained from Gaussian simulations and matches the analytical value calculated from Eq. \ref{eq:parityEvenVar}. As there is no bispectrum signal, the failure of this test implies that the variance is larger than is given by Eq. \ref{eq:parityEvenVar}.  In Figure \ref{fig:NoiseNullTests} we also plot the result from using the Commander dust temperature map and find that it also fails this test. There are two possible causes of this, either the linear term is insufficient or there are non-Gaussian contributions to the covariance matrix.
 
 To calculate the linear term we need to be able to evaluate the off-diagonal power spectrum term,  $\langle a_{\ell_1,m_1}a_{\ell_2,m_2}\rangle.$
 For even modest $\ell$ this matrix is very large and impractical to fully calculate. Instead the linear term is typically calculated through simulations as described in \citet{Smith2011} (or \citet{Bucher2016} for the binned equivalent). In analyses of extragalactic sources the main inhomogeneous terms arise from instrument noise and galactic sky cuts, the extragalactic signals tend to be homogeneous. As these sources of inhomogeneity can be easily simulated the linear term can be accurately calculated. In our case we have noise simulations of the 353 GHz maps and simulations of the 545 GHz noise. However the galactic signal itself is very anisotropic and the anisotropy of this signal is not included in our calculations as it is very difficult to simulate. This means the residual signal seen in Figure \ref{fig:NoiseNullTests}  could arise as we are missing the terms given in Eq. \ref{eq:inhomContr} for the dust signal itself. 
 
 There is a second possible explanation: as can be seen in Figure \ref{fig:dust1Dchisquared} the dust is very non-Gaussian. It would not be surprising then for there to be significant non zero higher point correlation functions. These would result in extra contributions to the variance of the form given in Eq. \ref{eq:nonGausContr}. It is very challenging to calculate these contributions to the covariance matrix.
 
 \subsection{Mitigation strategies}

 The approach we used to overcome this was to use the Planck 353 GHz maps and to mask the data. The Planck 353 GHz map has a slightly lower signal to noise than the 545 GHz or Commander dust map and so the contribution of the non-Gaussian variance terms should be suppressed. By applying a real space mask (as described in Section \ref{sec:pipeline}) and an harmonic space cut of low $\ell$ modes we reduce the inhomogeneity and non-Gaussianity. In Figure \ref{fig:NoiseNullTests} we plot the same null test as described above for this masked 353 GHz map, and we see that it passes this (and our other) null tests. In fact with these cuts (particularly the low $\ell$ cut) we find that the linear term has only a very small effect on our estimator and so we neglect it in this analysis. This can be seen in Figure \ref{fig:chisquaredAvSMICA} where we plot the average 2D chi-squared for 30 simulations of Planck SMICA noise without including the linear term. We find that mean is $\sim10\%$ higher than the expectation (of 1), which corresponds to a $\sim 10\%$ underestimation of the variance and is sufficiently accurate for this work. We also see that the variance of all the configurations seems to be equally increased and thus we do not expect that neglecting the linear term will bias any of our results.
 
As masking the signal is counter productive (as we wish to study the signal), we briefly note that an alternative method that will be explored more in future work. Instead of masking the signal we take an approach to reduce the inhomogeneity and higher point functions. As the Commander dust map is positive definite this can be achieved by taking the log of this map. In Figure  \ref{fig:NoiseNullTests}  we plot the same null test as described above for this transformed map, we find that it also passes this null test.  In Figure \ref{fig:chisquared_logT_vs_353} we plot the 1D chi-squared from correlations between the log Commander dust map with the Commander dust E and B maps. We see strong evidence of non-Gaussianity, at a similar level to that seen in using the masked 353 GHz map, suggesting that the log operation has not destroyed all the correlations and retains a high level of information. In future work we will further examine these correlations.

\bibliographystyle{aasjournal}
\bibliography{Planck_bib,foregrounds}
\end{document}